\documentclass[aps, prd, showkeys,superscriptaddress, nofootinbib, floatfix]{revtex4-2}

\usepackage{amsmath}
\usepackage{graphicx}
\usepackage{dcolumn}
\usepackage{bm}

\usepackage{amssymb}
\usepackage{dsfont}
\usepackage{url}
\usepackage{caption,subcaption}
\usepackage[colorlinks=true,allcolors=blue]{hyperref}

\begin{document}

\title{Hydrodynamic theories for a system of weakly 
self-interacting classical ultra-relativistic scalar particles: microscopic derivations and attractors}


\author{Gabriel S. Rocha}
\email{gabrielsr@id.uff.br}
\affiliation{Instituto de F\'{\i}sica, Universidade Federal Fluminense, Niter\'{o}i, Rio de Janeiro, 24210-346,
Brazil}
\affiliation{Institut für Theoretische Physik, Goethe-Universität, Max-von-Laue-Str.~1,  Frankfurt am Main, 60438, Germany}
\author{Caio V. P. de Brito}
\email{caio\_brito@id.uff.br}
\author{Gabriel S. Denicol}
\email{gsdenicol@id.uff.br}
\affiliation{Instituto de F\'{\i}sica, Universidade Federal Fluminense, Niter\'{o}i, Rio de Janeiro, 24210-346, Brazil}

\begin{abstract}
We derive and investigate several hydrodynamic formalisms that emerge from a system of classical, ultra-relativistic scalar particles self-interacting via a quartic potential. The specific form of the total cross-section enables the analytical computation of all transport coefficients that appear in Navier-Stokes (NS), Bemfica-Disconzi-Noronha-Kovtun (BDNK), and second-order transient hydrodynamic theories. We solve all these formalisms in a Bjorken flow scenario and show that NS and BDNK theories display unphysical features when gradients become sufficiently large. This implies that these hydrodynamic approaches may not be suitable to describe the early stages of heavy ion collisions.   
\end{abstract}

\maketitle

\section{Introduction}

Relativistic hydrodynamics is an effective theory constructed to describe the long-distance, long-time dynamics of macroscopic systems. It is widely applied in high energy nuclear physics, being employed to describe the hot nuclear matter produced in ultra-relativistic heavy ion collisions \cite{gale2013hydrodynamic, heinz2013collective, florkowski2018new} and the dense nuclear matter existing in the core of compact stars \cite{rezzolla2013relativistic,chabanov:21-general,Fujibayashi:2017puw}. Naturally, the main goal of these endeavours is to study how the properties of nuclear matter change when extreme temperatures and pressures are achieved \cite{rischke2004quark, yagi2005quark}. Nevertheless, such investigations, in particular in the context of heavy ion collisions, have also stimulated considerable research on how relativistic fluid dynamics emerges from a microscopic theory and what is its domain of applicability \cite{florkowski2018new, bemfica:18causality}.

The first to propose a relativistic formulation of dissipative fluid dynamics were Eckart \cite{Eckart:1940te}, in 1940, followed by Landau and Lifshitz \cite{landau:59fluid}, which provided an independent derivation in 1959. Both theories, often called first-order theories, are based on a covariant extension of the Navier-Stokes equations. However, it was demonstrated later that such formulations are ill-defined, since they contain intrinsic linear instabilities when perturbed around an \textit{arbitrary} global equilibrium state \cite{hiscock1987linear, hiscock1983stability, olson1990stability, denicol2008stability, Pu_2010}. Such linear instabilities were then shown to be related to the acausal nature of these theories \cite{hiscock1983stability, denicol2008stability, Pu_2010}.

Hydrodynamic theories that can be linearly causal and stable when perturbed around global equilibrium were only derived in the 1970s, by Israel and Stewart \cite{Israel:1976tn, israel1979jm}. The resulting equations are often referred to as Israel-Stewart theory or transient second-order theory and differ qualitatively from the Navier-Stokes equations by promoting the dissipative currents to independent dynamical variables. Such transient dynamics of the dissipative currents can only be incorporated into the description with the inclusion of terms that are of second order in gradients (Navier-Stokes theory only includes first-order terms) -- hence the name second-order theory. Israel-Stewart theory has been shown to be linearly causal and stable around global equilibrium, as long as their transport coefficients satisfy a set of fundamental constraints \cite{hiscock1983stability, olson1990stability, denicol2008stability, Pu_2010, Brito_2022, Sammet:2023bfo}. More general constraints for the causality of Israel-Stewart theory, valid also in the non-linear regime, were derived in Refs.~\cite{Bemfica:2019cop, Bemfica:2020xym}.

Recently, a novel theory of hydrodynamics was proposed by Bemfica, Disconzi, Noronha, and Kovtun (BDNK) \cite{bemfica:18causality,kovtun:19first,bemfica2019nonlinear,Hoult:2020eho,Bemfica:2020zjp}. This formulation is qualitatively different from Israel-Stewart's approach, but still provides hydrodynamic equations of motion that can be causal and linearly stable around global equilibrium. In this approach, causality is restored by considering constitutive relations for the dissipative currents that also include time-like derivatives of the hydrodynamic fields. In this case, conditions that ensure causality of the theory in the regime of vanishing net-charge were derived in Refs.~\cite{bemfica:18causality,bemfica2019nonlinear,DisconziBemficaRodriguezShaoSobolevConformal,DisconziBemficaGraber,Bemfica:2020zjp} and feasible conditions for linear stability around global equilibrium were proven to exist \cite{bemfica:18causality,kovtun:19first,bemfica2019nonlinear,Hoult:2020eho,Bemfica:2020zjp}. The drawback of this formalism is the requirement of imposing unorthodox definitions of the local equilibrium state, usually implemented by the so-called matching conditions \cite{Denicol:2021}.

The goal of this paper is to derive all of the aforementioned hydrodynamic theories from the Relativistic Boltzmann equation considering a system of classical weakly-interacting scalar fields in the ultra-relativistic regime. The self-interacting $\lambda \varphi^{4}$ scalar field theory is the most simple, yet not trivial, interacting system in high-energy physics \cite{Calzetta:1986cq,Peskin:1995ev,Kapusta:2006pm}. In the context of relativistic kinetic theory, the eigenvalues and eigenvectors of the linearized collision term for this interaction have been recently obtained in exact form in Ref.\ \cite{Denicol:2022bsq}, in the ultra-relativistic regime. We use these results to derive all hydrodynamic theories discussed so far, providing \textit{exact} and \textit{analytical} expressions for \textit{all} their transport coefficients. Furthermore, we study the solutions of these theories in a Bjorken flow scenario, providing a unique insight into their domain of validity at the early stages of a heavy ion collision.

The present paper is organized as follows: in Sec.\ \ref{sec:moments-and-hydro}, we summarize the basic aspect of relativistic hydrodynamics for arbitrary definitions of the local equilibrium state, i.e., for arbitrary matching conditions. In Sec.\ 
\ref{sec:eigenval}, we discuss basic features of the Boltzmann equation for classical weakly self-interacting scalar fields and provide microscopic expressions for the fundamental hydrodynamic variables. In Secs.\ 
\ref{Sec:NS} and \ref{sec:BDNK-th},  we derive the relativistic Navier-Stokes and BDNK equations, respectively, for arbitrary matching conditions and obtain, for the first time, \textit{exact} expressions for \textit{all} their transport coefficients. In Sec.\ \ref{sec:trans-hydro}, we perform the same task for transient second-order hydrodynamics, with the exception that we perform our derivation of the equations of motion and its respective transport coefficients assuming Landau matching conditions. In Sec.\ \ref{sec:eoms-bjorken}, we investigate the solutions of all these theories considering a highly symmetric flow configuration, the Bjorken flow \cite{Bjorken:1982qr}. We then demonstrate that Navier-Stokes and BDNK solutions display unphysical solutions when gradients are large, indicating that such theories may not be well suited to describe the early stages of heavy ion collisions. Solutions of transient second-order hydrodynamics do not display these unphysical features at large gradients, but are shown to break-down instead when the longitudinal pressure is initially negative.
In Sec.\ \ref{sec:concl}, we summarize our work and provide some conclusions and future perspectives. In Appendix \ref{app:collisional-moments}, we provide some details of our calculations while in Appendix \ref{appendix:Hilbert} we develop a derivation of the relativistic Hilbert theory. Finally, in Appendix \ref{app:stability-BDNK}, we derive some necessary conditions for the BDNK theory to be linearly stable around global equilibrium. We use $(+ - - -)$ as our metric signature and natural units, $\hbar = c = k_{B} = 1$.

\section{Hydrodynamics}
\label{sec:moments-and-hydro}
The fundamental equations in hydrodynamics are the continuity equations which describe the local conservation of particle number (net-charge, in general), energy and momentum,
\begin{equation}
\begin{aligned}
\label{eq:consv-eqns}
    \partial_{\mu}N^{\mu} = 0,\\
    \partial_{\mu}T^{\mu \nu} = 0,
\end{aligned}
\end{equation}
where $N^{\mu}$ is the particle 4-current and $T^{\mu \nu}$ is the energy-momentum tensor. Without any loss of generality, these tensors can be decomposed in terms of a time-like normalized 4-vector $u^{\mu}$, $u^{\mu}u_{\mu} = 1$, in the following way
\begin{equation}
\begin{aligned}
\label{eq:decompos-numu-tmunu}
   N^{\mu} &= n u^{\mu} + \nu^{\mu},  \\
    T^{\mu \nu} &= \varepsilon u^{\mu} u^{\nu} - P \Delta^{\mu \nu} + h^{\mu} u^{\nu} + h^{\nu} u^{\mu} + \pi^{\mu \nu},
\end{aligned}
\end{equation}
where $n$ is the total particle density, $\varepsilon$ is the total energy density, $P$ is the total isotropic pressure, $\nu^\mu$ is the particle diffusion 4-current, $h^\mu$ is the energy diffusion 4-current, and $\pi^{\mu\nu}$ is the shear-stress tensor. We further introduced the projection operator onto the 3-space orthogonal to $u^\mu$, $\Delta^{\mu \nu} \equiv g^{\mu \nu} - u^{\mu} u^{\nu}$. Each term introduced in this tensor decomposition can be expressed in terms of projections and contractions of the conserved currents, 
\begin{equation}
\begin{aligned}
\label{eq:definitions}
    n &\equiv u_{\mu}N^{\mu} , \, \, \varepsilon \equiv u_{\mu}u_{\nu}T^{\mu\nu}, \, \, P \equiv -\frac{1}{3}\Delta_{\mu\nu}T^{\mu\nu},  \\
    \nu^{\mu} &\equiv \Delta^{\mu}_{\nu} N^{\nu}, \, \, h^{\mu} \equiv \Delta^{\mu}_{\nu} u_{\lambda} T^{\nu\lambda}, \, \, \pi^{\mu\nu} \equiv \Delta^{\mu\nu}_{\alpha\beta} T^{\alpha\beta}.
\end{aligned}
\end{equation}
We note that we have introduced above the double-symmetric and traceless projection operator,
\begin{equation}
\Delta^{\mu \nu \alpha \beta} \equiv \frac{1}{2}\left( \Delta^{\mu \alpha } \Delta^{\nu \beta} + \Delta^{\nu \alpha } \Delta^{\mu \beta} \right) - \frac{1}{3}\Delta^{\mu \nu} \Delta^{\alpha \beta}.
\end{equation}

We now define a reference local equilibrium state \cite{israel1979annals} and separate the particle density, energy density, and isotropic pressure into equilibrium and non-equilibrium components. That is,
\begin{equation}
\label{eq:definitions2}
    n \equiv n_0(\alpha,\beta) + \delta n, \,\,\,
    \varepsilon \equiv \varepsilon_0(\alpha,\beta) + \delta \varepsilon, \,\,\,
    P \equiv P_0(\alpha,\beta) + \Pi,
\end{equation}
where $\alpha \equiv \mu/T$ is the thermal potential and $\beta \equiv 1/T$ is the inverse temperature, with $\mu$ being the chemical potential, of this fictitious local equilibrium state. The quantities $n_0$, $\varepsilon_0$ and $P_0$ are then determined by an equation of state, as if the system were in thermodynamic equilibrium, while $\delta n$, $\delta \varepsilon$, and $\Pi$ are the corresponding non-equilibrium corrections.

The local equilibrium variables $\alpha$, $\beta$, and $u^\mu$ introduced so far in the decomposition of the conserved currents must be properly defined. This task is usually performed by introducing matching conditions, with the most traditional matching conditions being constructed by Eckart \cite{Eckart:1940te} and Landau \cite{landau:59fluid}. Landau matching conditions define the fluid 4-velocity as a time-like and normalized eigenvector of $T^{\mu \nu}$, i.e., $T^{\mu}_{\ \nu} u^{\nu} \equiv \varepsilon u^{\mu}$, leading to the condition $h^\mu=0$. The Eckart matching conditions define the fluid 4-velocity as being parallel to the particle 4-current (net-charge, when the number of particles is not conserved), i.e., $N^\mu \equiv n u^\mu$, leading to the condition $\nu^\mu = 0$. Finally, in both Landau and Eckart matching conditions, the inverse temperature and thermal potential are defined assuming that the particle number and energy densities in the local rest frame are given by their respective thermodynamic values, i.e., $\delta n \equiv 0$ and $\delta \varepsilon \equiv 0$.
Naturally, more general matching conditions can also be considered, even though they can be very difficult to define outside the scope of kinetic theory.

Substituting the tensor decomposition given in Eqs.~\eqref{eq:decompos-numu-tmunu} into the conservation laws \eqref{eq:consv-eqns}, and projecting them into their components parallel and orthogonal to $u^\mu$, we obtain the following equations of motion, 
\begin{subequations}
 \label{eq:basic-hydro-EoM}
\begin{align}
 \label{eq:hydro-EoM-n0}
 Dn_{0}+D\delta n + (n_{0}+\delta n) \theta + \partial_{\mu} \nu^{\mu} &= 0, \\
\label{eq:hydro-EoM-eps}
 D\varepsilon_{0}+D\delta \varepsilon + (\varepsilon_{0}+\delta \varepsilon + P_{0} + \Pi) \theta - \pi^{\mu \nu} \sigma_{\mu \nu} + \partial_{\mu}h^{\mu} + u_{\mu} Dh^{\mu} &= 0, \\
\label{eq:hydro-EoM-umu}
(\varepsilon_{0} + \delta \varepsilon + P_{0} + \Pi)Du^{\mu} - \nabla^{\mu}(P_{0} + \Pi) + h^{\mu} \theta + h^{\alpha} \Delta^{\mu \nu} \partial_{\alpha}u_{\nu} +  \Delta^{\mu \nu} Dh_{\nu} + \Delta^{\mu \nu} \partial_{\alpha}\pi^{\alpha}_{ \ \nu} &= 0,
\end{align}
\end{subequations}
where $D = u^\mu \partial_\mu$ is the comoving time derivative, $\nabla^\mu = \Delta^{\mu\nu} \partial_\nu$ is the 4-gradient operator, $\theta = \partial_\mu u^\mu$ is the expansion rate, and  $\sigma^{\mu \nu} = \Delta^{\mu \nu \alpha \beta} \partial_{\alpha} u_{\beta} $ is the shear tensor.

Naturally, these equations are not closed and, in order to solve them, one must provide additional relations satisfied by the dissipative terms, $\delta n$, $\delta \varepsilon$, $\Pi$, $\nu^\mu$, $h^\mu$ and $\pi^{\mu\nu}$. The challenge then resides in consistently deriving such equations, that must be expressed solely in terms of the fields that appear in $N^\mu$ and $T^{\mu\nu}$. In the following sections, we shall perform this task assuming a weakly interacting gas of massless particles corresponding to scalar fields 
self-interacting via a $\lambda\varphi^4$ term. In this case, the relativistic Boltzmann equation can be used as the starting point for our derivation, which will be implemented for 3 different fluid-dynamical frameworks: relativistic Navier-Stokes theory, BDNK theory and transient second-order fluid dynamics. For the sake of completeness, we also derive relativistic Hilbert theory in Appendix \ref{appendix:Hilbert}. 

\section{Kinetic theory for $\lambda \varphi^{4}$ self-interacting particles}
\label{sec:eigenval}

The relativistic Boltzmann equation is a non-linear, integro-differential equation describing the time evolution of the single-particle distribution function, $f(x^{\mu},{\bf p}) \equiv f_{\bf p}$. 
In the classical limit and considering only binary elastic collisions, it reads  
\begin{equation}
\label{eq:EdBoltzmann}
p^{\mu }\partial_{\mu }f_{\mathbf{p}} = \frac{1}{2} \int dQ \ dQ^{\prime} \ dP^{\prime} W_{\mathbf{p}\mathbf{p}' \leftrightarrow \mathbf{q}\mathbf{q}'} (  f_{\mathbf{q}}f_{\mathbf{q}'} - f_{\mathbf{p}}f_{\mathbf{p}'} )   \equiv C\left[ f_{\mathbf{p}}\right]. 
\end{equation}
Above, we defined the Lorentz invariant integration measure for on-shell massless particles $dP \equiv d^{3}p/[(2\pi)^3p^{0}] = d^{3}p/[(2\pi)^3\vert {\bf p} \vert]$ and the Lorentz invariant transition rate
\begin{equation}
\begin{aligned}
& 
W_{\mathbf{p}\mathbf{p}' \leftrightarrow \mathbf{q}\mathbf{q}'}  = (2\pi)^{6} s \sigma(s,\Theta) \delta^{(4)}(p+p'-q-q'), \end{aligned}    
\end{equation}
where $\sigma(s,\Theta)$ is the differential cross section and we used the Mandelstam variables, $s \equiv (q^{\mu}+q'^{\mu})(q_{\mu}+q'_{\mu}) = (p^{\mu}+p'^{\mu})(p_{\mu}+p'_{\mu})$ and defined $\Theta$ such that
\begin{equation}
\begin{aligned}
&
\cos \Theta = \frac{(p-p') \cdot (q-q')}{(p-p')^{2}} = 
\left. \frac{\mathbf{p} \cdot \mathbf{q}}{\vert \mathbf{p} \vert   \vert \mathbf{q}\vert}
\right\vert_{\mathrm{CM}}.
\end{aligned}    
\end{equation}
Here, $\mathrm{CM}$ denotes that the expression is given in the center-of-momentum frame, i.e., the frame in which the total momentum of the collision vanishes, $p^{\mu}+p'^{\mu} = q^{\mu}+q'^{\mu} = (\sqrt{s}, \mathbf{0})$. The specific functional form of the cross-section $\sigma(s,\Theta)$ varies according to the microscopic interactions involved \cite{Peskin:1995ev}. In this work, we shall consider a system composed of massless scalar particles whose dynamics is given by the Lagrangian density,
\begin{equation}
\mathcal{L} = \frac{1}{2} \partial_{\mu} \varphi \ \partial^{\mu} \varphi
-
\frac{\lambda \varphi^{4}}{4!},
\end{equation}
providing, at leading order in $\lambda$, a differential cross-section that does not possess any angular dependence. The corresponding total cross section is \cite{Peskin:1995ev}, 
\begin{equation}
\label{eq:cross-sec-phi4}
\begin{aligned}
& \sigma_{T}(s) = \frac{1}{2} \int d\Phi d\Theta \, \sin\Theta \, \sigma(s, \Theta) = \frac{\lambda^2}{32 \pi s} \equiv \frac{g}{s},
\end{aligned}    
\end{equation}
where $\Phi$ is the azimuthal angle in the CM frame and $g \equiv \lambda^{2}/(32 \pi)$, as implied above. 

In the context of kinetic theory, the particle 4-current and the energy-momentum tensor are identified as the first and second moments of the single-particle distribution function, respectively,
\begin{equation}
\label{eq:currents_kin}
N^{\mu} = \int dP \, p^{\mu} f_{\textbf{p}}, \hspace{0.2cm}
T^{\mu\nu} = \int dP \, p^{\mu}p^{\nu} f_{\textbf{p}} .
\end{equation}
Applying the same tensor decomposition used in the previous section, Eq.\ \eqref{eq:decompos-numu-tmunu}, we obtain kinetic expressions for the hydrodynamic variables. For this purpose, we must also introduce a reference local equilibrium state and decompose the single-particle distribution function into an equilibrium part and a non-equilibrium one,\begin{equation}
\begin{aligned}
&
f_{\bf{p}} = f_{0\bf{p}}+\delta f_{\bf{p}} \equiv f_{0\bf{p}}(1+\phi_{\bf{p}}),
\end{aligned}    
\end{equation}
where we defined the deviation functions $\delta f_{\bf{p}} \equiv f_{\bf{p}} - f_{0\bf{p}} $ and $\phi_{\bf{p}} \equiv (f_{\bf{p}} - f_{0\bf{p}})/f_{0\bf{p}}$ and introduced the local equilibrium single-particle distribution function,
\begin{equation}
\begin{aligned}
&
f_{0\bf{p}} \equiv \exp(\alpha - \beta E_\mathbf{p}),
\end{aligned}    
\end{equation}
with $E_\mathbf{p} \equiv u_{\mu} p^{\mu}$ being the energy in the local rest frame. The inverse temperature $\beta$, thermal potential $\alpha$ and 4-velocity $u^{\mu}$ must be defined using matching conditions, as discussed in the previous section. In a kinetic theory framework, an ensemble of matching conditions can be formulated in the following way \cite{Rocha:2021lze,Rocha:2022ind,Denicol:2021},  
\begin{equation}
\label{eq:kinetic_match}
\begin{aligned}
    \int dP \, E^{q}_{\textbf{p}} \delta f_{\textbf{p}} \equiv 0, \, \, \, 
    \int dP \, E_{\textbf{p}}^s \delta f_{\textbf{p}} \equiv 0, \, \, \,
    \int dP \, E_{\textbf{p}}^{z} p^{\langle \mu \rangle} \delta f_{\textbf{p}} \equiv 0,
\end{aligned}
\end{equation}
where $p^{\langle \mu \rangle} \equiv \Delta^{\mu\nu} p_\nu$, while $q$, $s$, and $z$ are free parameters. The above conditions reduce to the Landau matching conditions when $q=1$, $s=2$, and $z=1$ and to the Eckart conditions when $q=1$, $s=2$, and $z=0$. Other values of $q$, $s$, and $z$ lead to novel matching conditions that often do not have any intuitive physical interpretation. 
Then, the hydrodynamic fields can be identified as the following contraction and/or projections of the conserved currents,
\begin{subequations}
\label{eq:def_kinetic-all}
\begin{align}
\label{eq:def_kinetic-1}
    n_0 &\equiv \int dP \, E_{\textbf{p}} f_{0\textbf{p}}, \, \, \delta n \equiv \int dP \, E_{\textbf{p}} \delta f_{\textbf{p}}, \\
    \varepsilon_0 &\equiv \int dP \, E_{\textbf{p}}^2 f_{0\textbf{p}}, \, \, \delta \varepsilon \equiv \int dP \, E_{\textbf{p}}^2 \delta f_{\textbf{p}}, \\
\label{eq:def_kinetic-2}
    \, \, P_{0} & \equiv -\frac{1}{3} \int dP \, \Delta_{\mu \nu}p^{\mu}p^{\nu} f_{0\textbf{p}}, \, \,
    \Pi \equiv -\frac{1}{3} \int dP \, \Delta_{\mu \nu}p^{\mu}p^{\nu}\delta f_{\textbf{p}}, \, \, \\
\label{eq:def_kinetic-3}
    \nu^{\mu} & \equiv \int dP \, p^{\langle \mu \rangle} f_{\textbf{p}}, \, \,
    h^{\mu} \equiv \int dP \, E_{\textbf{p}}p^{\langle \mu \rangle} f_{\textbf{p}}, \\ \pi^{\mu\nu} &\equiv \int dP \, p^{\langle \mu}p^{\nu \rangle} f_{\textbf{p}},
\end{align}
\end{subequations}
where $p^{\langle \mu}p^{\nu \rangle} \equiv \Delta^{\mu\nu\alpha\beta} p_\alpha p_\beta$ denotes the irreducible projection of $p^\mu p^\nu$.

\section{Navier-Stokes theory}
\label{Sec:NS}
We first consider the traditional relativistic Navier-Stokes theory. We derive it from the relativistic Boltzmann equation employing the well known Chapman-Enskog expansion \cite{chapman1916vi,enskog1917kinetische,deGroot:80relativistic,cercignani:02relativistic,Denicol:2021} -- a perturbative solution of the Boltzmann equation based on an expansion of the single-particle distribution function in space-like gradients of hydrodynamic variables. In practice, one converts the original Boltzmann equation into the following perturbative problem, introducing the book-keeping parameter $\epsilon$,
\begin{equation}
\label{eq:chap-ensk}
\begin{aligned}
\epsilon E_{\mathbf{p}} Df_{\mathbf{p}} + \epsilon p^\mu \nabla_\mu f_{\mathbf{p}}   = C[f_{\mathbf{p}}].
\end{aligned}
\end{equation}
We then consider a series solution for $f_{\mathbf{p}}$, 
\begin{equation}
\label{eq:chapman-expn-SPDF}
f_{\mathbf{p}} = \sum_{i=0}^{\infty} \epsilon^{i} f^{(i)}_{\mathbf{p}},    
\end{equation}
and also consider an expansion of the co-moving derivative, 
\begin{equation}
\label{eq:chapman-expn-SPDF}
Df_{\mathbf{p}} = \sum_{i=0}^{\infty} \epsilon^{i} D^{(i)}f_{\mathbf{p}},   
\end{equation}
where $D^{(n)}f_{\mathbf{p}}$ denotes the $n$-th order contribution in $\epsilon$ of the comoving derivative of $f_{\mathbf{p}}$. The latter expansion promotes a resummation of the gradient expansion so that only terms containing the space-like 4-gradient $\nabla_{\mu}$ appear in the constitutive relations. The Boltzmann equation is then solved order by order in $\epsilon$, with the solution of the original equation being recovered by setting $\epsilon=1$.  

The zeroth-order solution of this series satisfies $C[f^{(0)}_{\mathbf{p}}]=0$ and is given by a local equilibrium distribution function, $f^{(0)}_{\mathbf{p}}=f_{0\mathbf{p}}$, that is fully determined by fixing the matching conditions, as explained in the last section. The first-order solution for the deviation function $\phi_\mathbf{p}=f^{(1)}_\mathbf{p}/f_{0\mathbf{p}}$ is then obtained by solving the following equation,
\begin{equation}
 \frac{1}{4}L^{(3)}_{1\mathbf{p}}p_{\langle\mu\rangle}\nabla ^{\mu }\alpha -\beta 
p_{\langle \mu }p_{\nu \rangle}
\sigma^{\mu \nu } 
=\hat{L}\phi_{\mathbf{p}} , \label{CE}
\end{equation}
where, in deriving this equation, we used that \cite{Rocha:2022ind},
\begin{equation}
\label{eq:time-like-D-subst}
\begin{aligned}
D^{(0)} f_{\bf p}
=  \left[- \frac{\beta}{3} E_{\bf p} \theta + p_{\langle \mu \rangle}\left(\nabla^{\mu}\beta - \frac{\beta}{4} \nabla^{\mu}\alpha \right) \right]f_{0\bf{p}}.
\end{aligned}    
\end{equation}
For the sake of convenience, we have expressed the left-hand side of Eq.~\eqref{CE} in terms of associate Laguerre polynomials \cite{gradshteyn2014table}. We also note that the terms proportional to $\theta$ and $\nabla^{\mu}\beta$ in Eq.~\eqref{eq:time-like-D-subst} do not appear in Eq.\ \eqref{CE} since they cancel exactly with terms stemming from $\nabla_{\mu} f_{0 {\bf p}}$. On the right-hand side we have the \textit{linearized} collision term, $\hat{L}$, a linearization of the collision term with respect to a local equilibrium state. For the interaction considered in this work, it is given by,   
\begin{equation}
\label{Lhat}
\begin{aligned}
&
\hat{L}\phi_{\mathbf{p}} \equiv
\frac{g}{2}\int dQ \ dQ^{\prime} \ dP^{\prime} (2\pi)^{5} \delta^{(4)}(p+p'-q-q') f_{0\mathbf{p}'} (  \phi_{\mathbf{q}} + \phi_{\mathbf{q}'} - \phi_{\mathbf{p}} - \phi_{\mathbf{p}'}  ).
\end{aligned}    
\end{equation}
For massless particles in the classical limit, the spectrum of $\hat{L}$ has been recently determined analytically in Ref.\ \cite{Denicol:2022bsq}, i.e., the eigensystem of the linearized collision operator can be computed exactly. In fact, it was demonstrated that,
\begin{equation}
\label{eq:eigenvalues-lin-col}
\begin{aligned}
&
\hat{L}\left[ L^{(2\ell+1)}_{n{\bf p}} p^{\langle \mu_{1}} \cdots p^{\mu_{\ell} \rangle} \right]
=
\chi_{n\ell} 
L^{(2\ell+1)}_{n {\bf p}} p^{\langle \mu_{1}} \cdots p^{\mu_{\ell} \rangle},
 \\
&
\chi_{n\ell} = - \frac{g}{2} I_{0,0} \left(\frac{n+\ell-1}{n+\ell+1} + \delta_{n0}\delta_{\ell 0}\right),
\end{aligned}    
\end{equation}
where $L^{(\alpha)}_{n {\bf p}} \equiv L^{(\alpha)}_{n}(\beta E_\mathbf{p})$ denotes the $n$-th associated Laguerre polynomial \cite{gradshteyn2014table} and $p^{\langle \mu_{1}} \cdots p^{\mu_{\ell} \rangle} \equiv \Delta^{\mu_{1} \cdots \mu_{\ell}}_{\ \ \nu_{1} \cdots \nu_{\ell}} p^{\nu_{1}} \cdots p^{\nu_{\ell}}$ are irreducible tensors constructed from 4-momentum, with $\Delta^{\mu_{1} \cdots \mu_{\ell}}_{\nu_{1} \cdots \nu_{\ell}}$ being the $2\ell$--rank projection operator orthogonal to the fluid 4-velocity in every index. These tensors are constructed from combinations of $\Delta^{\mu\nu}=g^{\mu\nu} - u^{\mu}u^{\nu}$ so that they are symmetric under the exchange of the indices ($\mu_{1} \cdots \mu_{\ell}$) and ($\nu_{1} \cdots \nu_{\ell}$), separately, and also traceless in each subset of indices, for $\ell > 1$ \cite{deGroot:80relativistic,Denicol:2021}. We also made use of the thermodynamic integrals
$I_{n,q}$, 
\begin{equation}
\label{thermodynamic_integral}
 I_{n,q} = \frac{1}{(2q+1)!!} \int dP \left( -\Delta^{\lambda \sigma}p_{\lambda} p_{\sigma} \right)^{q} E_{\mathbf{p}}^{n-2q} f_{0\bf{p}}.
\end{equation}

The general first order solution for $\phi_{\mathbf{p}}$ is formally given by
\begin{equation}
\phi_{\mathbf{p}} = \phi^{\mathrm{hom}}_{\mathbf{p}} + \hat{L}^{-1}\left[\frac{1}{4}L^{(3)}_{1\mathbf{p}}p_{\langle\mu\rangle}\nabla ^{\mu }\alpha -\beta 
p_{\langle \mu }p_{\nu \rangle}
\sigma^{\mu \nu } \right],
 \label{CEsol}
\end{equation}%
where we denoted the homogeneous solution $\phi^{\mathrm{hom}}_{\mathbf{p}}=a + b_\mu p^\mu$, with the free parameters $a$ and $b_\mu$ being later determined by matching conditions \cite{Denicol:2021}. Since the eigenvalues and eigenfunctions of the linear operator $\hat{L}$ are known, the solution above can be trivially evaluated \cite{Denicol:2022bsq},
\begin{equation}
\phi_{\mathbf{p}} = a + b_\mu p^\mu + \frac{1}{4\chi_{11}}L^{(3)}_{1\mathbf{p}}p_{\langle\mu\rangle}\nabla ^{\mu }\alpha -\frac{\beta}{\chi_{02}} 
p_{\langle \mu }p_{\nu \rangle}
\sigma^{\mu \nu }.
 \label{CEsol}
\end{equation}
For the set of matching conditions introduced in Eq.~\eqref{eq:kinetic_match}, one fixes $a=0$ and $b^\mu=z\nabla^\mu \alpha / (4\chi_{11})$, with $z$ being one of the free parameters that is used to specify the matching condition.

Using this exact first order solution for $\phi_{\mathbf{p}}$, we can determine all irreducible moments of the nonequilibrium distribution function,
\begin{equation}
\label{eq:irreducible_moments}
\begin{aligned}
\rho^{\mu_{1} \cdots \mu_{\ell}}_{r} \equiv \int dP  E_{\mathbf{p}}^{r} p^{\langle \mu_{1}} \cdots p^{\mu_{\ell} \rangle} \delta f_{\mathbf{p}}. 
\end{aligned}    
\end{equation}
The irreducible tensors $p^{\langle \mu_{1}} \cdots p^{\mu_{\ell} \rangle}$ satisfy the following orthogonality relations \cite{deGroot:80relativistic, Denicol:2021},
\begin{equation}
\label{eq:polys-orthogonal}
\begin{aligned}
\int dP
p^{\langle \mu_{1}} \cdots p^{\mu_{\ell} \rangle}
p_{\langle \nu_{1}} \cdots p_{\nu_{m} \rangle}
H(E_{\mathbf{p}})
& = \frac{\ell!\delta_{\ell m}}{(2\ell + 1)!!}  \Delta^{\mu_{1} \cdots \mu_{\ell}}_{\nu_{1} \cdots \nu_{\ell}} \int dP \left(\Delta^{\mu \nu} p_{\mu} p_{\nu} \right)^{\ell} H(E_{\mathbf{p}}),   
\end{aligned}    
\end{equation}
where $H(E_{\mathbf{p}})$ is an arbitrary function of $E_{\mathbf{p}}$. These relations imply that any scalar irreducible moment, as well as those of rank higher than 2, will necessarily vanish at first order (in particular, the scalar moments vanish due to the assumption of massless particles). Meanwhile, the irreducible moments of rank 1 and 2 are given by,
\begin{equation}
\label{eq:NSlimit}
\begin{aligned}
&
\rho^{\mu}_{r} = \int dP  E^{r}_{\mathbf{p}} p^{\langle\mu\rangle} f_{0\mathbf{p}}\phi_{\mathbf{p}}=\frac{1}{4\chi_{11}} \nabla_{\nu}\alpha \int dP  E^{r}_{\mathbf{p}} p^{\langle\mu\rangle}p^{\langle\nu\rangle}(z+L^{(3)}_{1\mathbf{p}}) f_{0\mathbf{p}}
\equiv \kappa_{r}\nabla^{\mu}\alpha,
\\
&
\rho^{\mu\nu}_{r} = \int dP  E^{r}_{\mathbf{p}} p^{\langle\mu}p^{\nu\rangle} f_{0\mathbf{p}}\phi_{\mathbf{p}}=-\frac{\beta}{\chi_{02}} \sigma_{\alpha\beta} \int dP  E^{r}_{\mathbf{p}} p^{\langle\mu}p^{\nu\rangle}p^{\langle\alpha}p^{\beta\rangle} f_{0\mathbf{p}}
\equiv 2\eta_{r} \sigma^{\mu\nu},
\end{aligned}    
\end{equation}
where we used the orthogonality relation Eq.\ \eqref{eq:polys-orthogonal} and defined the transport coefficients
\begin{equation}
\label{eq:transportcoeffs}
\begin{aligned}
&
\kappa_{r}=-\frac{z-r}{12\chi_{11}} I_{r+2,0},
\\
&
\eta_{r}=-\frac{\beta}{15\chi_{02}} I_{r+4,0}.
\end{aligned}    
\end{equation}
Thus, we obtain the following constitutive relations,
\begin{equation}
\label{eq:transportcoeffs}
\begin{aligned}
&
\delta n= 0, \, \, \, \nu^\mu =z\frac{3}{g\beta^2} \nabla^{\mu}\alpha,
\\
&
\delta \varepsilon = 0, \, \, \,
h^\mu=(z-1)\frac{12}{g\beta^3} \nabla^{\mu}\alpha,
\\
&
\pi^{\mu \nu}=\frac{96}{g \beta^3} \sigma^{\mu \nu}.
\end{aligned}    
\end{equation}
As expected, $\nu^\mu$ vanishes when $z=0$ (Eckart matching condition) while $h^\mu$ vanishes when $z=1$ (Landau matching condition). We note that the expressions for the transport coefficients are \textit{exact} -- something extremely rare in these type of calculations.

Nevertheless, we remark that these constitutive relations for the dissipative currents render the fluid-dynamical equations acausal, which further leads to intrinsic unphysical instabilities \cite{Gavassino:2021owo}, regardless of the matching conditions imposed, i.e., regardless of the value of $z$. These issues render Navier-Stokes theory an unsuitable formalism to describe any relativistic fluid existing in nature. 

\section{BDNK theory}
\label{sec:BDNK-th}

In this section, we consider another formalism of relativistic hydrodynamic, the BDNK theory \cite{Bemfica:2017wps,Bemfica:2019knx,Kovtun:2019hdm,Bemfica:2020zjp}. This framework can in principle be constructed to be causal and stable, depending on the choice of matching conditions. We derive the BDNK equations from the Boltzmann equation following the procedure constructed in Ref.~\cite{Rocha:2022ind}. In this case, one considers a modification of the Chapman-Enskog expansion in which the perturbative procedure is implemented on moments of the Boltzmann equation and not on the Boltzmann equation itself. For this purpose, the equation is first multiplied by a complete basis and integrated in momentum and only then the perturbative parameter is inserted. In particular, it is convenient to choose the eigenfunctions of the linearized collision operator as basis elements, thus leading to
\begin{equation}
\label{eq:BDN0}
\epsilon \int dP L_{n, {\bf p}}^{(2 \ell + 1)} p_{\langle \mu_{1}} \cdots  p_{\mu_{\ell} \rangle}  p^{\mu}\partial_{\mu} f_{\bf{p}} 
= 
\int dP L_{n, {\bf p}}^{(2 \ell + 1)} p_{\langle \mu_{1}} \cdots  p_{\mu_{\ell} \rangle}C[f_{\bf{p}}].
\end{equation}
Once more, one considers the series solution%
\begin{equation}
\label{eq:chapman-expn-SPDF}
f_{\mathbf{p}} = \sum_{i=0}^{\infty} \epsilon^{i} f^{(i)}_{\mathbf{p}},   
\end{equation}
with the solution of the original equation being recovered by setting $\epsilon=1$. Note that if the basis elements correspond to $1,p^\mu$, the integral over the collision term vanishes and we obtain the usual conservation laws, with the perturbative parameter $\epsilon$ completly vanishing. These conservation laws are treated non-perturbatively \cite{Rocha:2022ind}, and, thus, from now on, we only consider the remaining moments in our analysis.

As before, the zeroth-order solution is given by the local equilibrium distribution, $f_{0\mathbf{p}}$ \cite{Rocha:2022ind}. The first-order solution, here denoted as $\phi_{\mathbf{p}} \equiv f^{(1)}_{\mathbf{p}}/f_{0\mathbf{p}}$, is then obtained by solving the following equation,
\begin{equation}
\label{eq:BDNK2}
\begin{aligned}
&
\int dP L_{n, {\bf p}}^{(2 \ell + 1)} p_{\langle \mu_{1}} \cdots  p_{\mu_{\ell} \rangle}  p^{\mu}\partial_{\mu} f_{0\bf{p}} 
= 
\int dP L_{n, {\bf p}}^{(2 \ell + 1)} p_{\langle \mu_{1}} \cdots  p_{\mu_{\ell} \rangle}f_{0\bf{p}}\hat{L}\phi_{\mathbf{p}},\end{aligned}
\end{equation}
where the values $(\ell, n)= (0,0), (0,1),$ and $ (1,0)$ are explicitly excluded from this equation, since they amount to linear combinations of conserved quantities. Also, $\hat{L}$ is the linearized collision term, already introduced in Eq.\ \eqref{Lhat}. Moreover, the derivative of the equilibrium distribution on left-hand side can be irreducibly expressed as
\begin{equation}
\label{df0}
\begin{aligned}
&
p^{\mu} \partial_{\mu} f_{0\bf{p}} = \left[ E_{\mathbf{p}} D\alpha - \beta E_{\mathbf{p}}^{2} \left( \frac{D\beta}{\beta} - \frac{\theta}{3} \right) +   p^{\langle \mu \rangle} \nabla_{\mu}\alpha - E_{\mathbf{p}} p^{\langle \mu \rangle} \left( \beta Du_{\mu} + \nabla_{\mu} \beta \right) - \beta p^{\langle \mu}p^{\nu \rangle} \sigma_{\mu \nu} \right] f_{0\mathbf{p}}.
\end{aligned}    
\end{equation}
%
Since the perturbative procedure was constructed using the moments of the Boltzmann equation, and not the Boltzmann equation itself, we are not required to exchange the time-like derivatives above in terms of space-like ones -- as occurs in the traditional Chapman-Enskog expansion. This may seem as a minor difference, but will end up being the decisive factor in deriving the BDNK equations.

In order to solve \eqref{eq:BDNK2}, we expand $\phi_\mathbf{p}$ using a complete basis of irreducible tensors and associated Laguerre polynomials,
\begin{equation}
\label{momentexp}
\begin{aligned}
&
\phi_{\bf p}
=
\sum_{\ell,n = 0}^{\infty}  \Phi_{n}^{\mu_{1} \cdots \mu_{\ell}} L_{n}^{(2 \ell + 1)} p_{\langle \mu_{1}} \cdots  p_{\mu_{\ell} \rangle},
\end{aligned}    
\end{equation}
where the Laguerre polynomials can be shown to obey the following orthogonality relation (in the massless limit),
\begin{equation}
\label{eq:orth-laguerre}
\begin{aligned}
&
\int dP \left( \Delta^{\mu \nu} p_{\mu} p_{\nu} \right)^{\ell} L_{n{\bf p}}^{(2 \ell + 1)} L_{m{\bf p}}^{(2 \ell + 1)} f_{0 {\bf p}}
=
(-1)^{\ell}\frac{e^{\alpha}}{2 \pi^{2} \beta^{2 \ell + 2}}
\frac{(n+2\ell+1)!}{n!}
\delta_{nm} 
\equiv
A^{(\ell)}_{n} \delta_{nm},
\end{aligned}    
\end{equation}
with $A_n^{(\ell)}$ defined as implied. Thus, combining Eq.~\eqref{eq:orth-laguerre} with the orthogonality relations satisfied by the irreducible tensors, Eq.~\eqref{eq:polys-orthogonal}, the expansion coefficients $\Phi_{n}^{\mu_{1} \cdots \mu_{\ell}}$ are expressed in terms of $\phi_\mathbf{p}$ as
\begin{equation}
\begin{aligned}
&
\frac{\ell !}{(2\ell+1)!!} A_{n}^{(\ell)}\Phi_{n}^{\mu_{1} \cdots \mu_{\ell}} 
= 
\int dP L^{(2\ell+1)}_{n} p^{\langle \mu_{1}} \cdots p^{ \mu_{\ell}\rangle} f_{0 {\bf p}} \phi_{\bf p} 
\equiv
\Hat{\Phi}_{n}^{\mu_{1} \cdots \mu_{\ell}},
\end{aligned}    
\end{equation}
where the tensors $\Hat{\Phi}_{n}^{\mu_1 \cdots \mu_\ell}$ are defined as implied and we note that $\Hat{\Phi}_{0}^{\mu \nu} = \pi^{\mu \nu}$ and $\Hat{\Phi}_{0}^{\mu} = \nu^{\mu}$. Replacing Eq.\ \eqref{momentexp} and Eq.\ \eqref{df0} into Eq.\ \eqref{eq:BDNK2}, and using the orthogonality relations, Eqs.\ \eqref{eq:polys-orthogonal} and \eqref{eq:orth-laguerre}, we obtain
\begin{equation}
\begin{aligned}
&
\Hat{\Phi}_{n}^{\alpha \beta} 
=
- \frac{2 \beta A^{(2)}_{0}}{15\chi_{02}} \sigma^{\alpha \beta} \delta_{n,0},
\quad
n=0,1, \cdots,
\\
&
\Hat{\Phi}_{n}^{\lambda} 
= \frac{A^{(1)}_{1}}{ 3\chi_{11}} \left(\frac{\nabla^{\lambda} \beta}{\beta} + D u^{\lambda} \right) \delta_{n,1}, \quad n=1,2,\cdots, \\
&
\Hat{\Phi}_{n} = - \frac{ 2A^{(0)}_{2}}{\beta \chi_{20}}
\left( \frac{D \beta}{\beta} - \frac{\theta}{3} \right) \delta_{n,2}, \quad n=2,3,\cdots,
\end{aligned}    
\end{equation}
where we also used the self-adjoint property of $\hat{L}$, i.e., $g_\mathbf{p}\hat{L}\phi_\mathbf{p}=\phi_\mathbf{p}\hat{L}g_\mathbf{p}$, and our knowledge of its spectrum (cf.~Eq.~\eqref{eq:eigenvalues-lin-col}) \cite{Denicol:2022bsq}. The coefficients $\Hat{\Phi}_{0}$, $\Hat{\Phi}_{1}$, $\Hat{\Phi}_{0}^{\lambda}$, are related to the zero modes or homogeneous solutions of $\Hat{L}$, and cannot be obtained from this inversion procedure. These quantities are calculated from the matching conditions, Eqs.~\eqref{eq:kinetic_match}, leading to
\begin{equation}
\begin{aligned}
&
\Hat{\Phi}_{0} = \frac{qs}{6} \Hat{\Phi}_{2},
\\
&
\Hat{\Phi}_{1} = \frac{1}{3} (q+s-1) \Hat{\Phi}_{2},
\\
&
\Hat{\Phi}_{0}^{\lambda} 
=
\frac{z}{4} 
\Hat{\Phi}_{1}^{\lambda}.
\end{aligned}    
\end{equation}

Now that the first-order solution for the moments of the distribution function is determined, we can derive the equations of motion for the dissipative terms by replacing this solution into their expressions \eqref{eq:def_kinetic-all}.  Then, we obtain,
\begin{equation}
\label{eq:const-rel-BDNK-chap-8}
\begin{aligned}
& \Pi 
=  \frac{\chi}{3} \left( \frac{D\beta }{\beta} -  \frac{\theta}{3} \right),
\ \
 \delta n 
= \xi \left( \frac{D\beta }{\beta} -  \frac{\theta}{3} \right),
\ \
 \delta \varepsilon 
= \chi  \left( \frac{D\beta }{\beta} -  \frac{\theta}{3} \right),\\
& \nu^{\mu} 
= \kappa \left( \frac{\nabla^{\mu} \beta}{\beta} +   D u^{\mu}\right), \ \
 h^{\mu} 
= \lambda \left( \frac{\nabla^{\mu} \beta}{\beta} +   D u^{\mu}\right),
    \\
&
\pi^{\mu \nu} = 2 \eta \sigma^{\mu \nu}.
\end{aligned}
\end{equation}
The transport coefficients 
 are given in \textit{analytical} form by
\begin{equation}
\label{eq:transp-coeffs-BDNK-cap8}
\begin{aligned}
&
\xi =  \frac{12}{g\beta ^2} (q-1) (s-1), \quad 
\chi = \frac{36}{g\beta^3}(q-2)(s-2),
\\
&
\kappa = \frac{12}{g\beta ^2}z, \quad 
\lambda 
= 
\frac{48}{g\beta ^3}(z-1),
\\
&
\eta
= \frac{48}{g\beta ^3}.
\end{aligned}    
\end{equation}
We note that, since $\xi$ and $\chi$ are in general non-zero, $\delta n$ and $\delta \varepsilon$ are also non-vanishing, in contrast to Navier-Stokes and Hilbert theories. A brief discussion on the linear stability of the BDNK theory derived above is developed in Appendix \ref{app:stability-BDNK}, considering a simplified scenario of homogeneous perturbations around global equilibrium. Then, we obtain the following \textit{necessary} conditions for linear stability, 
\begin{equation}
\frac{\lambda}{\varepsilon_{0}} > 0, \ \ \frac{\xi}{n_{0}} > \frac{\chi}{\varepsilon_{0}},
\end{equation}
which, using Eqs.~\eqref{eq:transp-coeffs-BDNK-cap8}, lead to the following constraints on the matching conditions,
\begin{equation}
\label{eq:stability-cond-BDNK}
z>1, \hspace{.45cm} (q-1)(s-1) > (q-2)(s-2).
\end{equation}
These constraints imply that the so-called exotic Eckart matching condition (which imposes $z=0$ and $q=1$), adopted in previous works \cite{Bemfica:2019knx,Rocha:2021lze,Rocha:2022ind}, renders the theory linearly unstable and, thus, is unphysical.

\section{Transient second-order hydrodynamics}
\label{sec:trans-hydro}

Causal and stable fluid-dynamical equations are traditionally derived from the Boltzmann equation using the method of moments \cite{Denicol:2012cn,Denicol:2021}, without a perturbative procedure, leading to second-order \textit{transient} fluid-dynamical theories. In this formalism, one derives the equations of motion for the irreducible moments of the non-equilibrium component of the single-particle distribution function, $\rho^{\mu_{1} \cdots \mu_{\ell}}_{r}$, defined in Eq. \eqref{eq:irreducible_moments}. These moment equations are then systematically truncated in order to derive a closed set of fluid-dynamical equations. This procedure will be outlined and implemented in this section, considering only the \textit{Landau matching conditions}. We note that the derivation of second-order transient hydrodynamics from kinetic theory for general matching conditions was performed in Ref.\ \cite{Rocha:2021lze}, by considering a generalization of the 14-moment approximation. However, extending this derivation to the interaction considered in this work is still very complicated and will be delegated to another publication.  

The relativistic moment equations were first derived in \cite{Denicol:2012cn} and their form in the \textit{massless limit} will be listed below, for irreducible moments of rank $1$ and $2$. The rank-1 irreducible moments obey the following equation of motion,
\begin{equation}
\label{eq:transient-l=1}
\begin{aligned}
& 
D\rho _{r}^{\langle \alpha \rangle} 
-
r \rho^{ \alpha \mu}_{r-1} Du_{\mu}
+
\frac{r+3}{3}   \rho_{r+1} Du^{\alpha}
+
\omega_{\mu}^{\ \alpha} \rho^{\mu}_{r}
+
\Delta^{\alpha}_{ \ \alpha'} \nabla_{\mu} \rho_{r-1}^{\alpha' \mu} 
-
(r-1)  \rho^{\alpha \mu \nu}_{r-2} \sigma_{\mu \nu}
-
\frac{1}{3} \nabla^{\alpha}\rho_{r+1} 
\\
&
+
\frac{r+3}{3} \rho_{r}^{\alpha} \theta
+
\frac{2r+3}{5} \sigma_{\mu}^{\ \alpha}  \rho_{r}^{\mu}  
-
\frac{\beta I_{r+2,1}}{(\varepsilon_{0} + P_{0})}   \Delta^{\alpha \beta} \partial_{\mu}\pi^{\mu}_{ \ \beta}
-
\alpha_{r}^{(1)} \nabla^{\alpha}\alpha
=
\int E_{\mathbf{p}}^{r-1} p^{\langle \alpha \rangle} C[f_{\mathbf{p}}]  
\equiv
\mathcal{C}_{r-1}^{\alpha}.
\end{aligned}    
\end{equation}
The equations of motion for the rank-2 irreducible moments are
\begin{equation}
\label{eq:transient-l=2}
\begin{aligned}
&  D\rho ^{\langle \alpha \beta \rangle}_{r}
-
r \rho^{ \alpha \beta \mu}_{r-1} Du_{\mu}
+
\frac{2}{5} (r+5) Du^{ \langle \alpha}  \rho_{r+1}^{\beta\rangle} 
+\Delta^{\alpha \beta}_{ \ \ \alpha' \beta'}
 \nabla_{\mu} \rho_{r-1}^{\alpha' \beta' \mu} 
-
(r-1) \sigma_{\mu \nu} \rho^{\mu \alpha \beta \nu}_{r-2}
-
\frac{2}{5} 
 \nabla^{\langle \alpha} \rho_{r+1}^{\beta \rangle}
+
\frac{r+4}{3} \rho_{r}^{\alpha \beta} \theta 
\\
& 
+
2 
 \omega_{\mu}^{ \ \langle \alpha} \rho^{\beta \rangle \mu}_{r} 
+ \frac{2}{7} (2r+5)
\sigma_{\mu}^{\ \langle \alpha}  \rho_{r}^{\beta \rangle \mu}
-
\frac{2}{15} (r+4) \rho_{r+2} \sigma^{\alpha \beta}
-
\alpha_{r}^{(2)} 
 \sigma^{\alpha \beta}
=
\int dP E_{\mathbf{p}}^{r-1} p^{\langle \alpha} p^{\beta \rangle} C[f_{\mathbf{p}}]  
\equiv
\mathcal{C}^{\alpha \beta}_{r-1}.
\end{aligned}    
\end{equation}
Irreducible moments of rank $\ell \geq 3$, and their corresponding equations of motion, are not considered in the derivation of second-order fluid-dynamical theories \cite{Denicol:2012cn, Brito:2021iqr, deBrito:2023tgb} since they only contain contributions for the dissipative currents that are at least of third-order. Furthermore, since we are considering a system of massless particles, the scalar irreducible moments also only contain contributions that are of third order or higher and therefore shall be neglected in this derivation. In Eqs.~\eqref{eq:transient-l=1} and \eqref{eq:transient-l=2}, we make use of the following definitions
\begin{equation}
\alpha_{r}^{(1)} \equiv 
I_{r+1,1}  - \frac{n_{0}}{\varepsilon_{0} + P_{0}} I_{r+2,1} , \quad
\alpha_{r}^{(2)} \equiv 2 \beta I_{r+3,2} ,
\end{equation}
which are expressed in terms of the thermodynamic integrals defined in Eq.~\eqref{thermodynamic_integral}. We further defined the fluid vorticity $\omega_{\mu \nu} \equiv (\nabla_{\mu}u_{\nu} - \nabla_{\nu}u_{\mu} )/2$. 

The quantities, $\mathcal{C}^{\alpha}_{r-1}$ and $\mathcal{C}^{\alpha \beta}_{r-1}$, are moments of the collision term, defined as implied. These terms give rise to the relaxation time scales within which the system evolves towards local equilibrium and are essential to explain how the system approaches the fluid-dynamical limit. In particular, such transport coefficients are essential to ensure the causality and stability of a fluid-dynamical formulation.

\subsection{Irreducible moments of the collision term}
\label{sec:col-mom}

The most complicated aspect of consistently deriving fluid dynamics 
is to calculate the moments of the collision term that appear in Eqs.\ \eqref{eq:transient-l=1} and \eqref{eq:transient-l=2}. In general, these terms can depend on all irreducible moments of the distribution function and render the equations of motion for such moments highly coupled. In the following, we shall demonstrate that (inspired in the results of Ref.\ \cite{Denicol:2022bsq}) these terms become considerably simpler when considering a system composed of massless scalar particles with a $\lambda \varphi^4$ self-interaction. In this case, all that we must consider are Eq.\ \eqref{eq:transient-l=1} with $r=2$ and Eq.\ \eqref{eq:transient-l=2} with $r=1$. This happens because the moments of the collision term appearing in such equations only depend on the dissipative currents contained in $N^{\mu}$ and $T^{\mu\nu}$. This will allow us to systematically derive fluid-dynamical equations without resorting to phenomenological approximations of the collision integral \cite{bhatnagar:54model,marle:69etab,marle:69-2-etablissement, andersonRTA:74,Rocha:2021zcw,Hu:2022mvl} or any truncation procedure \cite{Rocha:2021lze,Denicol:2012cn,Rocha:2022ind}. %
This calculation will be outlined in the following subsections, with further details of the derivation being described in Appendix \ref{app:collisional-moments}.


\subsubsection{Collision integral of rank 2}

We start with the rank-2 moment of the collision term, $\mathcal{C}_{0}^{\mu \nu}$. In this case, it is convenient to separate it into its loss and gain parts,
\begin{equation}
\label{eq:collisionrank2}
\begin{aligned}
\mathcal{C}_{0}^{\mu \nu}&=\frac{1}{2}\int dP dQ \ dQ^{\prime} \ dP^{\prime} \ p^{\langle \mu} p^{\nu \rangle}  W_{\mathbf{p}\mathbf{p}' \leftrightarrow \mathbf{q}\mathbf{q}'}  ( f_{\mathbf{q}}f_{\mathbf{q}'} 
-
f_{\mathbf{p}}f_{\mathbf{p}'}  )
\\
&= - \frac{g}{2} \int dP dP' f_{\mathbf{p}} f_{\mathbf{p}'} p^{\langle\mu} p^{\nu\rangle} \int dQ dQ'  (2\pi)^5 \delta^{(4)}(p^\alpha + p'^\alpha - q^\alpha - q'^\alpha) \\
&+ \frac{g}{2}\int dP dP' f_{\mathbf{p}}f_{\mathbf{p}'} \int dQ dQ' q^{\langle\mu}q^{\nu\rangle} (2\pi)^5 \delta^{(4)}(p^\alpha + p'^\alpha - q^\alpha - q'^\alpha) ,
\end{aligned}
\end{equation}
where, in the second equality, we replaced the expression for the transition rate given in Eq.~\eqref{eq:cross-sec-phi4}. Furthermore, in-going and out-going momentum labels were exchanged $\mathbf{p} \leftrightarrow \mathbf{q}$, $\mathbf{p}' \leftrightarrow \mathbf{q}'$ in the first term, where we used the time-reversal property of the transition rate $W_{\mathbf{p}\mathbf{p}' \leftrightarrow \mathbf{q}\mathbf{q}'}  = W_{\mathbf{q}\mathbf{q}' \leftrightarrow  \mathbf{p}\mathbf{p}'}$. 

The first integral can be immediately calculated using that \cite{Denicol:2012cn,Bazow:2015dha,Bazow:2016oky,Mullins:2022fbx}
\begin{eqnarray}
\label{eq:relation0}
\int dQ dQ'  (2\pi)^5 \delta^{(4)}(p^\alpha + p'^\alpha - q^\alpha - q'^\alpha) = 1.
\end{eqnarray}
Then, we obtain
\begin{eqnarray}
- \frac{g}{2} \int dP dP' dQ dQ' f_{\mathbf{p}} f_{\mathbf{p}'} p^{\langle\mu} p^{\nu\rangle} (2\pi)^5 \delta^{(4)}(p^\alpha + p'^\alpha - q^\alpha - q'^\alpha) &=&
- \frac{g}{2}(\rho_0 + I_{0,0})\pi^{\mu\nu}.
\label{eq:firstterm_rank2}
\label{first_term_rank2_final}
\end{eqnarray}
The second term can be calculated using that \cite{Denicol:2012cn}
\begin{eqnarray}
\label{eq:relation1}
 \int dQ dQ' q^{\langle\mu}q^{\nu\rangle} (2\pi)^5 \delta^{(4)}(p^\alpha + p'^\alpha - q^\alpha - q'^\alpha) 
 =  \frac{1}{3} Q_T^{\langle\mu} Q_T^{\nu\rangle}, 
\end{eqnarray}
where we have introduced the total 4-momentum of the collision, $Q^\mu_T \equiv p^\mu + p'^\mu$. This leads to the following exact result,
\begin{eqnarray}
 \frac{g}{2}\int dP dP' f_{\mathbf{p}}f_{\mathbf{p}'} \int dQ dQ' q^{\langle\mu}q^{\nu\rangle} (2\pi)^5 \delta^{(4)}(p^\alpha + p'^\alpha - q^\alpha - q'^\alpha)
 =  \frac{g}{3} (\rho_0 + I_{0,0}) \pi^{\mu\nu} + \frac{g}{3} \nu^{\langle\mu} \nu^{\nu\rangle}.
\label{eq:secondterm_rank2}
\end{eqnarray}
Combining the results from Eqs.~\eqref{eq:firstterm_rank2} and \eqref{eq:secondterm_rank2}, we establish that this irreducible moment of the collision term can be expressed as, 
\begin{equation}
\mathcal{C}^{\mu\nu}_0 = - \frac{g}{6} (\rho_0 + I_{0,0}) \pi^{\mu\nu} + \frac{g}{3} \nu^{\langle\mu} \nu^{\nu\rangle}. \label{collision_ranp2_cross}
\end{equation}
Therefore, we conclude that this collisional moment can be \textit{exactly} expressed solely in terms of hydrodynamic variables and the moment $\rho_0$. We note that this result is exact and no approximations related to the existence of a hydrodynamic limit were imposed. Relations 
\eqref{eq:relation0} and \eqref{eq:relation1} are derived in Appendix \ref{app:collisional-moments}, following the procedure outlined in Ref.\ \cite{Denicol:2012cn}.

\subsubsection{Collision integral of rank 1}

Now we proceed with the rank-1 moment of the collision term, $\mathcal{C}^{\mu}_1$. As before, we separate it in the following way,
\begin{eqnarray}
\mathcal{C}^{\mu}_1 & = & \frac{1}{2}\int dP dQ \ dQ^{\prime} \ dP^{\prime} E_{\mathbf{p}} \ p^{\langle \alpha} p^{\beta \rangle}  W_{pp' \leftrightarrow qq'} ( f_{\mathbf{q}}f_{\mathbf{q}'} - f_{\mathbf{p}}f_{\mathbf{p}'} ) \notag \\
&=& -\frac{g}{2}\int  dP dP' f_{\mathbf{p}}f_{\mathbf{p}'} E_{\mathbf{p}} p^{\langle\mu\rangle} \int dQ dQ'  (2\pi)^5 \delta^{(4)}(p^\alpha + p'^\alpha - q^\alpha q'^\alpha) \\
&+&\frac{g}{2}\int dP dP' f_{\mathbf{p}}f_{\mathbf{p}'} \int dQ dQ' E_{\mathbf{q}} q^{\langle\mu\rangle} (2\pi)^5 \delta^{(4)}(p^\alpha + p'^\alpha - q^\alpha - q'^\alpha), \notag \label{eq:collisionrank1}
\end{eqnarray}
where, in the second equality, we replaced the expression for the transition rate given in Eq.~\eqref{eq:cross-sec-phi4}. Also, similarly to the calculation of the rank-2 moment of the collision term, in-going and out-going momentum labels were exchanged in the second term, $\mathbf{p} \leftrightarrow \mathbf{q}$, $\mathbf{p}' \leftrightarrow \mathbf{q}'$, using the time-reversal property of the transition rate. 

Once again, we shall calculate the two terms on the right-hand side separately. The first term can be calculated directly using \eqref{eq:relation0} and simply reads
\begin{eqnarray}
- \frac{g}{2}\int  dP dP'dQ dQ' f_{\mathbf{p}}f_{\mathbf{p}'} E_{\mathbf{p}} p^{\langle\mu\rangle}  (2\pi)^5 \delta^{(4)}(p^\alpha + p'^\alpha - q^\alpha - q'^\alpha) & = &
- \frac{g}{2}\int  dP dP' f_{\mathbf{p}}f_{\mathbf{p}'} E_{\mathbf{p}} p^{\langle\mu\rangle} \notag \\
&=& (\rho_0+I_{0,0})\rho_1^{\mu} .
\label{second_term1_final}
\end{eqnarray}
This term vanishes due to the Landau matching conditions, in which the fluid 4-velocity is defined so that the energy diffusion current, $\rho_1^{\mu}$, is zero -- see Eq.\ \eqref{eq:kinetic_match}.  

The second term can be calculated using the relation \cite{Denicol:2012cn}, 
\begin{eqnarray}
\label{eq:relation2}
 \int dQ dQ' E_{\mathbf{q}} q^{\langle\mu\rangle} (2\pi)^5 \delta^{(4)}(p^\alpha + p'^\alpha - q^\alpha - q'^\alpha)
 =  \frac{1}{3} u_\alpha Q^\alpha_T Q^{\langle\mu\rangle}_T , 
\end{eqnarray}
where $Q^\mu_T$ is the total 4-momentum of the collision. This leads to,
\begin{eqnarray}
&& \frac{g}{2}\int dP dP'dQ dQ' f_{\mathbf{p}}f_{\mathbf{p}'}  E_{\mathbf{q}} q^{\langle\mu\rangle} (2\pi)^5 \delta^{(4)}(p^\alpha + p'^\alpha - q^\alpha - q'^\alpha)
 = \frac{g}{3}\left[(\rho_0+I_{0,0}) \rho_1^\mu+n_0 \nu^\mu\right]. \label{first_term1_final}
\end{eqnarray}

Therefore, using the results derived in Eqs.~\eqref{second_term1_final} and \eqref{first_term1_final}, it is possible to rewrite $\mathcal{C}_1^\mu$ as
\begin{equation}
\mathcal{C}_1^\mu = - \frac{g}{6} (\rho_0+I_{0,0}) \rho_1^\mu + \frac{g}{3}n_0 \nu^\mu. \label{eq:collision_rank1_final}
\end{equation}
As already mentioned, since we employ Landau matching conditions in this section, the first term of the equation above simply vanishes. Nevertheless, we kept it in the equation for the sake of completeness. Thus, as was the case with $\mathcal{C}_2^{\mu\nu}$, $\mathcal{C}_1^{\mu}$ can be expressed solely in terms of the dissipative currents appearing in the conserved currents. 

\subsection{Order of Magnitude truncation scheme}

In the derivation of hydrodynamics from the Boltzmann equation, it is essential to reduce the degrees of freedom appearing in the general moment equations to the hydrodynamic fields ($n, \varepsilon , u^{\mu}, \nu^{\mu},  \pi^{\mu \nu}$, in the present case). In the present work, inspired in methods developed in the non-relativistic regime \cite{struchtrup2004stable}, we will use the order of magnitude truncation scheme \cite{Fotakis:2022usk}. This method is based on a power-counting scheme that estimates the order of magnitude of a given moment using its asymptotic expression in a gradient expansion. In the end, this will lead to a power-counting scheme solely based on space-like gradients of the fluid-dynamical variables or, equivalently, in the Knudsen number \cite{Denicol:2021}. For instance, at first order in a gradient expansion, the irreducible moments \eqref{eq:irreducible_moments} of rank 1 and 2 are well described by their respective Navier-Stokes values,
\begin{equation}
\begin{aligned}
&
\rho_{r}^{\mu} = \kappa_{r} \nabla^{\mu}\alpha + \mathcal{O}(2),
\\
&
\rho_{r}^{\mu \nu} = 2 \eta_{r} \sigma^{\mu \nu} + \mathcal{O}(2),
\end{aligned}    
\end{equation}
where microscopic expressions for $\kappa_r$ and $\eta_r$ are given in \eqref{eq:transportcoeffs} and $\mathcal{O}(2)$ denotes terms that are of second order in gradients. These relations can be re-arranged to express, at first order in a gradient expansion, all irreducible moments of rank-1 and 2 solely in terms of $\nu^\mu$ and $\pi^{\mu\nu}$, respectively. This procedure leads to,
\begin{subequations}
\label{eq:OoM-relations}
\begin{align}
&
\label{eq:OoM-pdiff}
\rho_{r}^{\mu} = \frac{\kappa_{r}}{\kappa} \nu^{\mu} + \mathcal{O}(2)
\equiv
B_{r} \nu^{\mu} + \mathcal{O}(2),
\\
&
\label{eq:OoM-shear}
\rho_{r}^{\mu \nu} = \frac{\eta_{r}}{\eta} \pi^{\mu \nu} + \mathcal{O}(2)
\equiv
C_{r} \pi^{\mu \nu} + \mathcal{O}(2).
\end{align}    
\end{subequations}
These relations will be employed to reduce the equations of motion for the irreducible moments, Eqs.\ \eqref{eq:transient-l=1} and \eqref{eq:transient-l=2}, for $r=2$ and $r=1$, respectively, to closed equations of motion for the dissipative currents. The resulting equations will include all terms that are asymptotically of second-order in a gradient expansion and, thus, it is usually regarded as a second-order theory. We note that, in the massless and classical limits, the thermodynamic integrals can be evaluated
analytically and the coefficients $B_r$ and $C_r$ have the following simple form,
\begin{equation}
\label{eq:Cr-OoM}
\begin{aligned}
&
B_{r} = -\frac{1}{6\beta^{r}}(r-1) \Gamma(r+4) , \, \, \,
C_{r} = \frac{1}{120 \beta^{r}} \Gamma(r+6),
\end{aligned}    
\end{equation}
where $\Gamma(x)$ is the Gamma function \cite{gradshteyn2014table}. Now we collect these results and use them to simplify the second order terms in Eqs.~\eqref{eq:transient-l=1} and \eqref{eq:transient-l=2}. 

\subsubsection{Particle diffusion 4-current}

We start with the derivation of the equation of motion for the particle diffusion 4-current. As outlined above, we consider Eq.~\eqref{eq:transient-l=1}, with $r=2$, and employ the exact expression for the moment of the collision term, Eq.~\eqref{eq:collision_rank1_final}. We then use Eqs.~\eqref{eq:OoM-relations} to approximate all second-order terms in this equation, that will be then given solely in terms of $\nu^{\mu}$ and $\pi^{\mu\nu}$ and derivatives thereof. Since we apply this approximation to terms that are at least of order 2, the accuracy of the approximation, in this asymptotic power-counting scheme, will be of order 3. Thus, we derive
\begin{equation}
\label{eq:eom-l=1}
\begin{aligned}
& 
\tau_{\nu} D\nu^{\langle \lambda \rangle} 
+
\nu^{\lambda}  
=
\kappa \nabla^{\lambda}\alpha
-
\delta_{\nu \nu} \nu^{\lambda}\theta 
- (\lambda_{\nu \pi}
\nabla_{\mu} \alpha + \tau_{\nu \pi}
\nabla_{\mu} P_{0})\pi^{\lambda \mu} 
+
\ell_{\nu \pi}\Delta^{\lambda}_{ \ \alpha}
\nabla_{\mu}\pi^{\alpha \mu} 
-
\frac{7}{5}\tau_{\nu} \sigma_{\mu}^{\ \lambda}    \nu^{\mu} 
-
\tau_{\nu} \omega_{\mu}^{\ \lambda}  \nu^{\mu}
,
\end{aligned}    
\end{equation}
where we have employed the following relations,
\begin{equation}
\label{eq:Da-Db-Du}
\begin{aligned}
D \beta & = \frac{\beta}{3} \theta + \mathcal{O}(2), \, \, \, \, 
Du^{\mu}
 =
 \frac{1}{4P_{0}} \nabla^{\mu}P_{0} 
+ \mathcal{O}(2).
\end{aligned}
\end{equation}
In Eq.~\eqref{eq:eom-l=1}, the different transport coefficients can be expressed as the following functions of $\alpha$, $\beta$ and the coupling coefficient $g$, 
\begin{equation}
\begin{aligned}
&
\tau_{\nu} = - 
\frac{3B_{2}}{g n_{0}} = \frac{60}{g n_{0} \beta^{2}}, 
\\
&
\kappa
= 
\frac{\tau_{\nu}}{B_{2}} \left(
I_{3,1}  - \frac{\beta}{4} I_{4,1} \right) 
=
\frac{3}{g \beta^{2}},
\\
&
\lambda_{\nu \pi} = \tau_{\nu} \frac{\beta }{4 B_{2}}
\frac{\partial C_{1}}{\partial \beta}
=
\frac{3 \tau_{\nu} \beta}{40},
\\
&
\tau_{\nu \pi} 
= - \frac{\tau_{\nu}}{B_{2}}
\left(\beta \frac{\partial C_{1}}{\partial \beta}
+
2 C_{1} 
- \frac{\beta J_{4,1}}{4P_{0}}
\right)\frac{1}{4P_{0}}
=
\frac{\tau_{\nu} \beta}{80 P_{0}},
\\
&
\ell_{\nu \pi} = -\frac{\tau_{\nu}}{B_{2}} \left( C_{1} -
\frac{\beta J_{4,1}}{4P_{0}}
\right)
=
\frac{\tau_{\nu} \beta}{40},
\\
&
\delta_{\nu \nu} = \frac{\tau_{\nu}}{3} \left(5 + \frac{\beta}{B_{2}}\frac{\partial B_{2}}{\partial \beta}  \right)
=
\tau_{\nu}.
\end{aligned}
\end{equation}
We note that the particle diffusion coefficient behaves as $\kappa \sim T^2$ and agrees with the result derived in Sec.\ \ref{Sec:NS} and in Ref.~\cite{Denicol:2022bsq} for Navier-Stokes theory. We further remark that the relaxation time $\tau_{\nu}$, which sets the timescale for the decay of the particle diffusion 4-current, behaves as $\tau_\nu \sim (T \exp{\alpha})^{-1}$, displaying a dependence on the fugacity as well as on the inverse temperature. It is also readily seen that due to the fact that the transport coefficients are proportional to $1/g$, in the perturbative limit, all coefficients are large.

\subsubsection{Shear-stress tensor}

The next step is to obtain an equation of motion for the shear-stress tensor. We start from the equation of motion for the irreducible moment of rank 2, Eq.~\eqref{eq:transient-l=2}, with $r=1$, and make use of Eq.~\eqref{collision_ranp2_cross}. Once again, all second order terms in the equation are approximated in terms of the hydrodynamic variables and derivatives thereof using Eqs.~\eqref{eq:OoM-relations}. This will lead to an equation of motion that is accurate up to second order in a gradient expansion. The equation of motion for $\pi^{\mu\nu}$ then reads
\begin{equation}
\label{eq:eom-shear}
\begin{aligned}
& \tau_{\pi} D\pi^{\langle \lambda \mu \rangle}
+
\pi^{\lambda \mu} 
=
2 \eta  \sigma^{\lambda \mu}
+
\varphi_{8}  \nu^{\langle \lambda} \nu^{\mu \rangle}
-
\delta_{\pi \pi} \pi^{\lambda \mu} \theta
- 
\tau_{\pi \nu}  \nabla^{\langle \lambda} P_{0} \ \nu^{\mu \rangle} 
+
\ell_{\pi \nu}  
 \nabla^{\langle \lambda}  \nu^{\mu \rangle} 
+
\lambda_{\pi \nu}
\nabla^{\langle \lambda} \alpha \ \nu^{\mu \rangle} 
\\
&
-
2 \tau_{\pi}
 \omega_{\nu}^{\ \langle \lambda} \pi^{\mu \rangle \nu} 
- 
 \tau_{\pi\pi}
\sigma_{\nu}^{\ \langle \lambda}  \pi^{\mu \rangle \nu} 
,
\end{aligned}    
\end{equation}
where we have employed once again relations \eqref{eq:Da-Db-Du} and general thermodynamic relations. The transport coefficients appearing in Eq.~\eqref{eq:eom-shear} are calculated exactly and read,
\begin{equation}
\label{eq:transp-coeff-l=2}
\begin{aligned}
& \tau_{\pi} \equiv \frac{6 C_{1}}{g I_{0,0}} = \frac{72}{g n_{0} \beta^{2}}, 
\\
& \eta = \frac{\beta I_{4,2}}{C_{1}} \tau_{\pi}
=
\frac{48}{g\beta^{3}},
\\
&
\varphi_{8} = \frac{g \tau_{\pi}}{3 C_{1}}
=
\frac{4}{n_{0} \beta},
\\
&
\delta_{\pi \pi}
=
\tau_{\pi}\left(
\frac{5}{3}
+
 \frac{\beta}{3C_{1}} \frac{\partial C_{1}}{\partial \beta} 
 \right)
=
\frac{4}{3}\tau_{\pi},
\\
&
\tau_{\pi\pi} = 2 \tau_{\pi}, 
\\
&
\tau_{\pi \nu} = \frac{\tau_{\pi}}{C_{1}}
\frac{1}{10 P_{0}}\left(
6 B_{2} 
+ 
\beta
 \frac{\partial B_{2}}{\partial \beta} \right)
 = - \frac{4}{3} \frac{\tau_{\pi}}{n_{0}}, 
\\
&
\ell_{\pi \nu} = \frac{2B_{2}}{5C_{1}} \tau_{\pi}
= -\frac{4}{3} \frac{\tau_{\pi}}{\beta},
\\
&
\lambda_{\pi \nu} = \frac{2 \tau_{\pi}}{5C_{1}} 
\frac{n_{0}}{\varepsilon_{0}+P_{0}}
\frac{\partial B_{2}}{\partial \beta}
= 
\frac{2}{3} \frac{\tau_{\pi}}{\beta}. 
\end{aligned}    
\end{equation}
We note that the shear viscosity behaves as $\eta \sim T^3$ and agrees with the result derived in Sec.\ \ref{Sec:NS} and in Ref.~\cite{Denicol:2022bsq} for Navier-Stokes theory. The shear relaxation time behaves as $\tau_\pi \sim (T \exp{\alpha})^{-1}$, displaying also a dependence on the fugacity of the gas. Moreover, we note that the shear relaxation time is larger than the particle diffusion relaxation time, $\tau_\pi = (6/5) \tau_{\nu} > \tau_\nu$. This implies that the particle diffusion 4-current relaxes to its Navier-Stokes limit prior to the shear-stress tensor. 

The exact expressions obtained above display a few quantitative differences to traditional calculations that employ either the relaxation time or the 14-moment approximations \cite{Denicol:2010xn,Denicol:2012cn,Denicol:2014vaa}. First, the relaxation time, is related to the shear viscosity as 
\begin{equation}
\label{eq:rlx-time-t-eta}
\tau_\pi = \frac{6\eta}{\varepsilon_0 + P_0},
\end{equation}
with the traditional calculations cited above obtaining a factor 5 instead of 6. This modification does not affect the linear causality and stability of the theory around global equilibrium, since it increases the relaxation time. Also, we note that the transport coefficient $\tau_{\pi \pi}$ is given by $2\tau_\pi$  instead of $(10/7) \tau_\pi$. The coefficient $\delta_{\pi \pi}$ remains as $(4/3) \tau_{\pi}$, as expected of a conformal system. Finally, we note that the product of the transport coefficients that couple the dissipative currents, namely $\ell_{\nu \pi}$ and $\ell_{\pi \nu}$, is negative, which is in agreement with the second law of thermodynamics \cite{israel1979annals,Muronga:2003ta}. We remark that all the expressions for the transport coefficients derived in this section are exact -- something that was not accomplished before, without relying on toy models for the collision term. 

\section{Bjorken flow}
\label{sec:eoms-bjorken}

In this section, we analyze the several fluid-dynamical formulations we have derived throughout the past sections in a Bjorken flow scenario \cite{Bjorken:1982qr}. Inspired in the phenomenology of heavy ion collisions, Bjorken flow assumes that the system is longitudinally boost-invariant, homogeneous and isotropic in the transverse plane and invariant by reflection around the longitudinal axis. In this case, it is convenient to make use of hyperbolic coordinates, with $\tau = \sqrt{t^{2}-z^{2}}$ and $\eta = \tanh^{-1}(z/t)$ being the proper time and space-time rapidity, respectively. In this case, the line element of Minkowski space reads, $ds^{2} =  d\tau^{2} - dx^{2} - dy^{2} - \tau^{2} d \eta^{2}$. Moreover, the only non-zero components of the Christoffel symbols are $\Gamma^{\tau}_{\eta \eta} = \tau$, $\Gamma^{\eta}_{\tau \eta}=\Gamma^{\eta}_{\eta \tau} = 1/\tau$.
In this coordinate system, a trivial fluid 4-velocity, $u^{\mu} = (1,0,0,0)$, satisfies Eq.~\eqref{eq:hydro-EoM-eps} without leading to a stationary solution -- Eq.~\eqref{eq:hydro-EoM-umu}, on the other hand, is trivially satisfied. In addition, the expansion rate and shear tensor do not vanish and manifest the strong longitudinal expansion displayed by the system (contained in the metric tensor),
\begin{equation}
\theta = \frac{1}{\tau} , \, \, \, 
\sigma^{\mu}_{\ \nu} =  \text{diag}\left(0, -\frac{1}{3 \tau}, -\frac{1}{3\tau}, \frac{2}{3\tau} \right).
\end{equation}
Furthermore, the shear-stress tensor has only one independent component, which we take as $\pi^{\eta}_{\ \eta} \equiv \pi$, and can be expressed as
\begin{equation}
\pi^{\mu}_{\ \nu} = \text{diag}\left(0, - \frac{\pi}{2} , - \frac{\pi}{2}, \pi \right).
\end{equation}
We note that the reflection symmetries assumed further imply that 4-vectors orthogonal to $u^{\mu}$, such as $\nu^{\mu}$, $\nabla^{\mu} P_0$, and $\nabla^{\mu} \alpha$, all vanish identically.

\subsection{Navier-Stokes theory}

In Bjorken flow, the Navier-Stokes equations derived in Sec.\ \ref{Sec:NS} become,
\begin{subequations}
\label{eq:NS-EoM1-bjorken0-chap8}
\begin{align}
\label{eq:NS-EoM1-bjorken1-chap8}
&  \dot{n}_{0} 
 + 
 \frac{n_{0}}{\tau}
  = 0 , \\
&  \dot{\varepsilon}_{0} 
 + 
 \frac{4 \varepsilon_{0}}{3 \tau}
 -
 \frac{64}{g \beta^{3} \tau^{2}}
 = 0 .
 \label{eq:Hilb-EoM-eps-2-chap-8}
\end{align}
\end{subequations}
Using the solution of the equation of motion \eqref{eq:NS-EoM1-bjorken1-chap8}, $n_{0}(\tau) = n_{0}(\tau_{0}) (\tau_{0}/\tau)$ (where $\tau_{0}$ is the initial time) and the equation of state $\varepsilon_{0}(\tau) = 3n_{0}(\tau)/\beta(\tau)$, Eq.~\eqref{eq:Hilb-EoM-eps-2-chap-8} can be expressed as
\begin{equation}
\label{eq:eoms-rexp-NS}
\begin{aligned}
&  \frac{d\varepsilon_{0}}{d \Hat{\tau}} 
 + 
 \frac{4 \varepsilon_{0}}{3 \Hat{\tau}}
 - 
\frac{\varepsilon_{0}^{3}}{K_{\mathrm{NS}}^{2}} \Hat{\tau}   = 0,
\\
 &
K_{\mathrm{NS}}^{2} \equiv \frac{g}{64} \varepsilon_{0}(\tau_{0})^{3} \beta(\tau_{0})^{3} \tau_{0},   
\end{aligned}    
\end{equation}
where $K_{\mathrm{NS}}$ is a constant with dimension of energy density and we also introduced the dimensionless time variable $\Hat{\tau} = \tau/\tau_{0}$.
The above equation is analytically solved by
\begin{equation}
\label{eq:sol-NS-BJ}
\varepsilon_{0}(\tau) = \varepsilon_{\mathrm{0,id}}(\tau)  \left\{ 1 - 3 \left[ \frac{\varepsilon_{0}(\tau_{0})}{K_{\mathrm{NS}}}\right]^{2} \left(1 -\frac{1}{\Hat{\tau}^{2/3}} \right)   \right\}^{-1/2},
\end{equation}
where $\varepsilon_{\mathrm{0,id}}(\tau) = \varepsilon_{0}(\tau_{0})  (\tau_{0}/\tau)^{4/3}$ is the solution for the energy density of a perfect fluid in Bjorken flow. It is straightforward to see that this solution becomes imaginary, and thus unphysical, at sufficiently long times if
\begin{equation}
\label{eq:phys-cond-NS}
\begin{aligned}
&
\left[\frac{\varepsilon_{0}(\tau_{0})}{K_{\mathrm{NS}}}\right]^{2} > \frac{1}{3} 
\Rightarrow
\frac{g}{192} \varepsilon_{0}(\tau_{0}) \beta(\tau_{0})^{3} \tau_{0} < 1,
\end{aligned}    
\end{equation}
which may occur for sufficiently small values of the coupling constant, chemical potential and/or initial time and for sufficiently large values of temperature. This behavior is not observed for Hilbert theory, see Appendix \ref{appendix:Hilbert}.

\begin{figure}[!ht]
\centering
  \includegraphics[scale=0.3]{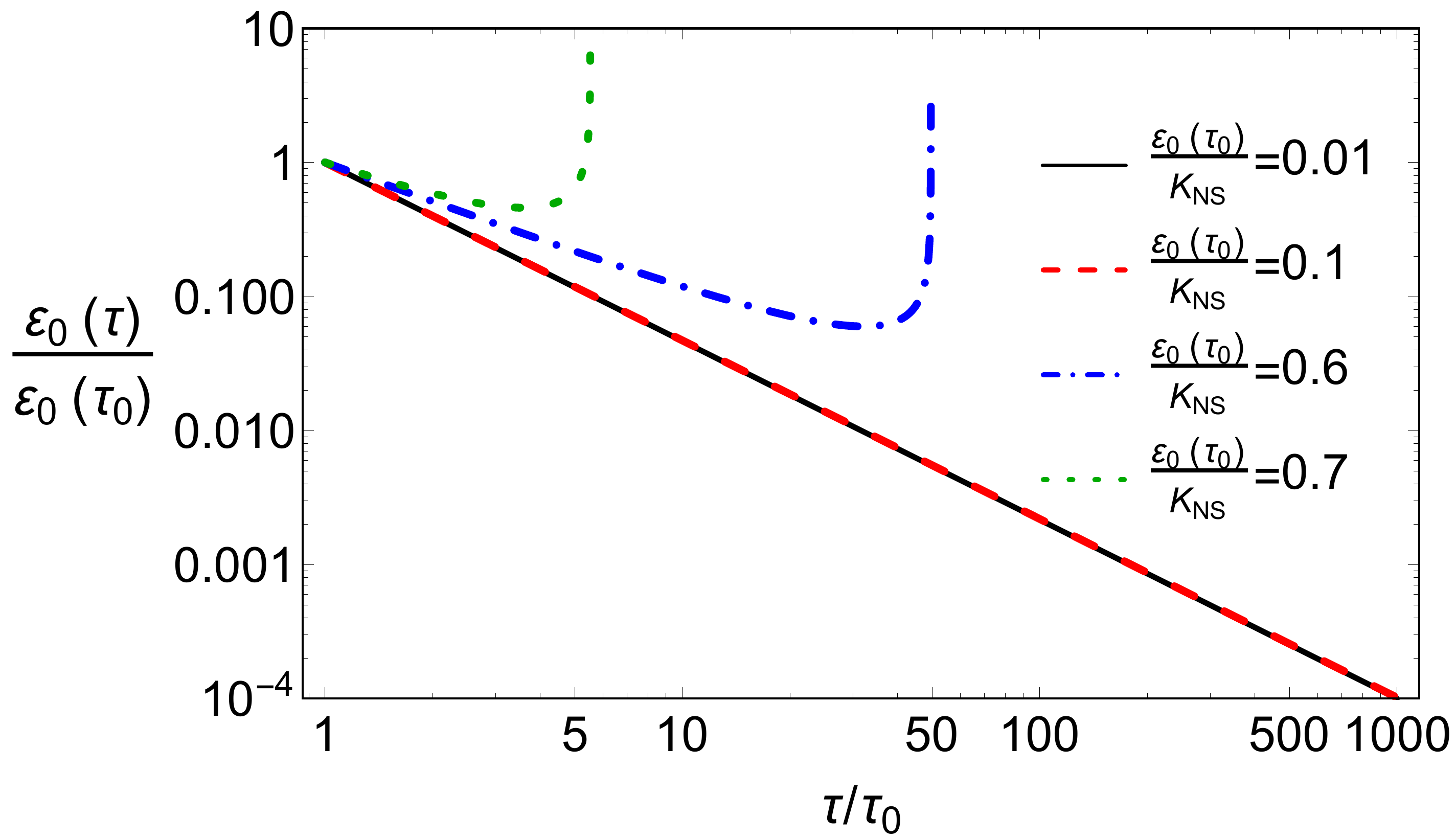}
\caption{(Color online) Solutions of the Navier-Stokes equations for some values of the dimensionless ratio $\varepsilon_{0}(\tau_{0})/K_{\mathrm{NS}}$.}
\label{fig:NS-solutions}
\end{figure}

Moreover, we remark that the ratio $[\varepsilon_{0}(\tau_{0})/K_{\mathrm{NS}}]^{2}$ is proportional to the initial value of the Knudsen number \cite{Denicol:2012cn}, which is usually defined as the ratio between the typical microscopic and macroscopic time/distance scales of the system. In Bjorken flow, all gradients of velocity behave as $1/\tau$ and thus the time coordinate $\tau$ can be used to estimate the characteristic macroscopic scale of the system. The characteristic microscopic scale, on the other hand, depends on the interaction and, in this case, can be estimated by shear viscosity coefficient as 
\begin{equation}
\tau_{\textrm{micro}} \sim \tau_\eta \equiv \frac{\eta}{\varepsilon + P}.
\end{equation}
Thus, we may estimate the Knudsen number using the following ratio $\textrm{Kn}(\tau) \sim \tau_{\eta}(\tau)/\tau$. We can then demonstrate that the initial value of the Knudsen number, $\textrm{Kn}(\tau_0)$, is proportional to
$[\varepsilon(\tau_{0})/K_{\mathrm{NS}}]^{2}$. Quantitatively, one obtains the relation,
\begin{equation}
\label{eq:Kn-KNS-relation}
\textrm{Kn}(\tau_0) \sim \frac{\tau_{\eta}(\tau_0)}{\tau_0} = 
\frac{9}{16}\left[\frac{\varepsilon_{0}(\tau_{0})}{K_{\mathrm{NS}}}\right]^{2}   .
\end{equation}
Thus, solutions of Navier-Stokes theory cease to exist whenever sufficiently large gradients are initially applied to the system. This feature can be seen in Fig.~\ref{fig:NS-solutions}, where the energy density is shown to diverge at a finite time if $\varepsilon_{0}(\tau_{0})/K_{\mathrm{NS}} > 1/\sqrt{3} \simeq 0.577$. In principle, solutions of Navier-Stokes theory would not be reliable under such circumstances anyway, but the fact the solutions themselves do not even exist is certainly not a good feature of the theory.

It is also interesting to analyze the late time expansion of the solution given by Eq.~\eqref{eq:sol-NS-BJ},
\begin{equation}
\label{eq:late-time-expn-NS}
\begin{aligned}
& \frac{\varepsilon_{0}(\tau)}{\varepsilon_{0}(\tau_{0})}
= 
\frac{1}{\Hat{\tau}^{4/3}}
\frac{1}{\sqrt{1 - 3 \left[ \frac{\varepsilon_{0}(\tau_{0})}{K_{\mathrm{NS}}}\right]^{2} }}
\left[ 1  - \frac{3}{2} \frac{1}{ \left[ \frac{K_{\mathrm{NS}}}{\varepsilon_{0}(\tau_{0})}\right]^{2} - 3} \frac{1}{\Hat{\tau}^{2/3}} + \cdots \right],
\end{aligned}    
\end{equation}
which is quite different from the structure seen for solutions obtained assuming the relaxation time approximation \cite{andersonRTA:74,Rocha:2021zcw} with a constant relaxation time (see e.g.~Eq.~(111) of Ref.~\cite{Rocha:2022ind}). For instance, the leading term of the expansion does not decay as $\sim \tau^{-1}$,  but rather as  $\sim \tau^{-2/3}$.  Furthermore, all terms of the expansion depend on the initial condition, whereas in the constant relaxation time case they did not. We remark that the relaxation time of a $\varphi^{4}$ self-interacting system  is proportional to the energy, $E_{\bf p}$ \cite{Denicol:2022bsq,Calzetta:1986cq}. Thus, the above-mentioned features serve as evidence that an energy-dependent relaxation time can radically change the dynamics of a system.

\subsection{BDNK theory}

\begin{figure}[!ht]
\centering
\begin{subfigure}{0.5\textwidth}
  \includegraphics[width=\linewidth]{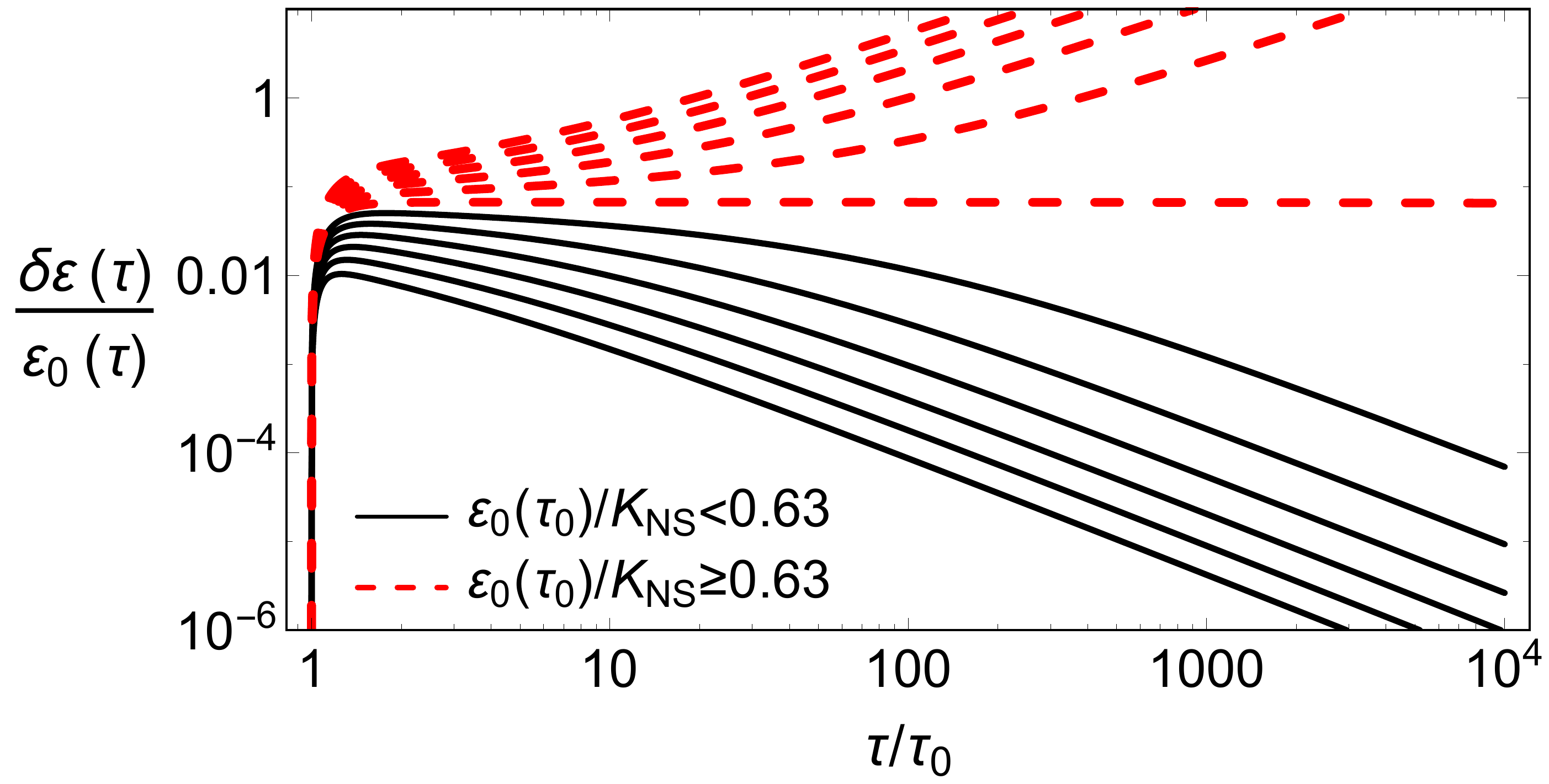}
  \caption{$s=3$}
  \label{fig:div-BDNK-1}
\end{subfigure}\hfil
\begin{subfigure}{0.5\textwidth}
  \includegraphics[width=\linewidth]{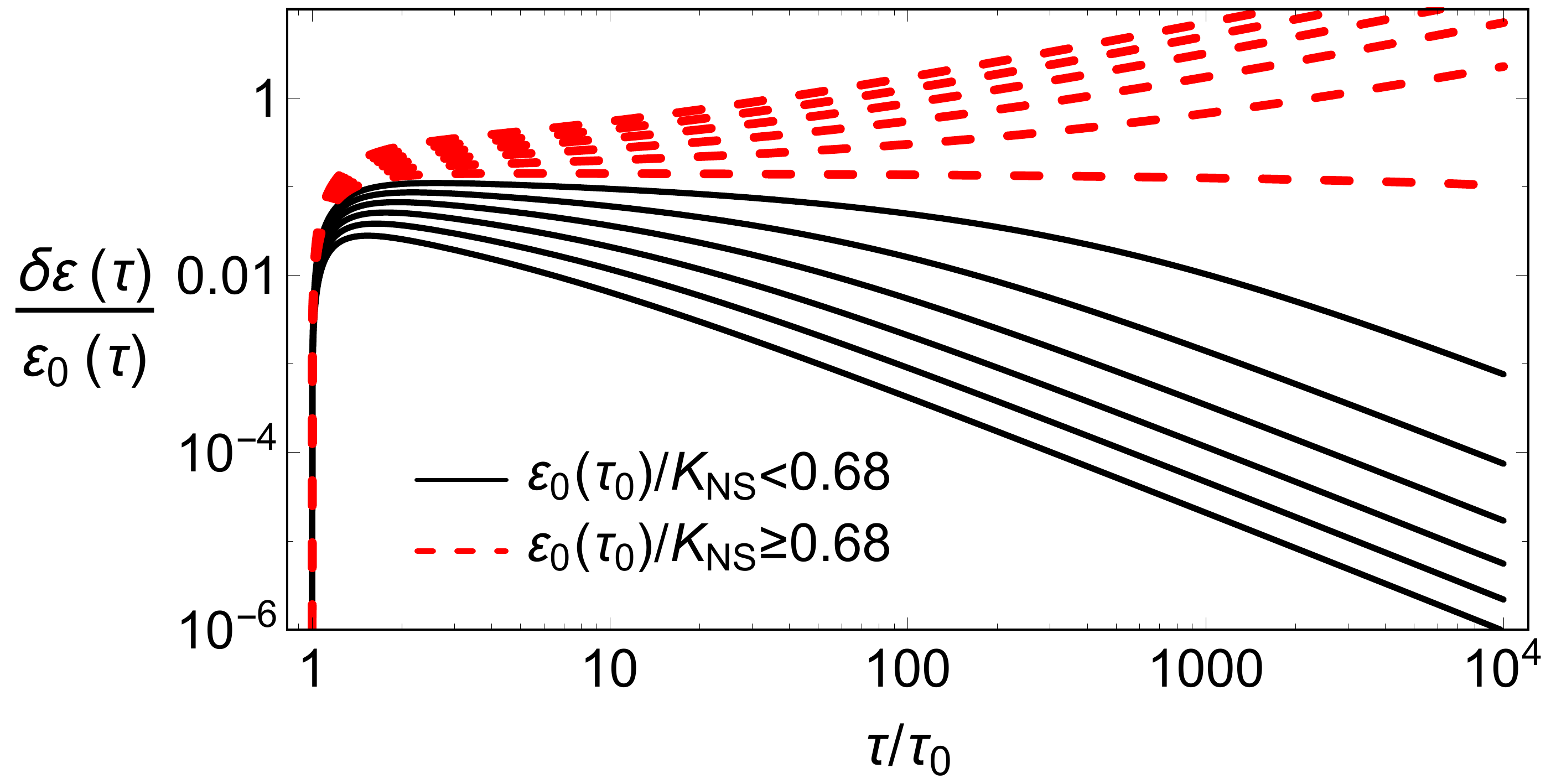}
  \caption{$s=4$}
  \label{fig:div-BDNK-2}
\end{subfigure}\hfil
\\
\begin{subfigure}{0.55\textwidth}
  \includegraphics[width=\linewidth]{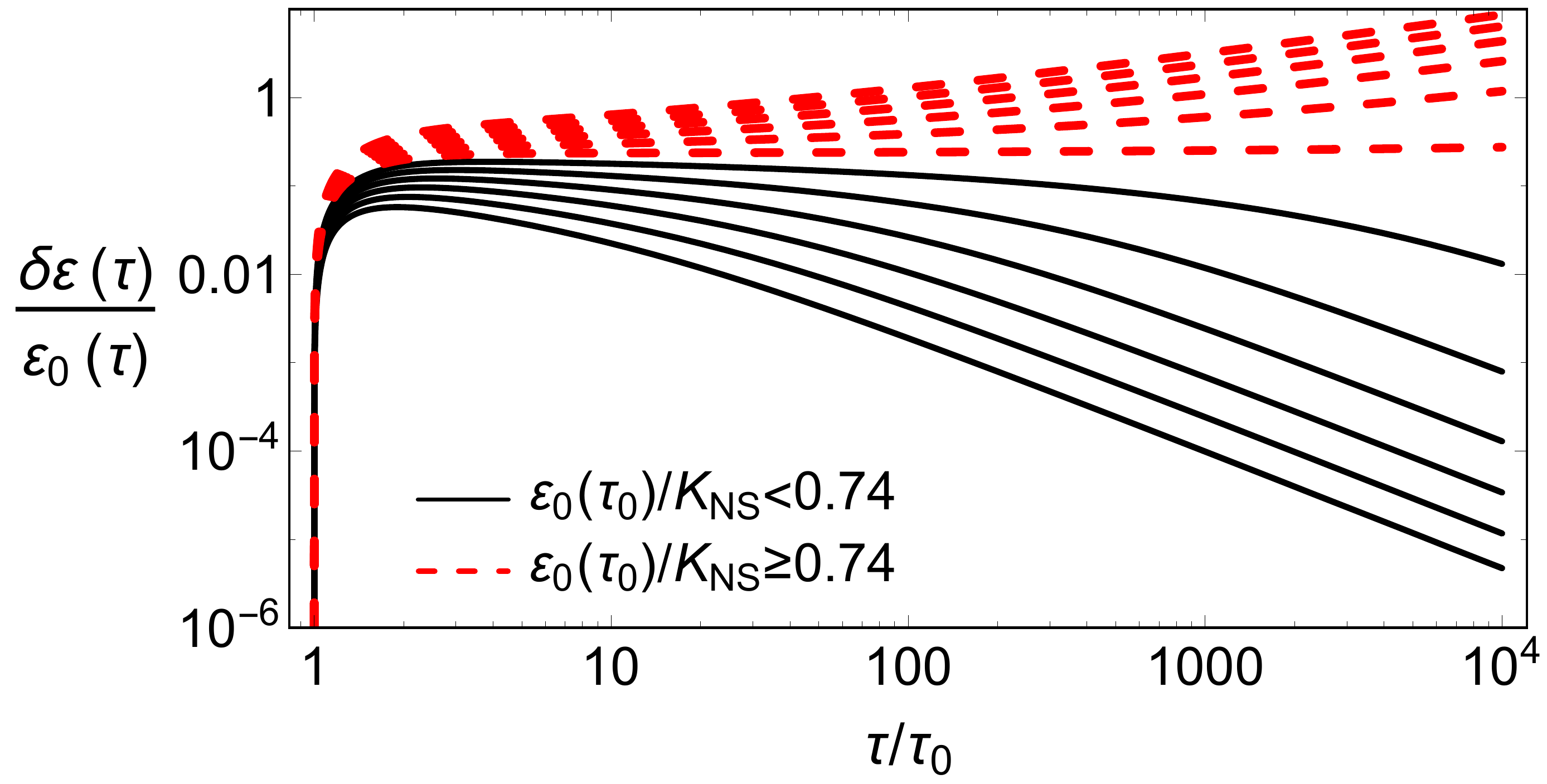}
  \caption{$s=5$}
  \label{fig:div-BDNK-3}
\end{subfigure}\hfil

\caption{(Color online) Solutions of Eq.~\eqref{eq:BDNK-coupled-eoms} for several initial conditions for (a) $s=3$ (b) $s=4$ (c) $s=5$. In each plot, solutions that evolve towards equilibrium at late times (black solid curves), and runaway solutions (red dashed curves) are displayed.}
\label{fig:div-BDNK-all}
\end{figure}

Now we discuss solutions of the BDNK theory derived in Sec.\ \ref{sec:BDNK-th} in Bjorken flow. To this end, we employ matching conditions so that $\delta n \equiv 0$ ($q=1$) and $\delta \varepsilon \neq 0$ ($s \neq 2$) -- these choices satisfy the necessary linear stability conditions shown in Eq.\ \eqref{eq:stability-cond-BDNK} as long as $s>2$. Then, collecting the results of Sec.~\ref{sec:BDNK-th} and the symmetry assumptions of Bjorken flow, we derive the following equations of motion
\begin{subequations}
\begin{align}
&
\label{eq:BDNK-n0-eom}
\dot{n}_{0} 
 + 
 \frac{n_{0}}{\tau}
  = 0 , \\
&
\label{eq:BDNK-e0-eom}
\Dot{\varepsilon}_{0} + \Dot{\delta \varepsilon} + \frac{4}{3 \tau} (\varepsilon_{0} + \delta \varepsilon) - \frac{64}{g \beta^{3} \tau^{2}}  = 0,  \\
&
\label{eq:BDNK-de-eom}
\delta \varepsilon = - \frac{36}{g\beta^3}(s-2) \left( \frac{\Dot{\beta}}{\beta} - \frac{1}{3 \tau} \right).
\end{align}    
\end{subequations}
Once again, using the solution of the equation of motion \eqref{eq:BDNK-n0-eom}, and the equation of state $\varepsilon_{0}(\tau) = 3n_{0}(\tau)/\beta(\tau)$, Eqs.~\eqref{eq:BDNK-e0-eom} and \eqref{eq:BDNK-de-eom} can be expressed as 
\begin{equation}
\label{eq:BDNK-coupled-eoms}
\begin{aligned}
&  \frac{d \Tilde{\delta \varepsilon}}{d \Hat{\tau}}
+
\frac{16}{9(s-2) \Hat{\tau}^{3}} \frac{\Tilde{\delta \varepsilon}}{\Hat{\varepsilon}_{0}^{2}} (1 + \Tilde{\delta \varepsilon})  
 - 
\Hat{\varepsilon}_{0}^{2} \Hat{\tau},
=
0\\
&
\frac{d\Hat{\varepsilon}_{0}}{d \hat{\tau}} + \frac{4}{3 \Hat{\tau}} \Hat{\varepsilon}_{0} 
-
\frac{16}{9(s-2) \Hat{\tau}^{3}} \frac{\Tilde{\delta \varepsilon}}{\Hat{\varepsilon}_{0}} 
=
0,
\end{aligned}    
\end{equation}
where we defined $\Tilde{\delta \varepsilon} \equiv \delta \varepsilon(\Hat{\tau})/\varepsilon_{0}(\Hat{\tau})$ and $\Hat{\varepsilon}_{0} = \varepsilon_{0}/K_{\mathrm{NS}}$.  Solving these coupled equations of motion, we have evidence that BDNK theory is also not amenable to large Knudsen number configurations. This is portrayed in Fig.~\ref{fig:div-BDNK-all}, which shows numerical solutions for equilibrium initial conditions, i.e., $\Tilde{\delta \varepsilon}(\Hat{\tau}_0) = 0$, considering several initial values of Knudsen number. For initial conditions with smaller values of Knudsen number, $\Tilde{\delta \varepsilon}$ decays to equilibrium with a late-time behavior of $\sim 1/\Hat{\tau}$. Meanwhile, for a sufficiently large initial value of the Knudsen number, $\Tilde{\delta \varepsilon}$ diverges at late-times and the system does not evolve towards equilibrium. For a given critical initial Knudsen number, which depends on the choice of matching condition, $\Tilde{\delta \varepsilon}$ evolves to a constant value at late times. The divergence of $\Tilde{\delta \varepsilon}$ at late times happens for $\varepsilon_{0}(\tau_{0})/K_{\mathrm{NS}} \gtrsim 0.63$ for $s=3$, $\varepsilon_{0}(\tau_{0})/K_{\mathrm{NS}} \gtrsim 0.68$ for $s=4$, and
$\varepsilon_{0}(\tau_{0})/K_{\mathrm{NS}} \gtrsim 0.74$ for $s=5$, which are of the same order of magnitude as the value for which Navier-Stokes theory produces imaginary solutions, $\varepsilon_{\mathrm{NS}}(\tau_{0})/K_{\mathrm{NS}} \gtrsim 0.577$ (cf.~\eqref{eq:phys-cond-NS}). 

\begin{figure}[!ht]
\centering
\begin{subfigure}{0.5\textwidth}
  \includegraphics[width=\linewidth]{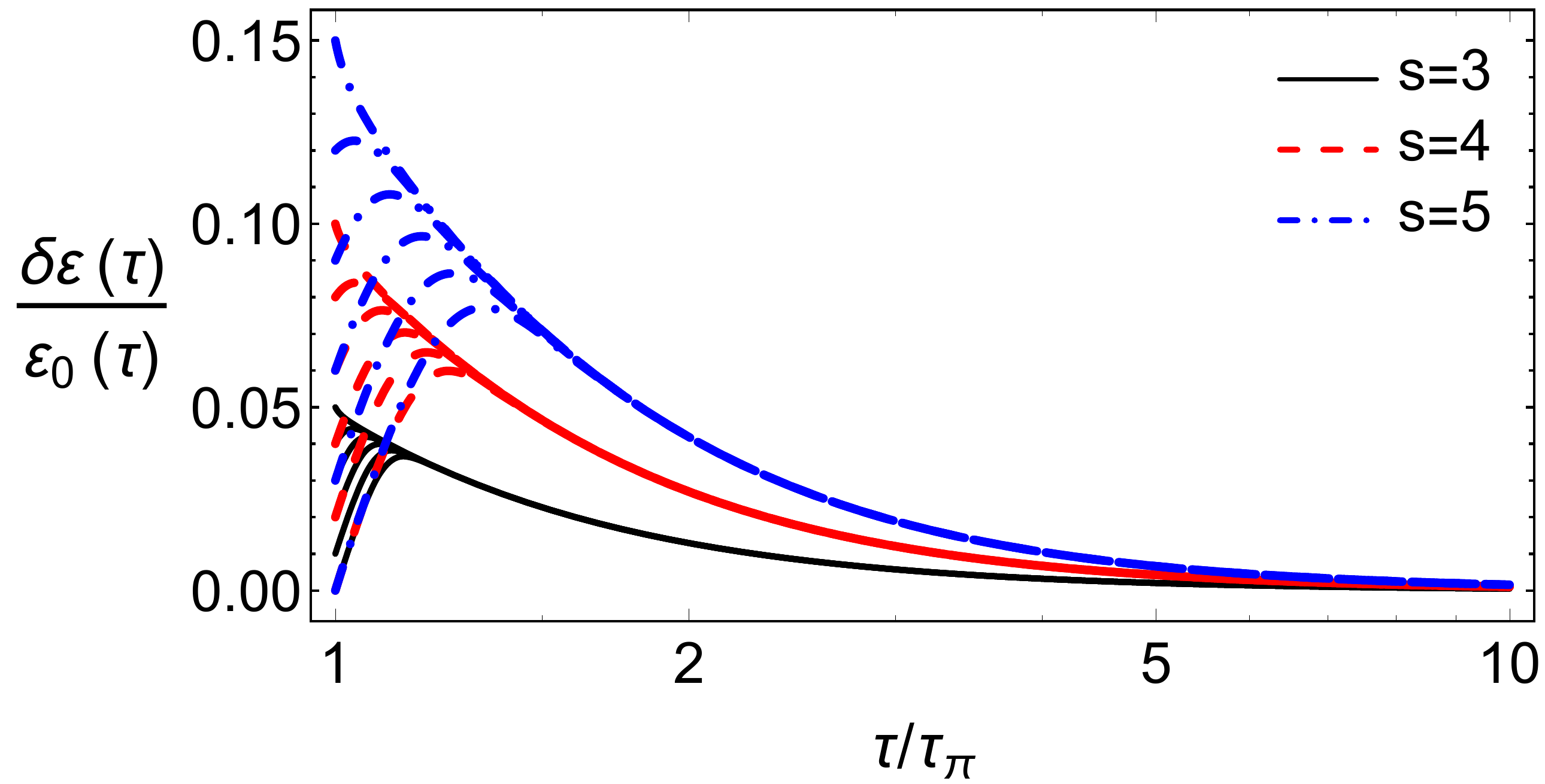}
  \caption{Attractor solutions}
  \label{fig:attractor-1-BDNK-s=3}
\end{subfigure}\hfil
\begin{subfigure}{0.5\textwidth}
  \includegraphics[width=\linewidth]{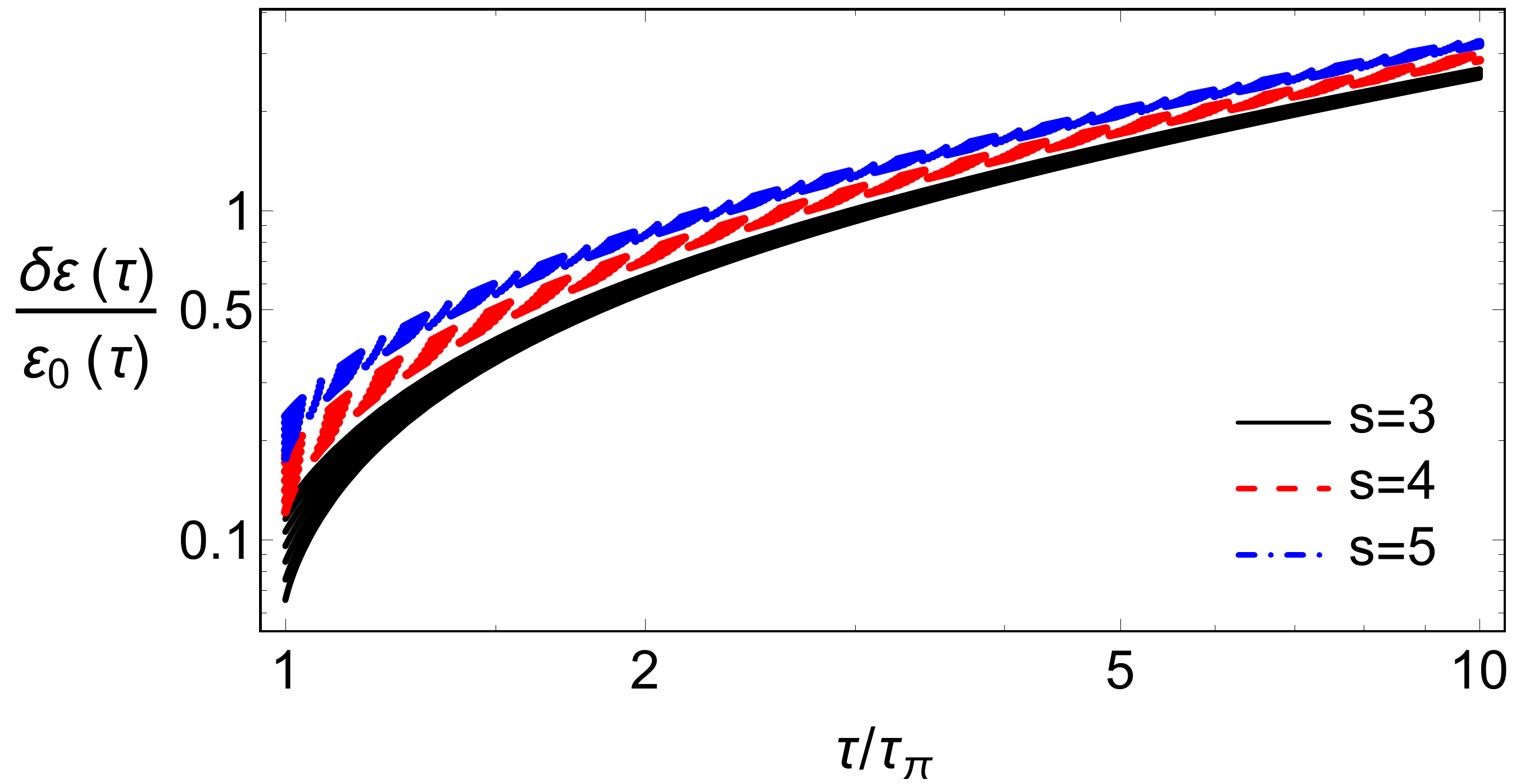}
  \caption{Runaway solutions}
  \label{fig:runaway-bdnk-s=3}
\end{subfigure}\hfil
\\
\begin{subfigure}{0.5\textwidth}
  \includegraphics[width=\linewidth]{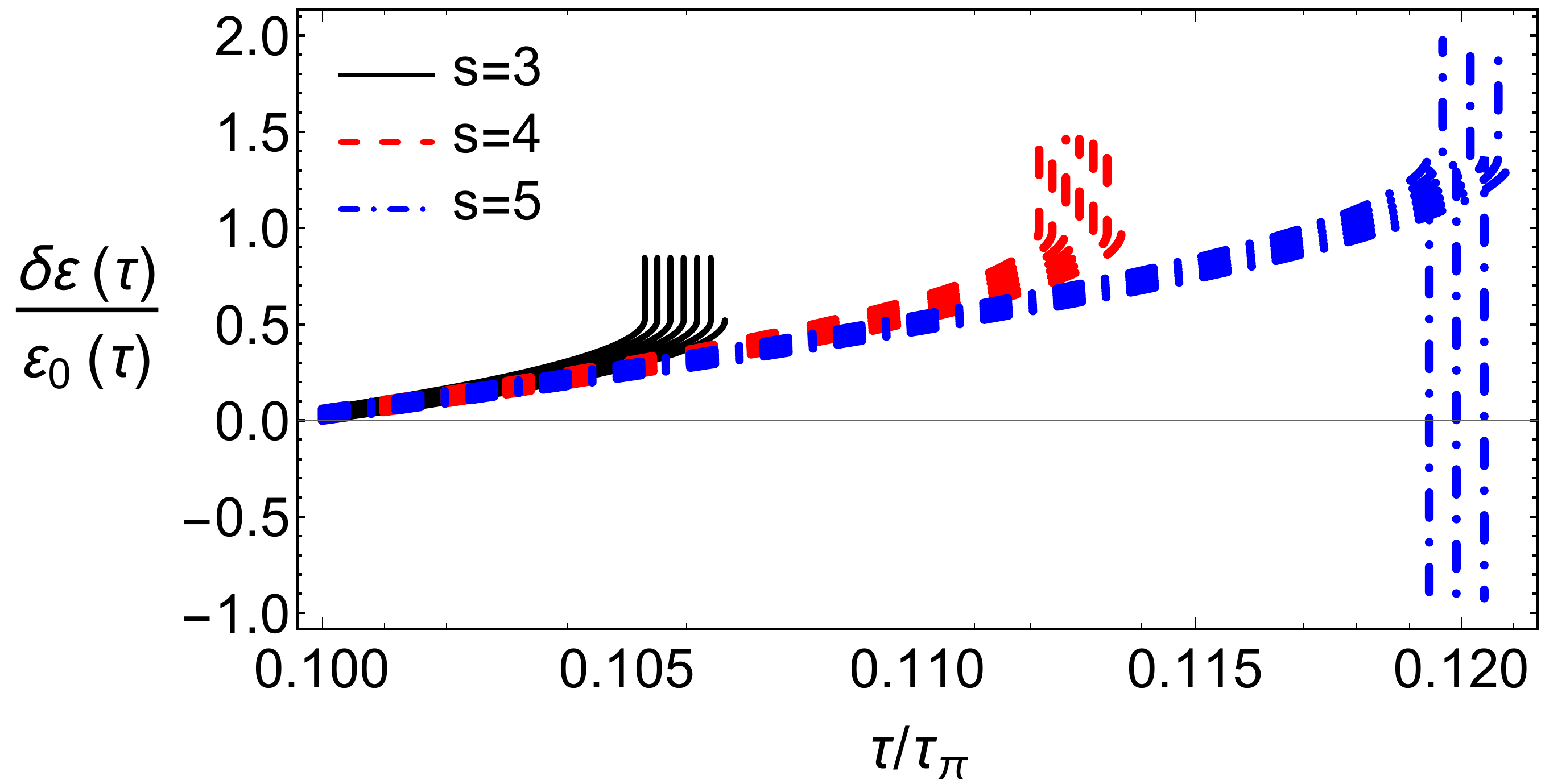}
  \caption{Large gradient initial conditions}
  \label{fig:early-attractor-BDNK-s=3}
\end{subfigure}\hfil

\caption{(Color online) Solutions of Eq.~\eqref{eq:del-eps-eom-norm} for several initial conditions at $\tilde{\tau}_{0}=1$, for panels (a) and (b), and $\tilde{\tau}_{0}=0.1$ for panel (c). (a) Attractor solutions with initial conditions such that $\Tilde{\delta \varepsilon}(\Tilde{\tau}_{0}) = 0, \cdots, 0.05$ 
for $s=3$, $\Tilde{\delta \varepsilon}(\Tilde{\tau}_{0}) = 0, \cdots, 0.10$ 
for $s=4$, and $\Tilde{\delta \varepsilon}(\Tilde{\tau}_{0}) = 0, \cdots, 0.15$ 
for $s=5$. (b) Runaway solutions with initial conditions such that $\Tilde{\delta \varepsilon}(\Tilde{\tau}_{0}) = 0.065, \cdots, 0.125$ 
for $s=3$, $\Tilde{\delta \varepsilon}(\Tilde{\tau}_{0}) = 0.121, \cdots, 0.181$ 
for $s=4$, and $\Tilde{\delta \varepsilon}(\Tilde{\tau}_{0}) = 0.177, \cdots, 0.237$ 
for $s=5$. (c) Divergent solutions with initial conditions such that $\Tilde{\delta \varepsilon}(\Tilde{\tau}_{0}) = 0, \cdots, 0.05$ 
for $s=3$, $\Tilde{\delta \varepsilon}(\Tilde{\tau}_{0}) = 0, \cdots, 0.10$ 
for $s=4$, and $\Tilde{\delta \varepsilon}(\Tilde{\tau}_{0}) = 0, \cdots, 0.15$ 
for $s=5$.}
\label{fig:attrac-BDNK-s=3-0}
\end{figure}

Alternatively, the BDNK equations of motion can be expressed as
\begin{equation}
\label{eq:del-eps-eom-norm}
\begin{aligned}
&
\frac{1}{3}\left[ \frac{1}{3} - \frac{6}{s-2} \Tilde{\tau}  \Tilde{\delta \varepsilon} \right]  \frac{d \Tilde{\delta \varepsilon}}{d \Tilde{\tau}}
+
\left(\Tilde{\delta \varepsilon}+1\right)
\frac{\Tilde{\delta \varepsilon}}{s-2} 
=
\frac{4}{81 \Tilde{\tau}^{2}}, 
\end{aligned}    
\end{equation}
%
%
in terms of the normalized variables $\delta \varepsilon \equiv  \delta \varepsilon(\Tilde{\tau})/\varepsilon_{0}(\Tilde{\tau})$ and $\Tilde{\tau} = \tau/\tau_{\pi} =   \tau/(6\tau_{\eta})$, see Eq.~\eqref{eq:rlx-time-t-eta}. This particular time re-scaling is convenient for comparisons with solutions of second order theories, which will be discussed in the next section. Solutions of Eq.~\eqref{eq:del-eps-eom-norm} are shown in Fig.~\ref{fig:attrac-BDNK-s=3-0}. From Fig.~\ref{fig:attractor-1-BDNK-s=3}, it is possible to conclude that attractor solutions do exist for $\Tilde{\delta \varepsilon}$. In contrast to Sec.~VI.C.1 of Ref.~\cite{Rocha:2022ind}, it is possible to see that the attractor solution is strongly dependent on the parameter $s$, which defines the matching condition. This is expected, since the normalized equations of motion explicitly depend on such parameter. From Fig.~\ref{fig:runaway-bdnk-s=3} it is also shown that, for sufficiently large $\Tilde{\delta \varepsilon}(\Tilde{\tau}_{0})$, the equations of motion \eqref{eq:del-eps-eom-norm} lead to runaway solutions. This can be understood analyzing the term multiplying the derivative on the left-hand side of Eq.~\eqref{eq:del-eps-eom-norm} -- for runaway solutions, $\Tilde{\tau}_{0}\Tilde{\delta \varepsilon}(\Tilde{\tau}_{0}) > (s-2)/18$ and then the corresponding term becomes negative, leading to solutions that grow with time.  

In Fig.~\ref{fig:early-attractor-BDNK-s=3}, we show that the solutions breakdown when sufficiently small values of the initial time are considered. This can be understood by noting that the rescaled time variable is inversely proportional to the Knudsen number, $\Tilde{\tau} \sim \tau/\tau_\eta \sim \textrm{Kn}^{-1}$. This implies that the constraint on the initial values of the Knudsen number found in our previous analyses is manifested here as a bound on the initial value of $\Tilde{\tau}$. As a matter of fact, if the initial value of $\Tilde{\tau}$ is too small, the time derivative term in the left hand side of the equations of motion change sign during the evolution of the system and the equation breaks down. This shows, in another manner, that BDNK theory is not amenable to large gradients.

\subsubsection*{Perturbative solutions}

\begin{figure}[!ht]
\centering
\begin{subfigure}{0.5\textwidth}
  \includegraphics[width=\linewidth]{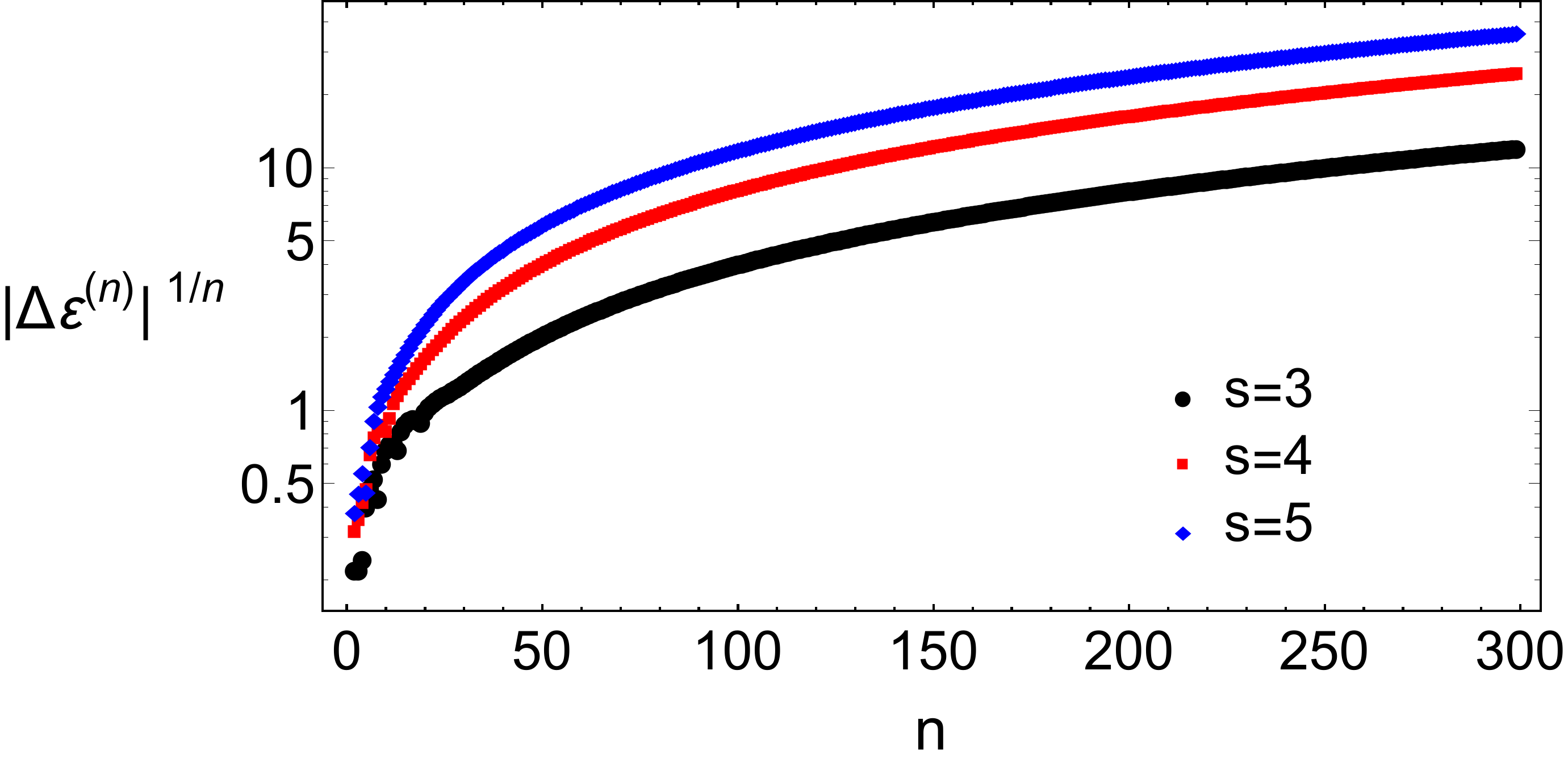}
  \caption{Series coefficients generated by recurrence relations \eqref{eq:recur-relat-BDNK}.}
  \label{fig:grad-expn-coeffs}
\end{subfigure}\hfil
\begin{subfigure}{0.5\textwidth}
  \includegraphics[width=\linewidth]{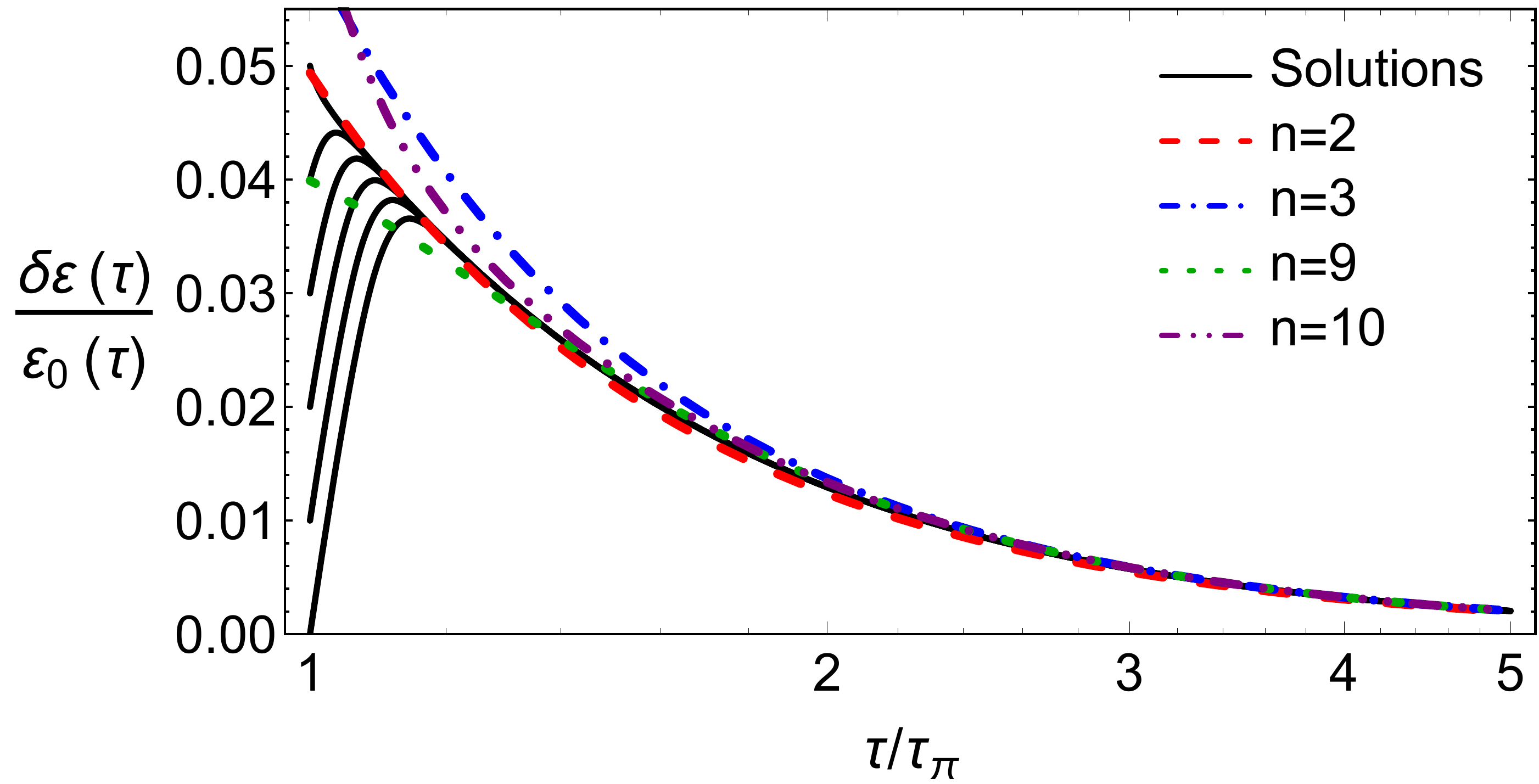}
  \caption{$s=3$}
  \label{fig:grad-expn-s=3}
\end{subfigure}\hfil
\\
\begin{subfigure}{0.5\textwidth}
  \includegraphics[width=\linewidth]{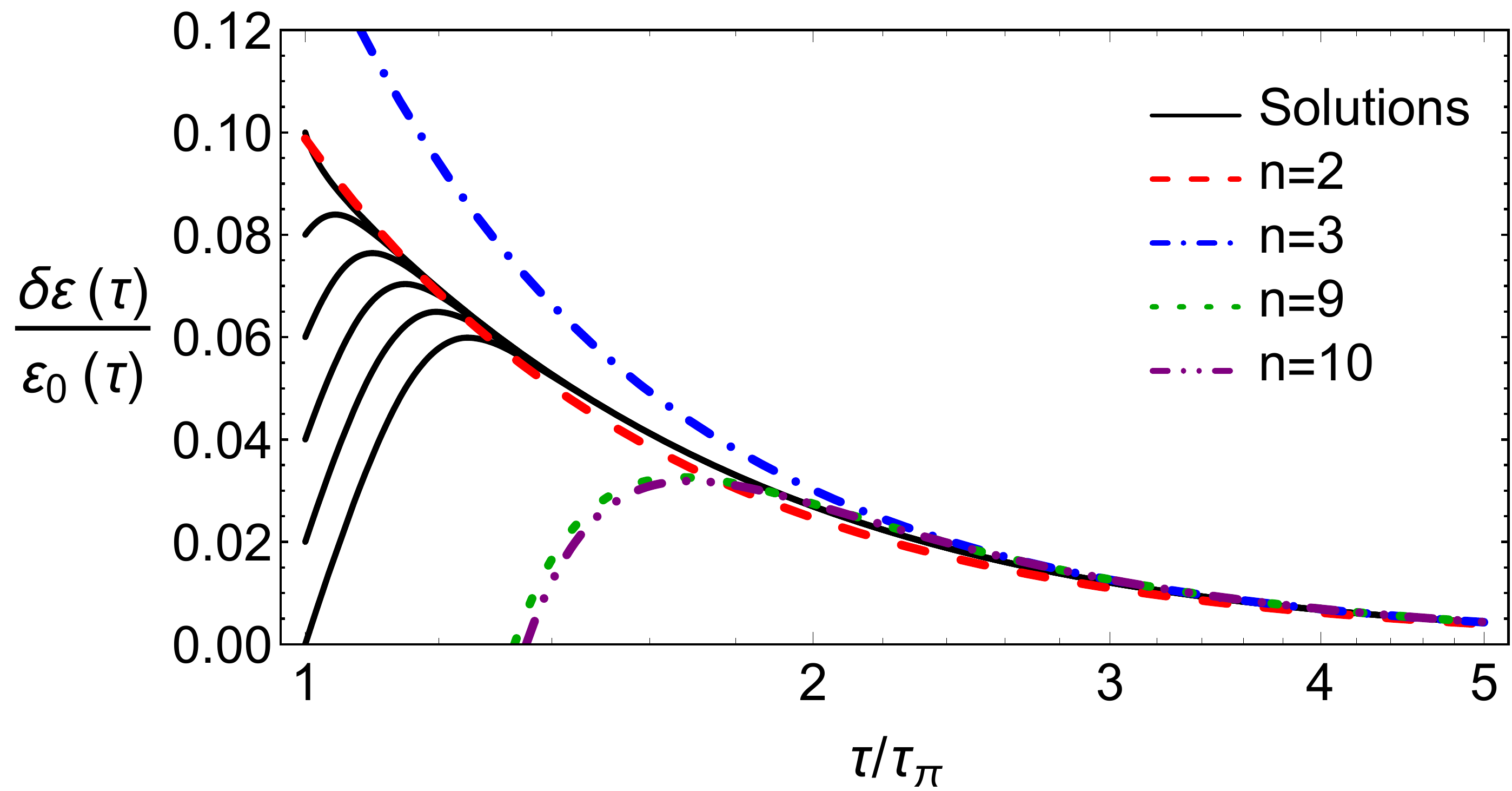}
  \caption{$s=4$}
  \label{fig:grad-expn-s=4}
\end{subfigure}\hfil
\begin{subfigure}{0.5\textwidth}
  \includegraphics[width=\linewidth]{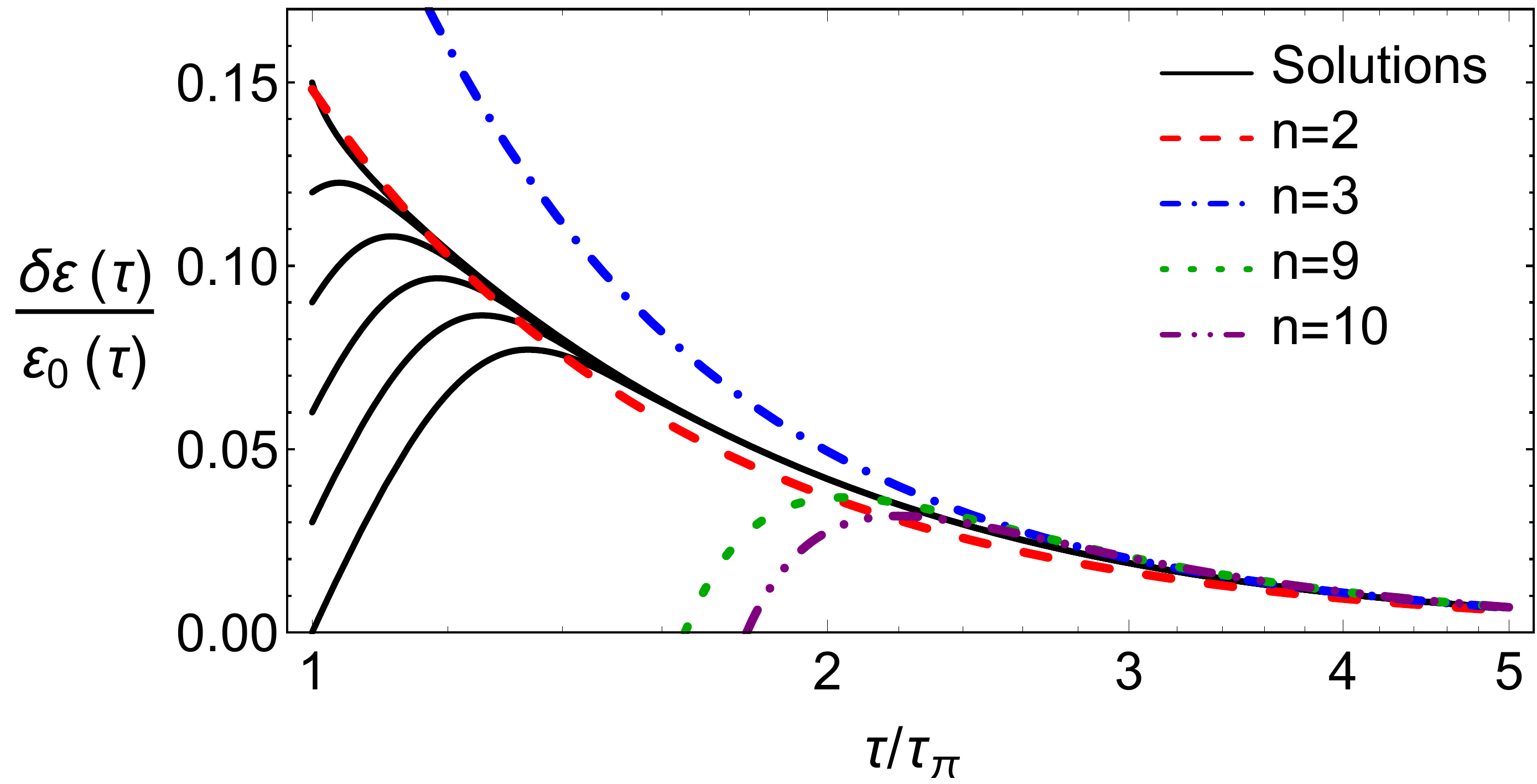}
  \caption{$s=5$}
  \label{fig:grad-expn-s=5}
\end{subfigure}\hfil

\caption{(Color online) (a) Series coefficients, $\Delta \varepsilon^{(n)}$, $n$-th root as a function of $n$. Solutions of Eq.~\eqref{eq:del-eps-eom-norm} for several initial conditions (black solid lines) in comparison to successive truncations of the gradient series for (b) $s=3$, (c) $s=4$, (d) $s=5$.}
\label{fig:attrac-BDNK-s=3}
\end{figure}

The existence of attractor solutions, that display a nontrivial dependence on $\Tilde{\tau}$, shows that BDNK theory is not a first order theory in the usual sense, i.e., it also contains terms that are of higher order in gradients. We now attempt to characterize such an attractor solution by performing a formal gradient expansion for $\Tilde{\delta \varepsilon}(\Tilde{\tau})$. In the Bjorken flow scenario, this is equivalent to the late-$\Tilde{\tau}$ expansion \cite{Denicol:2018pak,Denicol:2021},
\begin{equation}
\label{eq:grad-expn-1/t-BDNK}
\begin{aligned}
&
\Tilde{\delta \varepsilon}(\Tilde{\tau}) 
= \sum_{n=0}^{\infty} \frac{\Delta \varepsilon^{(n)}}{\Tilde{\tau}^{n}}.
\end{aligned}    
\end{equation}
Substituting this in the equations of motion \eqref{eq:del-eps-eom-norm}, we are able to derive the following recurrence relations for the coefficients $\Delta \varepsilon^{(n)}$,
\begin{equation}
\label{eq:recur-relat-BDNK}
\begin{aligned}
&\Delta \varepsilon^{(0)}\left(1 + \Delta \varepsilon^{(0)}\right) = 0, \quad \Delta \varepsilon^{(1)} = 0, \\
&
\left(\frac{2n+2}{s-2} \Delta \varepsilon^{(0)} + \frac{1}{s-2} \right)\Delta \varepsilon^{(n)} 
=
\frac{1}{9}(n-1) \Delta \varepsilon^{(n-1)}
-
\frac{1}{s-2}
\sum_{m=0}^{n-1} (2m+1)\Delta \varepsilon^{(m)} \Delta \varepsilon^{(n-m)}
+
\frac{4}{81} \delta_{n,2}
, \quad n \geq 2.
\end{aligned}    
\end{equation}
%
%
We see that the zero-th order expansion coefficient, $\Delta \varepsilon^{(0)}$, possesses two possible solutions: the usual  $\Delta \varepsilon^{(0)} = 0$, and $\Delta \varepsilon^{(0)} = -1$. The latter is excluded from further analysis since it implies in configurations that have a constant and negative late-time $\Tilde{\delta \varepsilon}$, which are not physically relevant since the corresponding system would not evolve towards equilibrium. We thus start the recurrence relations using $\Delta \varepsilon^{(0)} = 0$ and calculate the remaining coefficients for several values of $s$, which are plotted in Fig.~\ref{fig:grad-expn-coeffs}. 

We see that the expansion coefficients display factorial growth for large values of $n$ for all matching conditions considered, thus evidencing that Eq.~\eqref{eq:grad-expn-1/t-BDNK} is an asymptotic series. In Figs.~\ref{fig:grad-expn-s=3}--\ref{fig:grad-expn-s=5}, successive truncations of the gradient expansion are compared with numerical solutions of Eq.~\eqref{eq:del-eps-eom-norm} considering different values of $s$ and several initial conditions. Here we note a common feature of a divergent series: the presence of an optimal truncation, after which successive truncations worsens the agreement with the attractor solution. In all of the cases considered, $n=2$ was found to be the optimal truncation order. 

In Refs.\ \cite{Liddle_1994,Heller:2015dha,Denicol:2017lxn,Denicol:2018pak}, another perturbative solution was considered to describe the attractor solution: the slow-roll expansion. This series can be constructed by inserting a book-keeping parameter $\epsilon$ in the time derivative term of the equation of motion, Eq.~\eqref{eq:del-eps-eom-norm}, that is, 
\begin{equation}
\label{eq:del-eps-eom-slow-roll}
\begin{aligned}
&
\frac{1}{3} \epsilon \left[ \frac{1}{3} - \frac{6}{s-2} \Tilde{\tau}  \Tilde{\delta \varepsilon} \right]  \frac{d \Tilde{\delta \varepsilon}}{d \Tilde{\tau}}
+
\left(\Tilde{\delta \varepsilon}+1\right)
\frac{\Tilde{\delta \varepsilon}}{s-2} 
=
\frac{4}{81\Tilde{\tau}^{2}}, 
\end{aligned}    
\end{equation}
%
and considering a power-series solution for the rescaled energy density equilibrium deviation,
\begin{equation}
\label{eq:slow-roll-expn}
\begin{aligned}
&
\Tilde{\delta \varepsilon}(\Tilde{\tau}) 
= \sum_{n=0}^{\infty} \epsilon^{n} \widehat{\Delta \varepsilon}_{n}(\Tilde{\tau}).
\end{aligned}    
\end{equation}

\begin{figure}[!ht]
\centering
\begin{subfigure}{0.5\textwidth}
  \includegraphics[width=\linewidth]{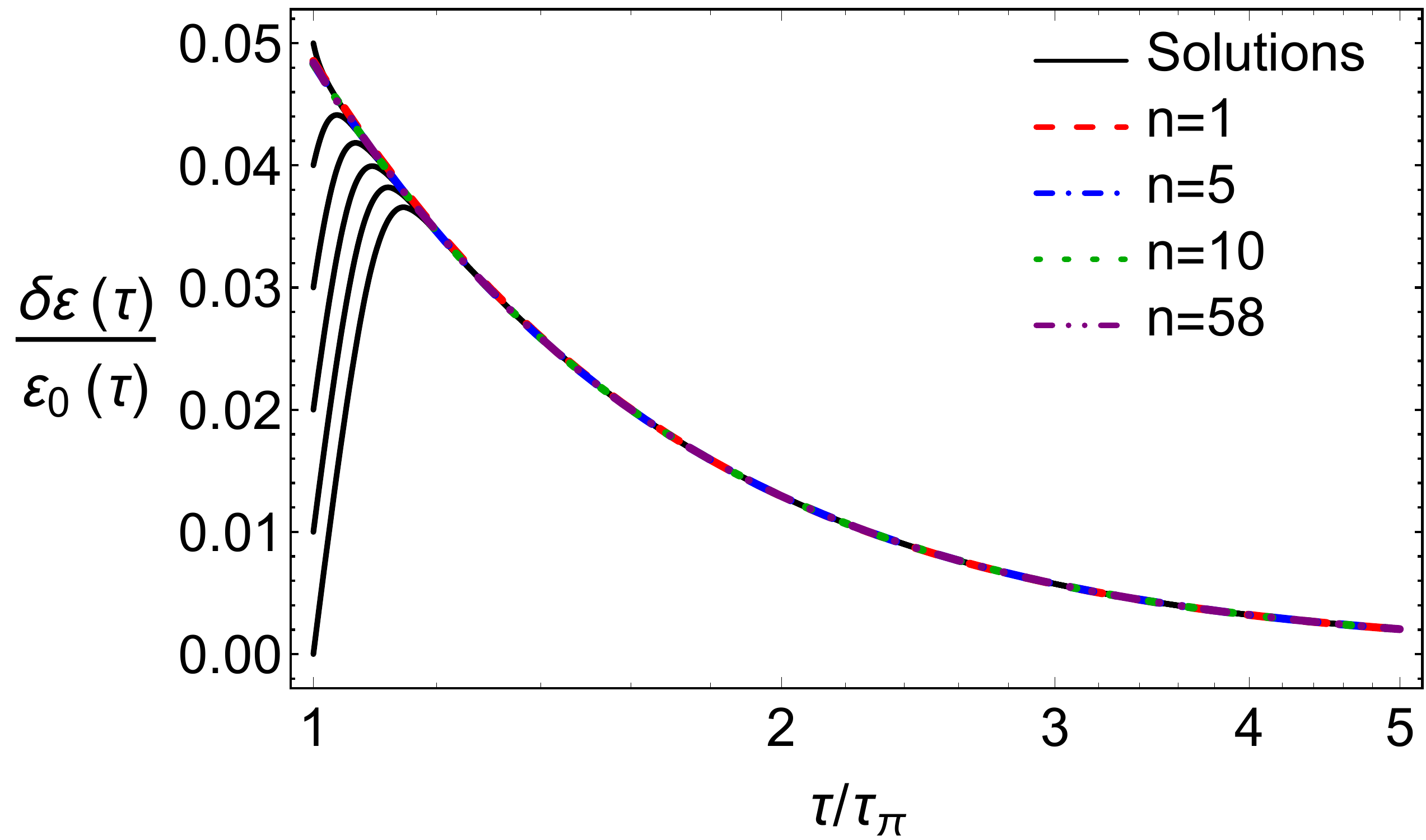}
  \caption{$s=3$}
  \label{fig:SR-expn-s=3}
\end{subfigure}\hfil
\begin{subfigure}{0.5\textwidth}
  \includegraphics[width=\linewidth]{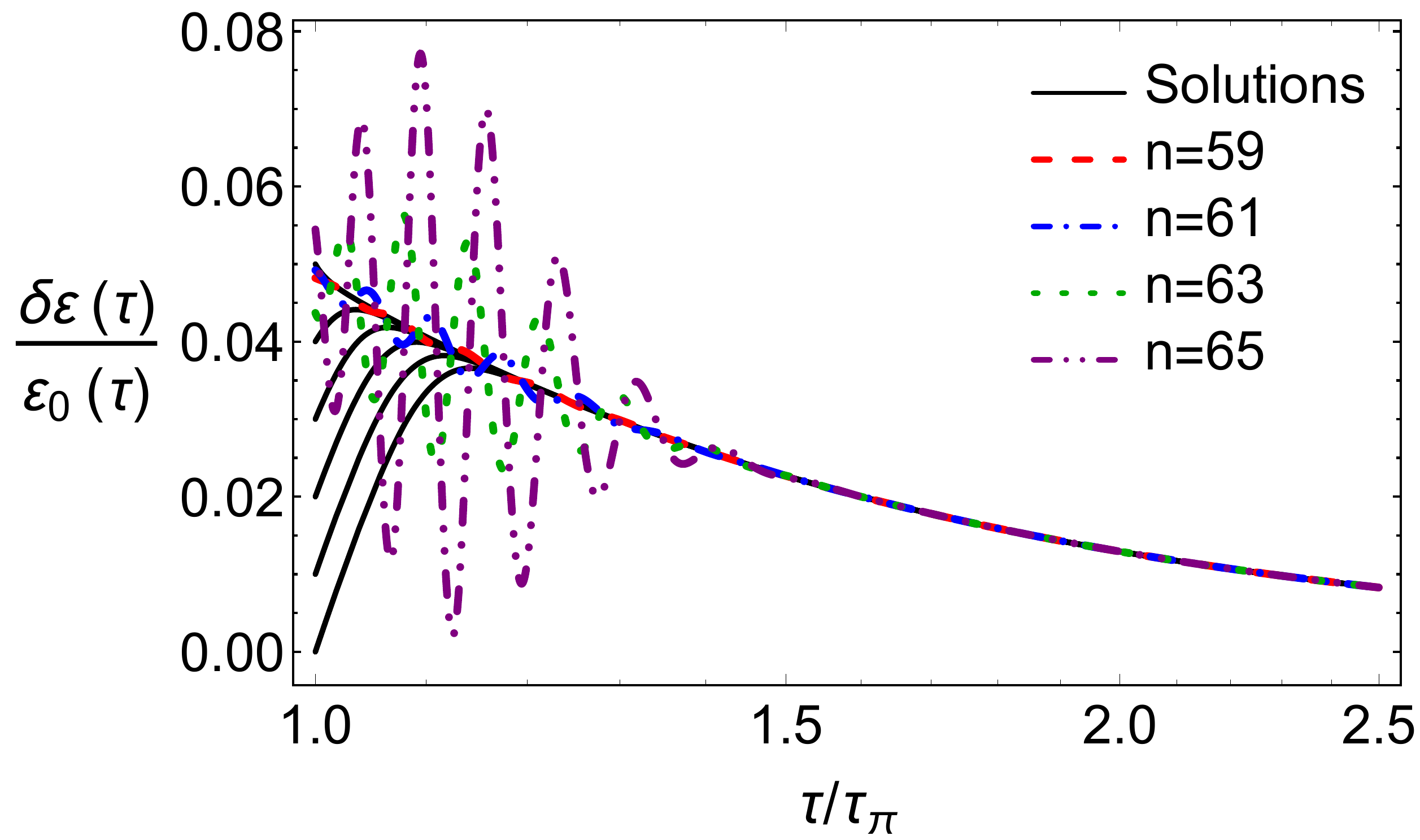}
  \caption{$s=3$}
  \label{fig:SR-expn-s=3-o}
\end{subfigure}\hfil
\\
\begin{subfigure}{0.5\textwidth}
  \includegraphics[width=\linewidth]{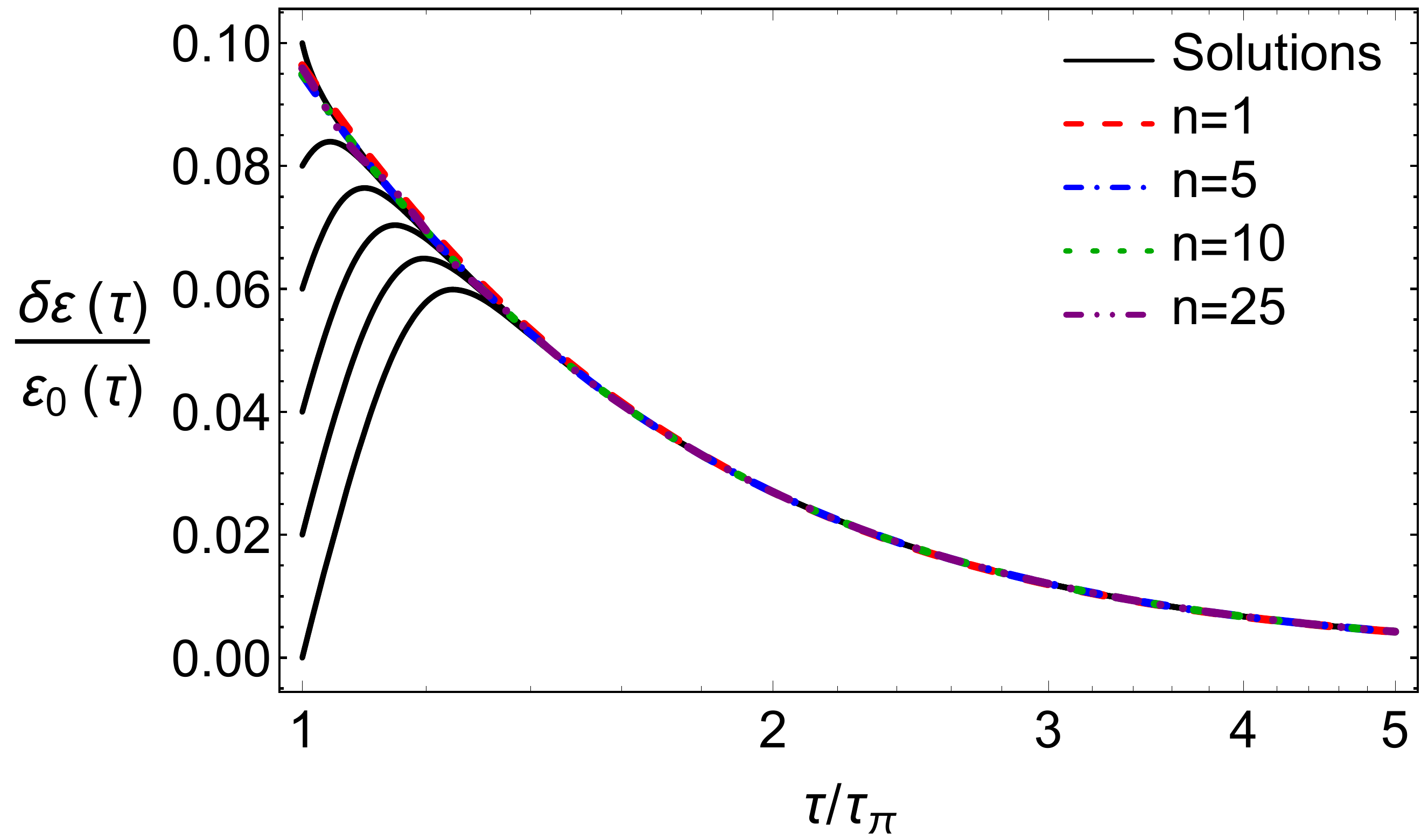}
  \caption{$s=4$}
  \label{fig:SR-expn-s=4}
\end{subfigure}\hfil
\begin{subfigure}{0.5\textwidth}
  \includegraphics[width=\linewidth]{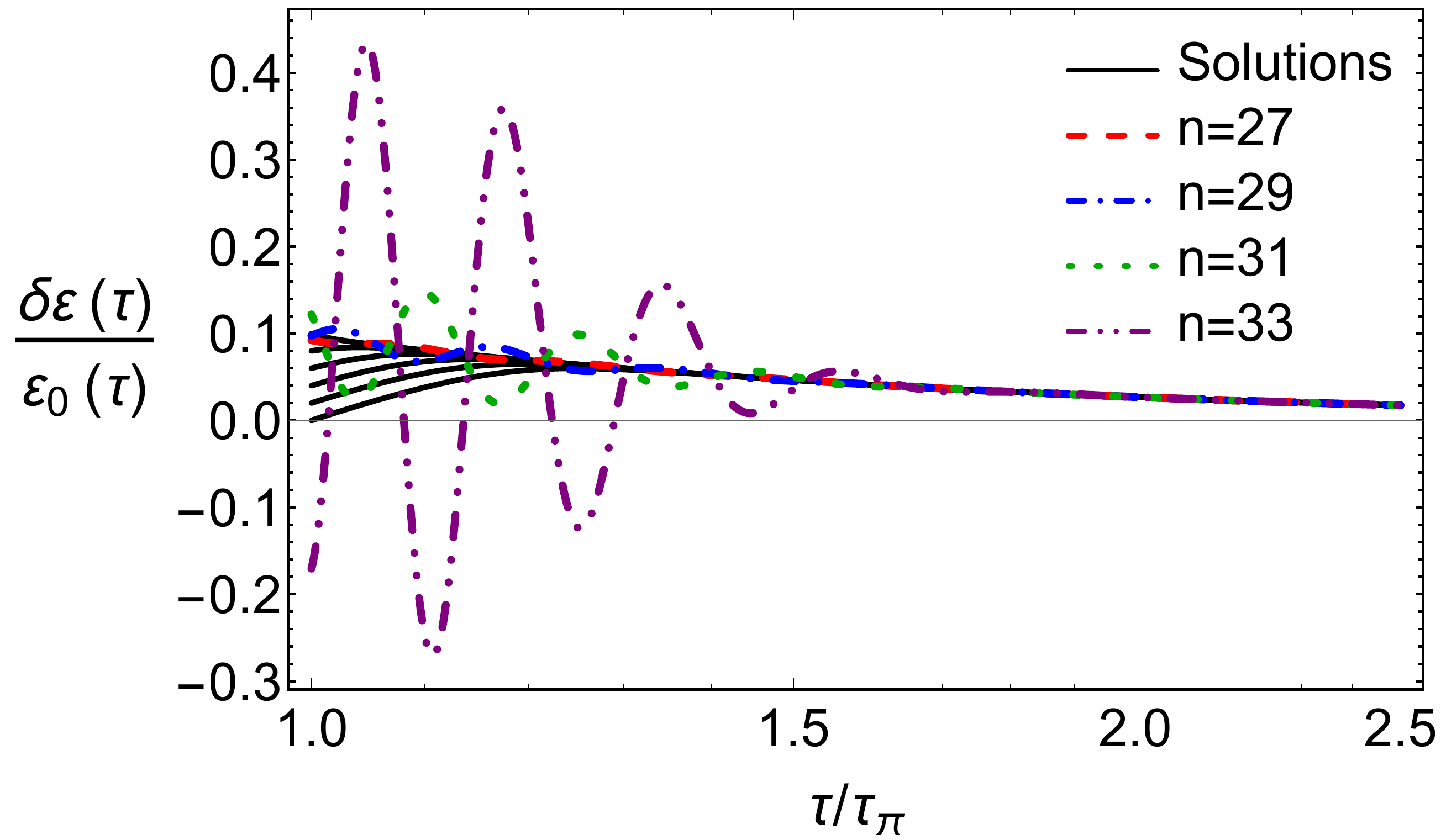}
  \caption{$s=4$}
  \label{fig:SR-expn-s=4-o}
\end{subfigure}\hfil
\\
\begin{subfigure}{0.5\textwidth}
  \includegraphics[width=\linewidth]{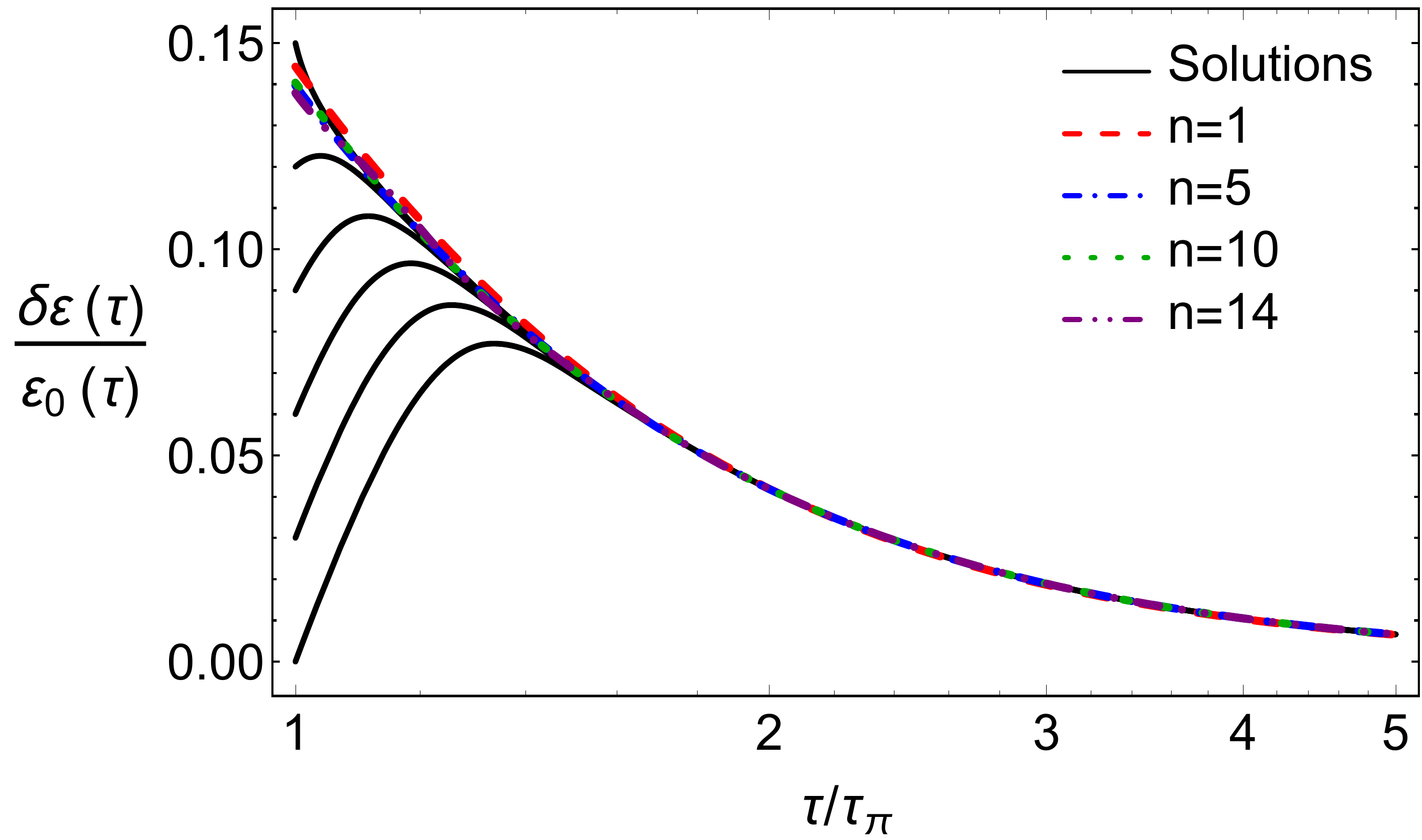}
  \caption{$s=5$}
  \label{fig:SR-expn-s=5}
\end{subfigure}\hfil
\begin{subfigure}{0.5\textwidth}
  \includegraphics[width=\linewidth]{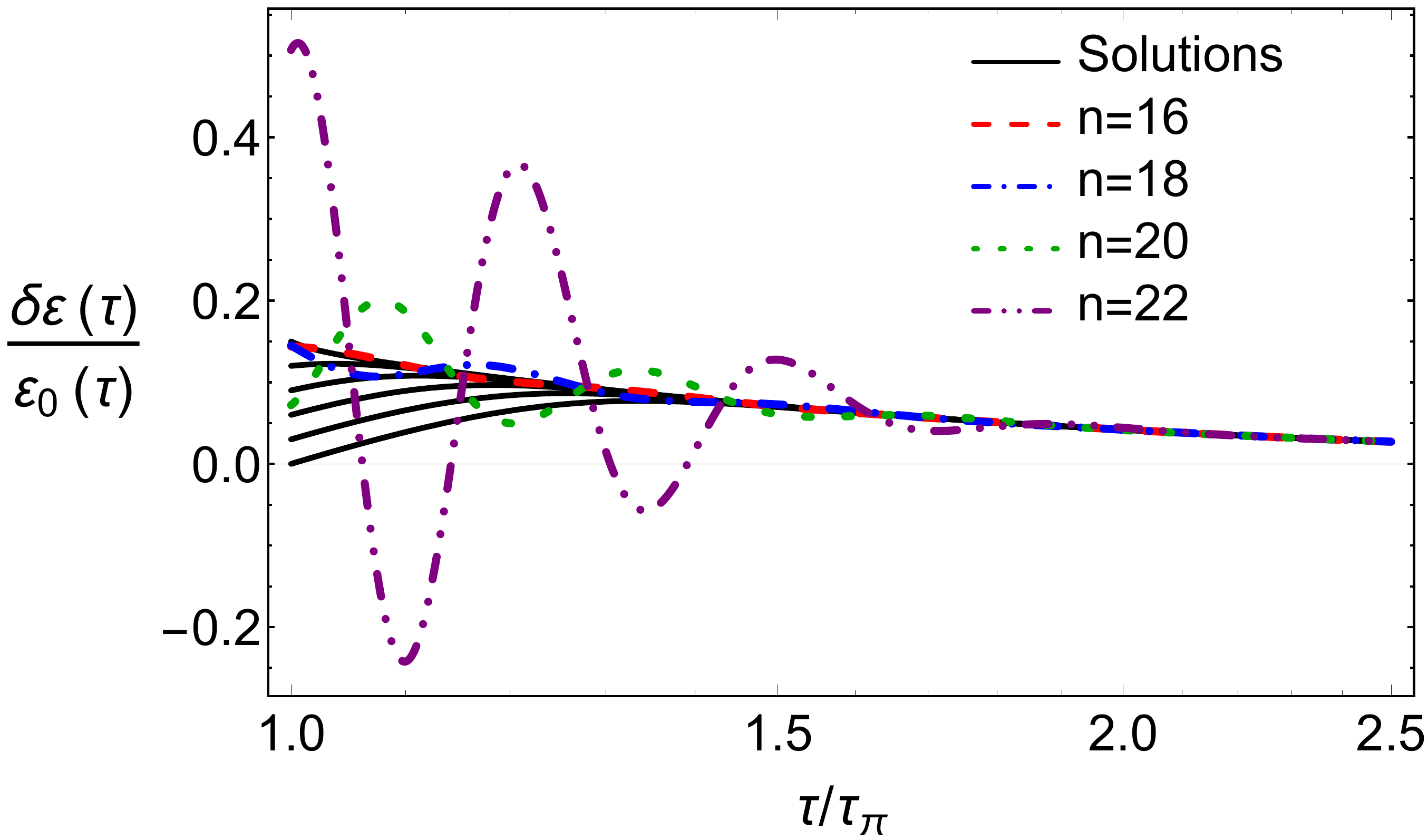}
  \caption{$s=5$}
  \label{fig:SR-expn-s=5-o}
\end{subfigure}\hfil

\caption{(Color online) Solutions of Eq.~\eqref{eq:del-eps-eom-norm} for several initial conditions (black  solid lines) in comparison to successive truncations of the slow-roll series for (a) $s=3$, $n \leq 58$, (b) $s=3$, $n \geq 58$, (c) $s=4$, $n \leq 25$, (d) $s=4$, $n > 25$, (e) $s=5$, $n \leq 14$, (f) $s=5$, $n > 14$.}
\label{fig:SR-expn-BDNK}
\end{figure}

Next, the terms at each order in $\epsilon$ are equated leading to the recursion relations 
\begin{equation}
\label{eq:del-eps-slow-roll-rec-rel}
\begin{aligned}
&
\left(\widehat{\Delta \varepsilon}_{0}+1\right)
\frac{\widehat{\Delta \varepsilon}_{0}}{s-2} 
=
\frac{4}{81 \Tilde{\tau}^{2}}, \\
&
\frac{1}{9} \frac{d \widehat{\Delta \varepsilon}_{n-1}}{d \Tilde{\tau}} - \frac{2}{s-2}  \Tilde{\tau}  \sum_{m=0}^{n-1}\widehat{\Delta \varepsilon}_{n-m-1}\frac{d \widehat{\Delta \varepsilon}_{m}}{d \Tilde{\tau}}  
+
\frac{1}{s-2}
\widehat{\Delta \varepsilon}_{n}
+
\frac{1}{s-2}
\sum_{m=0}^{n}
\widehat{\Delta \varepsilon}_{n-m} \widehat{\Delta \varepsilon}_{m}
=
0,
n \geq 1,
\end{aligned}    
\end{equation}
%
which should recover the attractor solution to the equation of motion \eqref{eq:del-eps-eom-norm} by taking $\epsilon = 1$. From Eq.~\eqref{eq:del-eps-slow-roll-rec-rel}, it is readily seen that, similar to what occurred when solving the gradient expansion, there are also two zeroth order solutions,
\begin{equation}
\begin{aligned}
&
\widehat{\Delta \varepsilon}_{0}^{\pm} = \frac{1}{2} \left[- 1 \pm \sqrt{1 + \frac{16 (s-2)}{81 \Tilde{\tau}^{2}}} \right].
\end{aligned}    
\end{equation}
%
%
One of such solutions yields a decaying late-time behavior, whereas the other yields a constant late-time behavior, just as in the gradient expansion case. Iterating to higher orders considering $\widehat{\Delta \varepsilon}_{0}^{+}$, we have the results portrayed in Fig.~\ref{fig:SR-expn-BDNK}. We can see in Figs.~\ref{fig:SR-expn-s=3}, \ref{fig:SR-expn-s=4} and \ref{fig:SR-expn-s=5} that the slow-roll series truncations perform better than the gradient expansion, since it can describe extremely accurately the early-time behavior of the attractor. Nevertheless, we found that the slow-roll expansion also diverges in BDNK theory, which can be seen in Figs.~\ref{fig:SR-expn-s=3-o}, \ref{fig:SR-expn-s=4-o}, and \ref{fig:SR-expn-s=5-o}. However, the strong oscillations that signal the divergence of the series only appear at very high orders, when compared to the gradient expansion. The optimal truncation order is also found to be strongly dependent on the matching parameter $s$: for $s=3,4$ and $5$, we find that the optimal truncation is  $58, 25$ and $14$, respectively. 

We note that the attractor can be identified as the solution for which $\Tilde{\delta \varepsilon}(\Tilde{\tau}_{0}) \to (s-2)/(18 \Tilde{\tau}_{0})$ from below. One is reminded that this is the initial condition for which the derivative term in Eq.~\eqref{eq:del-eps-eom-norm} vanishes, thus separating regular from runaway solutions.

\subsection{Transient hydrodynamic theory}

Now, we proceed to investigate the solutions of second-order transient fluid dynamics in Bjorken flow. As previously stated, we only consider second-order theories with Landau matching conditions. In this case, the system of equations of motion stemming from the local conservation laws become,
\begin{subequations}
\label{eq:consv-eqns-bjo}
\begin{align}
&
\label{eq:consv-eqns-bjo-1}
\dot{n}_{0} + \frac{n_{0}}{\tau} = 0, \\
&
\label{eq:consv-eqns-bjo-2}
\dot{\varepsilon}_{0} + \frac{4}{3}\frac{\varepsilon_{0}}{\tau} - \frac{\pi}{\tau} = 0.
\end{align}    
\end{subequations}
As stated above, any vector orthogonal to the 4-velocity vanishes because of the symmetries assumed for the system and any contribution due to diffusion can be neglected. Since we also limited our discussion to a gas of massless particles,
the only dissipative contribution will be from the shear-stress tensor. Using Eq.~\eqref{eq:eom-shear} we obtain that the only independent component of the shear-stress tensor obeys
\begin{equation}
\label{eq:normalized-shear}
\begin{aligned}
&
\dot{\pi}
+
\frac{\pi}{\tau_{\pi}}  
=
-2\frac{\pi}{\tau} 
+
\frac{8}{27\tau} \varepsilon_{0},
\end{aligned}    
\end{equation}
where $\tau_{\pi}^{-1} = (g/72)\beta^{2} n_{0}$ defines the temperature-dependent relaxation time. Equation \eqref{eq:normalized-shear} can be more conveniently expressed in terms of the  dimensionless variable $\Tilde{\pi} \equiv \pi/(\varepsilon_{0} + P_{0}) = 3 \pi/(4 \varepsilon_{0})$ and, after iterated use of Eqs.~\eqref{eq:consv-eqns-bjo}, we can derive
\begin{equation}
\label{eq:shear-bj-1}
\begin{aligned}
&
\dot{\Tilde{\pi}}
+
\frac{4}{3 \tau} \Tilde{\pi}^{2}
+
\frac{2}{3 \tau}
\Tilde{\pi}
+
\frac{\Tilde{\pi}}{\tau_{\pi}}  
=
\frac{2}{9\tau}. 
\end{aligned}    
\end{equation}

In the small coupling limit, $g \to 0$, and/or at early times\footnote{The shear relaxation time diverges at $\tau \to 0$ as long as the temperature decays with time faster than the power law $\tau^{-1/2}$. This will be the case for all the solutions we discuss.}, $\tau \to 0$, the relaxation time diverges and the last term on the left-hand side becomes asymptotically small (this is analogous to the cold plasma limit discussed in Ref.\ \cite{Marrochio:2013wla}, in the context of Gubser flow). As a result,  Eq.~\eqref{eq:shear-bj-1} decouples from Eqs.~\eqref{eq:consv-eqns-bjo} and admits an analytical solution
\begin{equation}
\label{eq:shear-no-rlx}
\begin{aligned}
&
\Tilde{\pi}(\tau)  
=
\frac{A \tau^{\sigma}+ B b_1}{ \tau ^{\sigma} + b_1},
 \\
A = \frac{1}{12}\left(\sqrt{33}-3\right) \simeq 0.2287, \quad & B = -\frac{1}{12}\left(\sqrt{33}+3\right) \simeq -0.7287, \quad
\sigma = \frac{2}{3} \sqrt{\frac{11}{3}} \simeq 1.27657,
\end{aligned}    
\end{equation}
with $b_{1}$ being an integration constant. When $\tau \to 0$, there are two universal solutions, $\tilde \pi = B$ for any $b_1 \neq 0$ and $\tilde \pi = A$ for $b_1 = 0$. This implies that all solutions, except the one determined by $b_1 = 0$, will approach $\tilde \pi = B$ as $\tau$ goes to zero -- this type of solution is referred to as a pullback attractor \cite{Behtash:2019txb}. Thus, the boundary condition $\tilde \pi (\tau \to 0)= A$ determines a unique solution of the theory, given by $b_1 = 0$, which is also the universal solution at late times, $\tau \to \infty$ -- that is, this is an attractor solution of the equations. Naturally, such attractor solution will deviate from $A$ at late times, when the relaxation time ceases to be large and the approximation performed above fails. In this regime, one expects the attractor solution to start approaching the Navier-Stokes limit of the theory. Finally, substituting Eq.~\eqref{eq:shear-no-rlx} in Eq.~\eqref{eq:consv-eqns-bjo-2}, we may also determine an analytical solution for the energy density, valid at sufficiently early times,
\begin{equation}
\begin{aligned}
&
\varepsilon_{0}(\tau) = C \tau ^{-\frac{4}{3}(1-B)} \left(b_{1}+\tau ^{\sigma }\right)^{\frac{4}{3 \sigma}(A-B) },
\end{aligned}    
\end{equation}
where $C$ is another integration constant. We note that the above solution decays as $\tau^{(4/3)(A-1)} \simeq \tau^{-1.02838} $ for large $\tau$. This is to be contrasted with the solution in the ideal case, where $\varepsilon_{0}(\tau) \propto \tau^{-4/3}$.

For finite values of the relaxation time, Eq.~\eqref{eq:shear-bj-1} is coupled to the conservation equations \eqref{eq:consv-eqns-bjo}. In order to decouple the equation for $\Tilde{\pi}$, we re-write Eq.~\eqref{eq:shear-bj-1} in terms of the normalized time variable $\Tilde{\tau} = \tau/\tau_{\pi}$. As $\tau_{\pi}$ depends on temperature and chemical potential, we again resort to Eqs.~\eqref{eq:consv-eqns-bjo} to derive a closed equation of motion for $\Tilde{\pi}$
\begin{equation}
\label{eq:shear-abel-1}
\begin{aligned}
&
\frac{2}{3}\left(1- 4\Tilde{\pi}\right)\partial_{\Tilde{\tau}}\Tilde{\pi}
+
\frac{4}{3 \Tilde{\tau}} \Tilde{\pi}^{2}
+
\frac{2}{3 \Tilde{\tau}}
\Tilde{\pi}
+
\Tilde{\pi} 
=
\frac{2}{9 \Tilde{\tau}}.
\end{aligned}    
\end{equation}
In the study of non-linear ordinary differential equations, the equation above is identified as an Abel equation of the second kind \cite{zaitsev2002handbook}. Analytical solutions for this type of equation are limited to some specific functional forms of the non-constant coefficients. Further non-linear transformations can be performed in order to express Eq.~\eqref{eq:shear-abel-1} in the Abel standard form \cite{zaitsev2002handbook,panayotounakos2005exact}. Indeed, changing the dependent variable to $w(\tilde{\tau}) = (\Tilde{\pi} - 1/4) \tilde{\tau}^{-1/2}$, and the independent variable to $\varphi = \int d\Tilde{\tau} \left[(1/2) \Tilde{\tau}^{-3/2} + (3/8) \Tilde{\tau}^{-1/2}\right] = -\Tilde{\tau}^{-1/2} + (3/4)\Tilde{\tau}^{1/2}$, we obtain the simpler form
\begin{equation}
\begin{aligned}
& w \partial_{\varphi} w 
=
w + F(\varphi), \\
&
F(\varphi) = \frac{3}{64} \frac{8 \varphi( \varphi +\sqrt{\varphi ^2+3})  +13}{\left(\sqrt{\varphi ^2+3}+\varphi \right)^2 \sqrt{\varphi ^2+3}}.
\end{aligned}    
\end{equation}
A solution to this equation can only be found implicitly \cite{panayotounakos2005exact}. As the discussion of this solution is lengthy and elusive to the physics involved, it will be omitted in this work.

\begin{figure}[ht]
\centering
\begin{subfigure}{0.5\textwidth}
  \includegraphics[width=\linewidth]{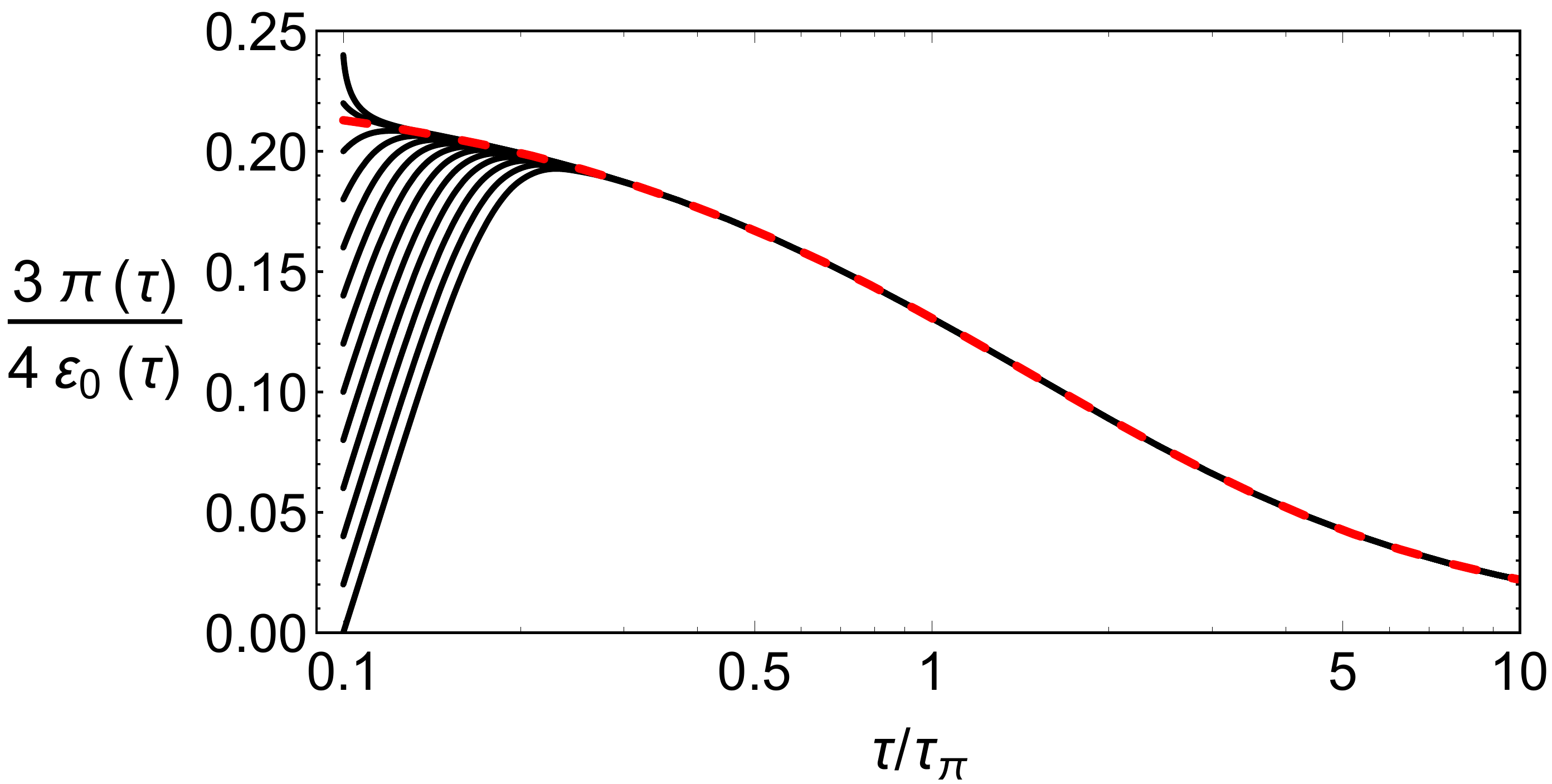}
  \caption{Attractor solution}
  \label{fig:attractor-1}
\end{subfigure}\hfil
\begin{subfigure}{0.5\textwidth}
  \includegraphics[width=\linewidth]{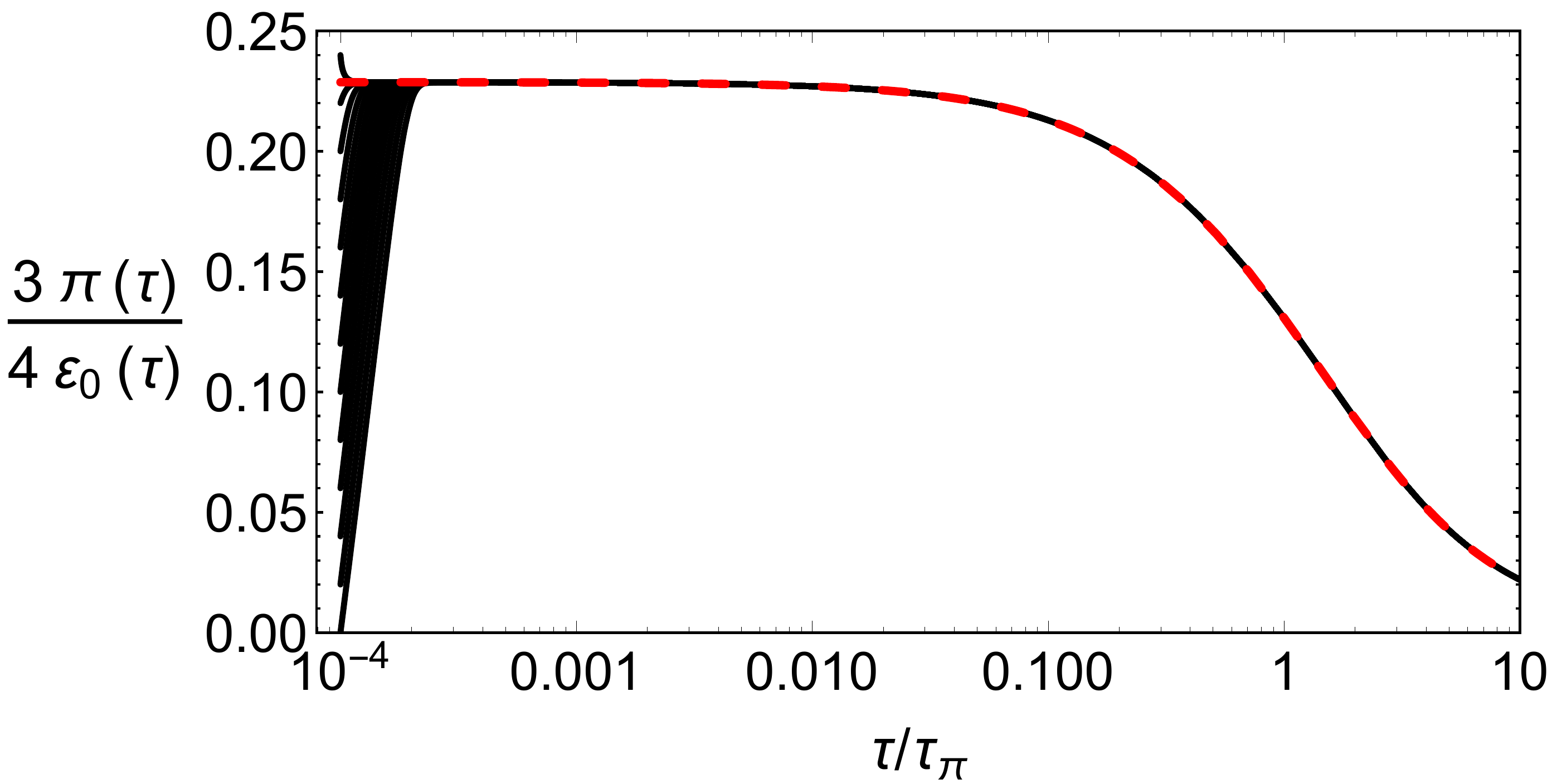}
  \caption{Attractor for small $\Tilde{\tau}_{0}$}
  \label{fig:early-attractor}
\end{subfigure}\hfil
\\
\begin{subfigure}{0.5\textwidth}
  \includegraphics[width=\linewidth]{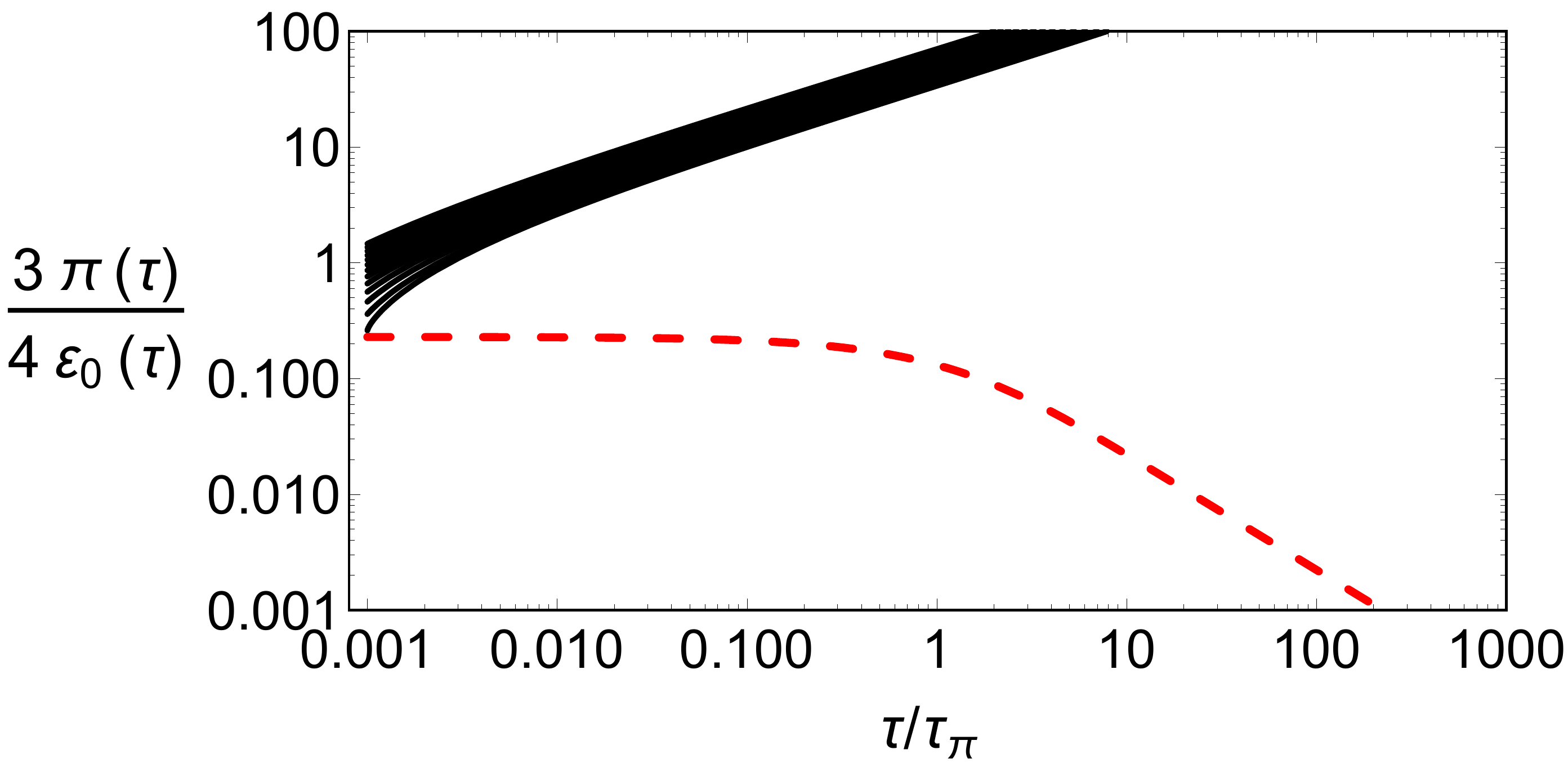}
  \caption{Runaway solutions}
  \label{fig:runaway}
\end{subfigure}\hfil
\begin{subfigure}{0.5\textwidth}
  \includegraphics[width=\linewidth]{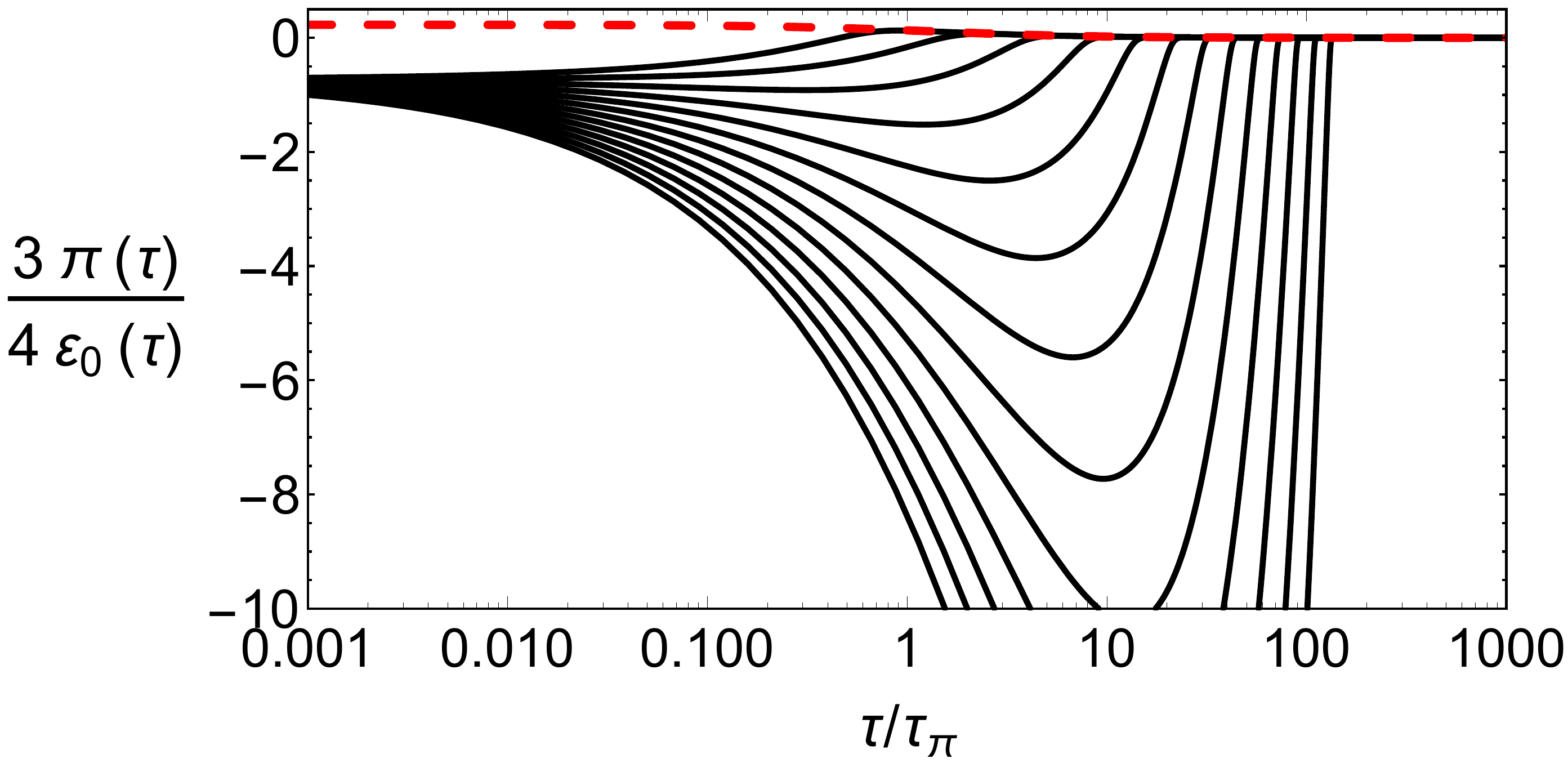}
  \caption{Attractor for negative $\Tilde{\pi}(\Tilde{\tau}_{0})$}
  \label{fig:cavitation}
\end{subfigure}\hfil
\caption{(Color online) Solutions of Eq.~\eqref{eq:shear-abel-1} for several initial conditions (black lines) in comparison with the attractor (solution of Eq.~\eqref{eq:shear-abel-1} with $\Tilde{\pi}(\Tilde{\tau_{0}} = 10^{-6}) = A \simeq 0.2287$, dashed curves). (a) Solutions for initial conditions $\Tilde{\pi}(\Tilde{\tau}_{0}=0.1) = 0, \cdots, 0.24$ 
(b) solutions for earlier $\Tilde{\tau}_{0}$ with initial conditions $\Tilde{\pi}(\Tilde{\tau}_{0}=0.1) = 0, \cdots, 0.24$ 
, (c) runaway solutions for the initial conditions $\Tilde{\pi}(\Tilde{\tau}_{0} = 10^{-3}) = 0.26, \cdots, 1.46$ 
,  (d) solutions for negative initial conditions, $\Tilde{\pi}(\Tilde{\tau}_{0}=10^{-3}) = -1, \cdots, -0.7$ 
. }
\label{fig:attrac}
\end{figure}

In the following, we analyze numerical solutions of Eq.~\eqref{eq:shear-abel-1}, which are shown in Fig.~\ref{fig:attrac}. In Figs.~\ref{fig:attractor-1} and \ref{fig:early-attractor}, we see that the system possesses an attractor solution, as expected. Here, we note that the attractor can be identified as the solution of Eq.~\eqref{eq:shear-abel-1} for $\Tilde{\pi}(\Tilde{\tau}_{0} \to 0) = A \simeq 0.2287$, which is displayed as the red dotted lines in all panels of this figure. We note that this late-time solution only emerges if the initial conditions are such that $\Tilde{\pi} < 1/4$. Otherwise, as seen in Fig.~\ref{fig:runaway}, the late-time solution diverges asymptotically. This can be understood directly from Eq.~\eqref{eq:shear-abel-1}, since for $\Tilde{\pi} > 1/4$ the term proportional to $\partial_{\Tilde{\tau}} \Tilde{\pi}$ becomes negative, thus making the equation yield an exponentially growing solution. This shows that the equations of motion yield nonphysical results when $\pi > P_{0} $, i.e., when the longitudinal pressure becomes negative. In Fig.~\ref{fig:cavitation}, it is shown that large negative initial values of $\Tilde{\pi}$ do not lead to runaway solutions as large positive values do. Indeed, as seen in Fig.~\ref{fig:cavitation}, even for $\Tilde{\pi}(\Tilde{\tau}_{0}) = -1.1$ the evolution leads to the same hydrodynamic attractor at very late times. However, for intermediate times,  $\Tilde{\pi}$ can reach very large, negative values.  

\subsubsection*{Perturbative solutions}

Similarly to the last section, we attempt to characterize the attractor solution by performing a formal gradient expansion for the rescaled shear-stress tensor. In the Bjorken flow scenario, this is equivalent to the late-$\Tilde{\tau}$ expansion \cite{Denicol:2018pak,Denicol:2021},
\begin{equation}
\label{eq:grad-expn-1/t}
\begin{aligned}
&
\Tilde{\pi}(\Tilde{\tau}) 
= \sum_{n=0}^{\infty} \frac{\Tilde{\pi}^{(n)}}{\Tilde{\tau}^{n}}, 
\end{aligned}    
\end{equation}
which, when substituted in Eq.~\eqref{eq:shear-abel-1}, leads to the following recurrence relations
\begin{equation}
\label{eq:recur-relat}
\begin{aligned}
&\Tilde{\pi}^{(0)} = 0, \quad \Tilde{\pi}^{(1)} = \frac{2}{9}, \\
&
\Tilde{\pi}^{(n)} 
=
\frac{2}{3}(n-2) \Tilde{\pi}^{(n-1)}
-
\frac{4}{3} \sum_{m=0}^{n-1} (2m+1)\Tilde{\pi}^{(n-m-1)} \Tilde{\pi}^{(m)}, \quad n \geq 2.
\end{aligned}    
\end{equation}
A distinctive feature of the above equations is that the second order coefficient, $\Tilde{\pi}^{(2)}$, exactly vanishes. 

\begin{figure}[ht]
\centering
\begin{subfigure}{0.5\textwidth}
  \includegraphics[width=\linewidth]{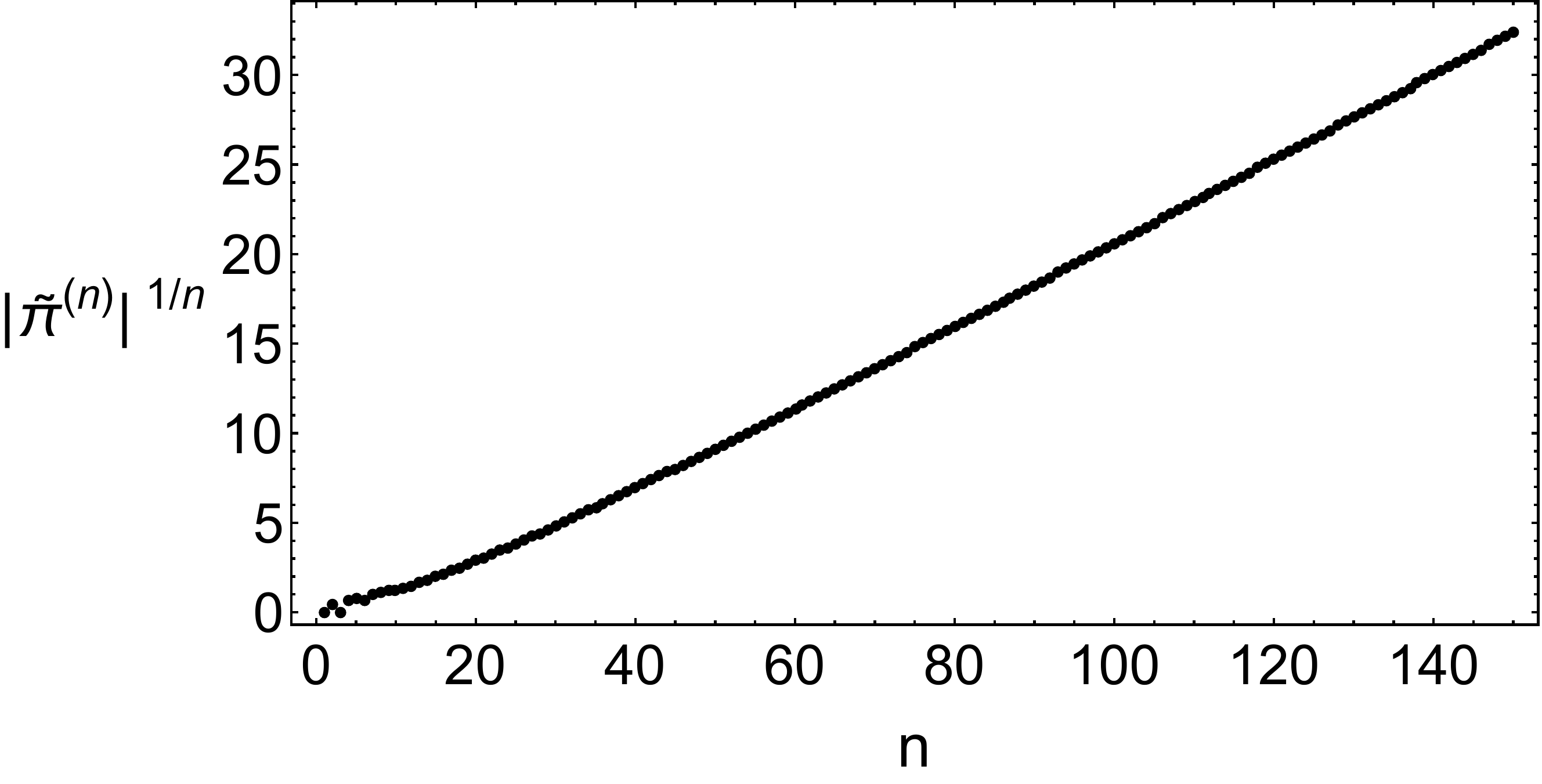}
  \caption{Series coefficients generated by recurrence relations \eqref{eq:recur-relat}.}
  \label{fig:grad-expn-2}
\end{subfigure}\hfil
\begin{subfigure}{0.5\textwidth}
  \includegraphics[width=\linewidth]{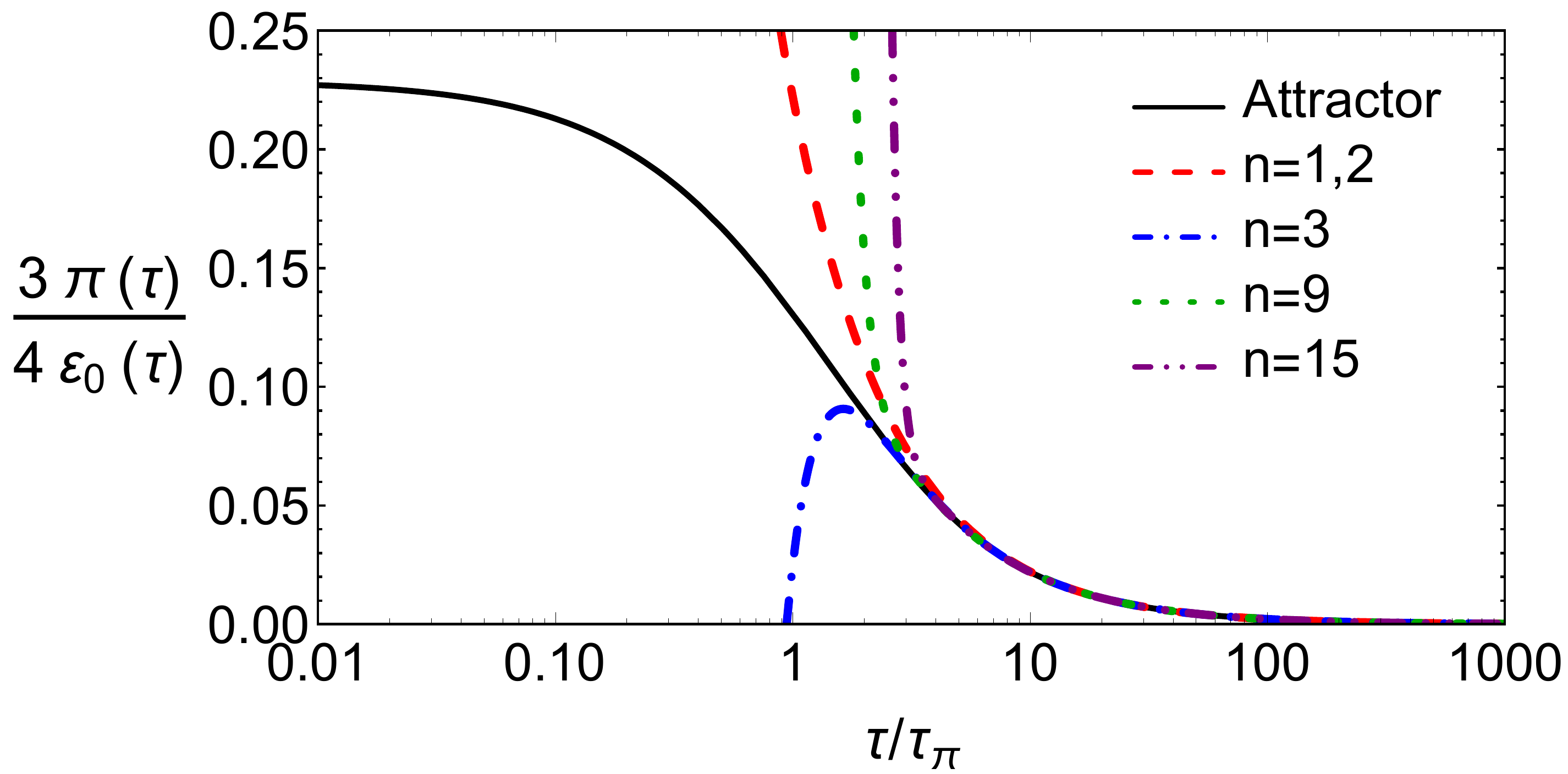}
 \caption{Attractor solution and  truncations of the gradient expansion.}
  \label{fig:grad-expn-1}
\end{subfigure}\hfil
\caption{(Color online) (a) Series coefficients, $\Tilde{\pi}^{(n)}$, $n$-th root as a function of $n$. (b) Attractor solution Eq.~\eqref{eq:del-eps-eom-norm} (black solid lines) in comparison to successive truncations of the gradient series.}
\label{fig:grad-expn}
\end{figure}

We solve Eq.\ \eqref{eq:recur-relat} numerically up to order 150 and display the results in Fig.~\ref{fig:grad-expn}. In Fig.~\ref{fig:grad-expn-2}, we see that the expansion coefficients $\Tilde{\pi}^{(n)}$ display factorial growth, $\Tilde{\pi}^{(n)} \sim n!$, which implies that the series diverges. We note that this is a common feature of the gradient expansion of second-order fluid dynamics and was first discovered in Ref.\ \cite{Heller:2015dha}. In Fig.~\ref{fig:grad-expn-1} we show consecutive truncations of the series \eqref{eq:grad-expn-1/t} compared with the attractor solution obtained in the last section, which is expected to be the resummed version of the gradient expansion. In the present case, it is seen that the optimal truncation order is $n=3$.

\begin{figure}[ht]
\centering
\begin{subfigure}{0.5\textwidth}
  \includegraphics[width=\linewidth]{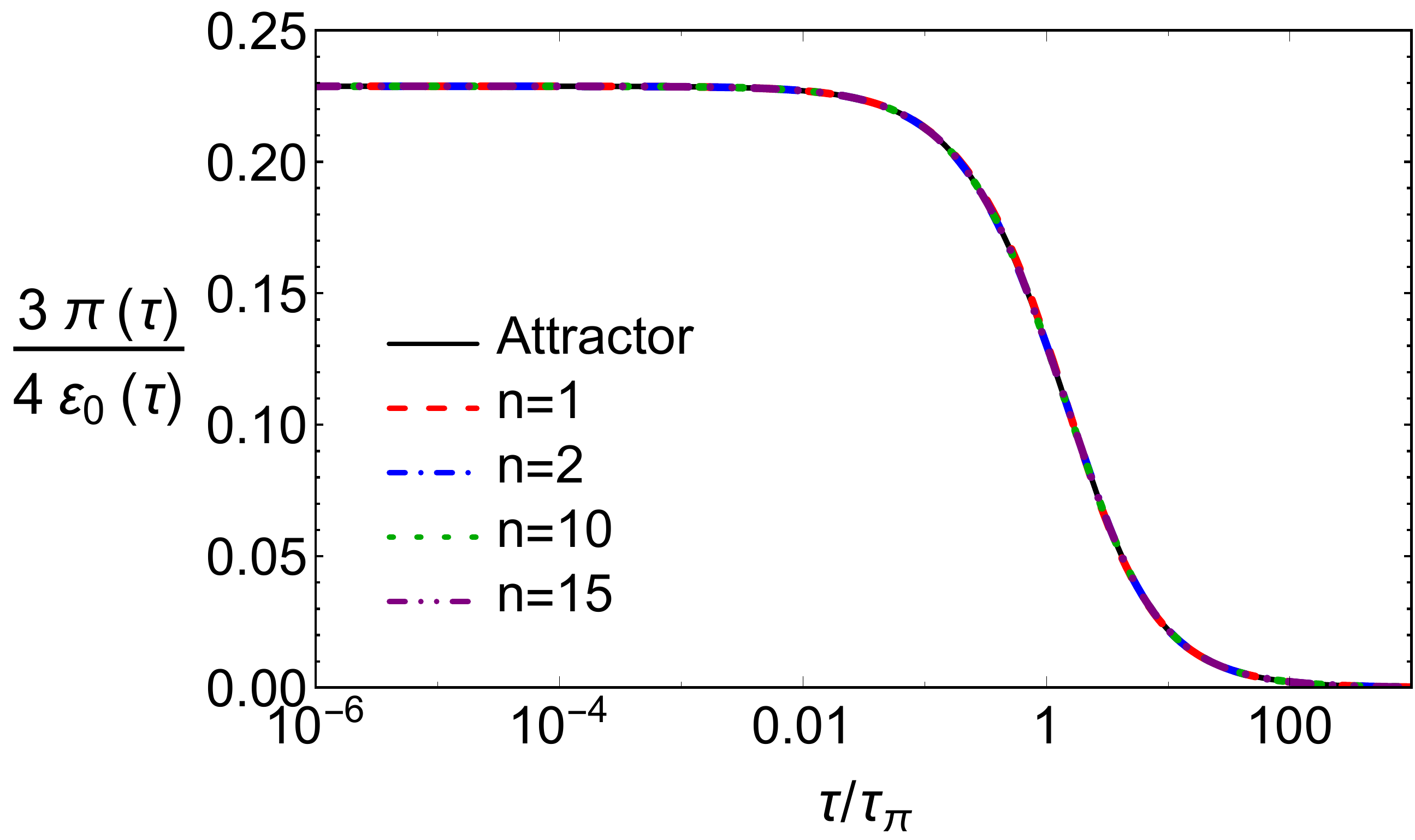}
  \caption{Attractor solution and  truncations of the slow-roll expansion.}
  \label{fig:sr-expn-2}
\end{subfigure}\hfil
\begin{subfigure}{0.5\textwidth}
  \includegraphics[width=\linewidth]{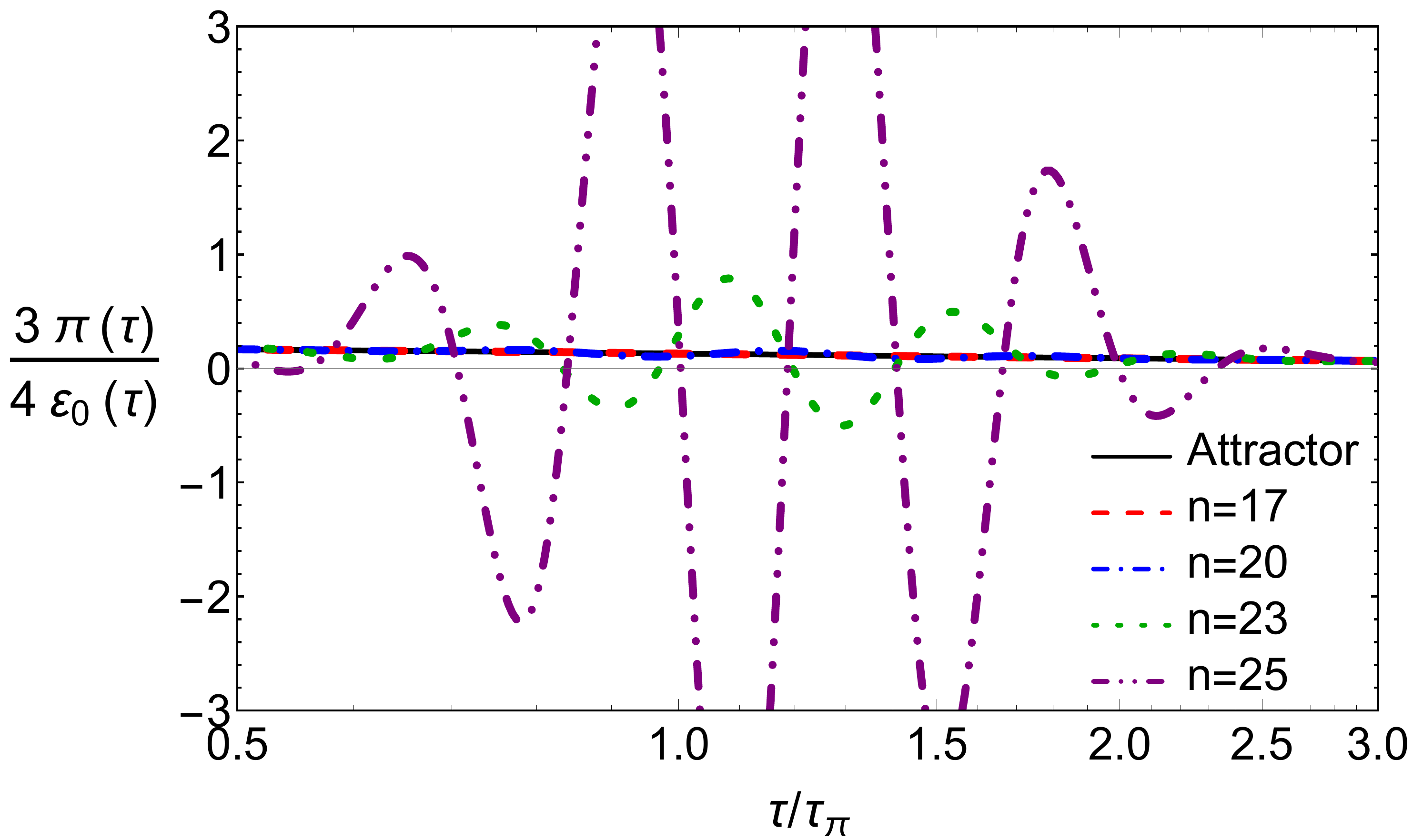}
 \caption{Attractor solution and truncations of the slow-roll expansion.}
  \label{fig:sr-expn-1}
\end{subfigure}\hfil
\caption{(Color online) Comparison between the attractor solution and truncations of the slow-roll series for (a) $n \leq 15$ and (b) $n > 15$.}
\label{fig:sl-rol-expn}
\end{figure}

Now we turn our attention to the slow-roll expansion of Eq.~\eqref{eq:shear-abel-1}, defined by 
\begin{equation}
\label{eq:shear-abel-1-slow-roll}
\begin{aligned}
&
\frac{2}{3} \epsilon \left(1- 4\Tilde{\pi}\right)\partial_{\Tilde{\tau}}\Tilde{\pi}
+
\frac{4}{3 \Tilde{\tau}} \Tilde{\pi}^{2}
+
\frac{2}{3 \Tilde{\tau}}
\Tilde{\pi}
+
\Tilde{\pi} 
=
\frac{2}{9 \Tilde{\tau}},
\end{aligned}    
\end{equation}
where a solution of $\Tilde{\pi}$ as a series in $\epsilon$ is assumed,
\begin{equation}
\label{eq:slow-roll-expn}
\begin{aligned}
&
\Tilde{\pi}(\Tilde{\tau}) 
= \sum_{n=0}^{\infty} \epsilon^{n} \Hat{\pi}^{(n)}(\Tilde{\tau}),
\end{aligned}    
\end{equation}
and Eq.~\eqref{eq:shear-abel-1-slow-roll} is solved order by order in $\epsilon$. A solution of the original equation is then recovered by setting $\epsilon=1$. This procedure leads to the following recurrence relation 
\begin{equation}
\label{eq:slow-roll-rec-rel}
\begin{aligned}
&
\frac{4}{3 \Tilde{\tau}} (\Hat{\pi}^{(0)})^{2}
+
\Tilde{\pi}^{(0)} \left(1 +
\frac{2}{3 \Tilde{\tau}}\right)
=
\frac{2}{9 \Tilde{\tau}}, \\
&
\frac{2}{3} \partial_{\Tilde{\tau}}\Hat{\pi}^{(n-1)} 
-
\sum_{m=0}^{n-1}
\frac{8}{3}\Hat{\pi}^{(n-m-1)} \partial_{\Tilde{\tau}}\Tilde{\pi}^{(m)}
+
\sum_{m=0}^{n}\frac{4}{3 \Tilde{\tau}} \Hat{\pi}^{(n-m)} \Hat{\pi}^{(m)}
+
\Hat{\pi}^{(n)} \left(
1 +\frac{2}{3 \Tilde{\tau}}
\right)
=
0
, n \geq 1.
\end{aligned}    
\end{equation}
It is readily seen that there are two solutions for the zeroth order coefficient. Namely,
\begin{equation}
\begin{aligned}
&
\Hat{\pi}^{(0) \pm} = \frac{1}{8} \left[
-(2 +3 \tau)  \pm \sqrt{ (2 +3 \tau)^{2} + \frac{32}{3}}\right].
\end{aligned}    
\end{equation}
The solution $\Tilde{\pi}^{(0) -}$ diverges as $\tau \to \infty$ whereas $\Tilde{\pi}^{(0) +}$ becomes identical to the Navier-Stokes solution in the same limit. Also, when $\tau \to 0$, $\Tilde{\pi}^{(0) +}=A$ and $\Tilde{\pi}^{(0) -}=B$, with $A$ and $B$  being defined in Eq.\ \eqref{eq:shear-no-rlx}. These features imply that the solutions $\Tilde{\pi}^{(0) +}$ and $\Tilde{\pi}^{(0) -}$ should be identified with the hydrodynamic attractor and the pullback attractor, respectively. In the following, we solve the recurrence relations \eqref{eq:slow-roll-rec-rel} up to order 25 using $\Tilde{\pi}^{(0) +}$ as the starting point of the series. The results for successive truncations can be seen in Fig.~\ref{fig:sl-rol-expn}. Solutions up to order 15 display features of convergence and become essentially identical to the attractor solution, as can be seen in Fig.\ \ref{fig:sr-expn-2}. However, considering a sufficiently high truncation order, $n>15$, it is possible to verify that the series actually diverges, as can be seen in Fig.\ \ref{fig:sr-expn-1}. The optimal truncation for this series is $n=15$. The divergence of the slow-roll expansion was already discussed in Ref.\ \cite{Denicol:2018pak}, but considering another type of microscopic interaction. Nevertheless, the optimal truncation of the slow-roll expansion provides a considerably better description of the attractor solution than the optimal truncation of the gradient expansion -- a feature also observed when solving BDNK theory.

\section{Conclusions}
\label{sec:concl}

In this work, we analyzed various hydrodynamic theories emerging from the classical scalar self-interacting $\lambda \varphi^{4}$ theory in the ultra-relativistic regime. The particular form of the cross-section for this system allows the computation of the corresponding transport coefficients in exact form, with the only approximation stemming from the power-counting procedure used to derive the corresponding fluid-dynamical theory. For Navier-Stokes theory, we reproduced the results obtained in Ref.~\cite{Denicol:2022bsq} and extended them to a more general class of matching conditions. 
Afterwards, using the generalized Chapman-Enskog expansion \cite{Rocha:2022ind}, BDNK theory was derived. And, finally, second-order transient fluid dynamics was derived following the method of moments \cite{Denicol:2021}. In all cases discussed, \textit{exact} expressions for the transport coefficients were obtained.

For Navier-Stokes and BDNK theories, the analytical expressions for the transport coefficients were derived considering a wide set of matching conditions. With the exception of the shear viscosity coefficient, all transport coefficients derived exhibited an explicit and significant dependence on the choice of matching condition. Furthermore, we have also derived necessary, yet not sufficient, conditions for the linear stability of BDNK theory around global equilibrium. These conditions exclude a wide set of matching conditions, including the so-called Exotic-Eckart matching condition \cite{Bemfica:2019knx,Rocha:2021lze,Bemfica:2020zjp,Rocha:2022ind}. 
Finally, for BDNK theory it was also demonstrated that  transport coefficients related to derivatives of the thermal potential all vanish exactly.

The derivation of second-order transient hydrodynamics was also developed, considering only Landau matching conditions, and employing the traditional method of moments \cite{Denicol:2012cn,Denicol:2021}. In this case, we calculated the moments of the collision term appearing in the moment equations and then truncated these equations using the order of magnitude scheme \cite{struchtrup2004stable,Fotakis:2022usk,Wagner:2022ayd}. The shear viscosity and particle diffusion coefficients were shown to be identical to those derived for Navier-Stokes theory. All second-order transport coefficients were also obtained exactly and were found
to be qualitatively similar to those obtained in previous calculations using the relaxation time \cite{Denicol:2014vaa} or the 14-moment \cite{Denicol:2012cn} approximations. Nevertheless, there were significant quantitative differences.

Solutions for all these hydrodynamic formulations were then obtained and analyzed in a simplified Bjorken flow scenario. We observed that both Navier-Stokes and BDNK theories exhibit unphysical solutions whenever the gradients of velocity are initially large. The issue is that, when sufficiently large initial gradients are imposed, these theories do not evolve towards equilibrium at late times. In contrast, second-order transient hydrodynamics does not display this pathological behavior with respect to the initial values of the gradients, but, instead, breaks down when the longitudinal pressure is initially negative. 

We further demonstrated that both BDNK (restricted to small initial gradients) and second-order (restricted to positive initial longitudinal pressure) theories give rise to attractor solutions, that we studied perturbatively via gradient \cite{deGroot:80relativistic,cercignani:02relativistic} and slow-roll \cite{Heller:2015dha,Liddle_1994} expansions. It was seen that the slow-roll expansion provides a better asymptotic representation of the attractor than the gradient series. For BDNK theory, the attractor solution was shown to depend significantly on the matching conditions employed.

As already noted, in this work second-order transient fluid dynamics was only derived for Landau matching conditions. Performing this task for more general matching conditions and still obtain exact expressions for all its transport coefficients is a nontrivial and cumbersome calculation that will be left to future work. Furthermore, we also plan to analyze the linear stability and causality of all the theories derived in this work, generalizing the calculations of Appendix \ref{app:stability-BDNK} to arbitrary perturbations and also to second-order theories. Finally, we plan to obtain exact solutions of the Boltzmann equation for the dissipative quantities and quantitatively assess the domain of validity of the derived theories. Solutions on more generic flow configurations will also be assessed.

\section*{Acknowledgments}

The authors also thank J.~Noronha and M.~Shokri for fruitful discussions. G.~S.~R. is partly funded by Coordenação de Aperfeiçoamento de Pessoal de Nível Superior (CAPES) Finance code 001, award No. 88881.650299/2021-01  and by Conselho Nacional de Desenvolvimento Científico e Tecnológico (CNPq), grant No.~142548/2019-7. C.~V.~P.~B. is also funded by CNPq, grant No.~140453/2021-0. G.~S.~D.~also acknowledges CNPq as well as Fundação Carlos Chagas Filho de Amparo à Pesquisa do Estado do Rio de Janeiro (FAPERJ), grant No.~E-26/202.747/2018.

\appendix

\section{Details of collisional moments calculation}
\label{app:collisional-moments}

In this appendix, we derive Eqs.~\eqref{eq:relation0}, \eqref{eq:relation1}, and \eqref{eq:relation2} which appear in the computations of the collisional moments $\mathcal{C}_{0}^{\mu \nu}$ and $\mathcal{C}_{1}^{\mu}$ in Sec.~\ref{sec:col-mom}. Starting with Eq.~\eqref{eq:relation0}, we see that it can be expressed as 
\begin{eqnarray}
\int dQ dQ'  (2\pi)^5 \delta^{(4)}(p^\alpha + p'^\alpha - q^\alpha - q'^\alpha) & = & \int \frac{d^3\mathbf{q}}{(2\pi)^3 q^0} \frac{d^3\mathbf{q}'}{(2\pi)^3 q'^0}  (2\pi)^5 \delta^{(4)}(p^\alpha + p'^\alpha - q^\alpha - q'^\alpha) \notag \\
& = & \int \frac{d^3q}{(2\pi)^3 \vert \mathbf{q}\vert} \frac{d^3 q'}{(2\pi)^3 \vert \mathbf{q}'\vert}  (2\pi)^5 \delta^{(4)}(p^\alpha + p'^\alpha - q^\alpha - q'^\alpha).
\end{eqnarray}
The second equality is only valid since we are considering massless particles, and thus $(k^0)^2=\mathbf{k}^2\equiv k^2$. Moreover, this is a scalar integral, and therefore it can be calculated in any reference frame. For the sake of convenience, we take the center of mass frame, in which the total 3-momentum of the system is zero. It is then convenient to define the total 4-momentum as
\begin{equation}
p^\mu+p'^\mu=Q^\mu_T=(\sqrt{s},0,0,0),
\end{equation}
and it follows that $Q^\mu_T Q_{T,\mu}=s$. Therefore,
\begin{eqnarray}
\int dQ dQ'  (2\pi)^5 \delta^{(4)}(p^\alpha + p'^\alpha - q^\alpha - q'^\alpha) & = & \int \frac{d^3q}{(2\pi)^3 \vert \mathbf{q}\vert} \frac{1}{(2\pi)^3 \vert \mathbf{q}\vert}  (2\pi)^5 \delta(\sqrt{s}-2\vert \mathbf{q}\vert) \notag \\
&=& \int d(2q)  \delta(\sqrt{s}-2q) = 1. \label{eq:scalar_int_1}
\end{eqnarray}

The derivation of Eqs.~\eqref{eq:relation1}, and \eqref{eq:relation2} can be done simultaneously, with the realization that they are the projections of the tensor
\begin{equation}
\label{eq:tensor-qq-int}
 \int dQ dQ' q^{\alpha}q^{\beta} (2\pi)^5 \delta^{(4)}(p^\alpha + p'^\alpha - q^\alpha - q'^\alpha)     
\end{equation}
with $u_{\alpha} \Delta^{\mu \beta}$ and $\Delta^{\mu \nu}_{\alpha \beta}$, respectively. The presence of the delta-function and the integration over four-momenta $q$ and $q'$ imply that the integral above can only depend on the \textit{total} momentum, and, in this context, it is convenient to define the normalized total momentum as
\begin{equation}
\hat{Q}^\mu_T=\frac{Q^\mu_T}{\sqrt{s}}=(1,0,0,0) \Longrightarrow \hat{Q}^\mu_T \hat{Q}_{T,\mu}=1,
\end{equation}
thus the tensor in Eq.~\eqref{eq:tensor-qq-int} can be decomposed in terms of a parallel and orthogonal components with respect to the normalized total momentum, 
\begin{eqnarray}
\int dQ dQ' q^{\alpha}q^{\beta} (2\pi)^5 \delta^{(4)}(p^\mu + p'^\mu - q^\mu - q'^\mu) 
& = & \int dQ dQ' q^\alpha q^\beta (2\pi)^5 \delta(\sqrt{s}-q-q')\delta^{(3)}(\mathbf{q}+\mathbf{q}') \notag \\
& = &  \left[ F \hat{Q}^\alpha_T\hat{Q}^\beta_T+G \Delta^{\alpha \beta}_{Q} \right], \label{eq:decomp_AB}
\end{eqnarray}
where $\Delta^{\alpha \beta}_{Q} = g^{\alpha\beta}-\hat{Q}^\alpha_T\hat{Q}^\beta_T $ is the projector in the 3-space orthogonal to $\hat{Q}^\alpha_T$. This is the most general rank-2 tensor structure built only with $\hat{Q}_{T}^{\alpha}$ and $g^{\alpha \beta}$. The parameters $A$ and $B$ can be obtained by projecting the above equation with $\hat{Q}^\alpha_T\hat{Q}^\beta_T$ and $\Delta^{\alpha \beta}_{Q}$. Indeed,  
\begin{eqnarray}
F & = & \hat{Q}_{T,\alpha}\hat{Q}_{T,\beta}\int dQ dQ' q^\alpha q^\beta (2\pi)^5 \delta(\sqrt{s}-q-q')\delta^{(3)}(\mathbf{q}+\mathbf{q}') \notag \\
&=&\int dQ dQ' (q_\alpha \hat{Q}^\alpha_T)^2 (2\pi)^5 \delta(\sqrt{s}-q-q')\delta^{(3)}(\mathbf{q}+\mathbf{q}') \notag \\
&=&\int \frac{d^3q}{(2\pi)^3 \vert \mathbf{q}\vert} \frac{d^3q'}{(2\pi)^3 \vert \mathbf{q}'\vert} \vert \mathbf{q}\vert^2 (2\pi)^5 \delta(\sqrt{s}-\vert \mathbf{q}\vert-\vert \mathbf{q}'\vert)\delta^{(3)}(\mathbf{q}+\mathbf{q}') \notag \\
&=&\frac{1}{4}\int (2\vert \mathbf{q}\vert)^2 d(2\vert \mathbf{q}\vert) \delta(\sqrt{s}-2\vert \mathbf{q}\vert) \notag \\
&=&\frac{s}{4}.
\end{eqnarray}
Moreover, $G$ is obtained as
\begin{eqnarray}
G &=&\frac{1}{3} \Delta^{\alpha \beta}_{Q} \int dQ dQ' q_\alpha q_\beta (2\pi)^5 \delta(\sqrt{s}-q-q')\delta^{(3)}(\mathbf{q}+\mathbf{q}') \notag \\
&=&-\frac{1}{3}\int dQ dQ' (q_\alpha \hat{Q}^\alpha_T)^2 (2\pi)^5 \delta(\sqrt{s}-q-q')\delta^{(3)}(\mathbf{q}+\mathbf{q}') = - \frac{A}{3} = -\frac{s}{12},
\end{eqnarray}
where we have used the on-shell condition to obtain the second equality, $g_{\alpha \beta} p^\alpha p^\beta = 0$. Furthermore, it is necessary to include the factor $1/3$ since $\Delta^{\alpha \beta}_{Q} \Delta_{Q \alpha \beta} = 3$. Then, 
\begin{equation}
 \int dQ dQ' q^{\alpha}q^{\beta} (2\pi)^5 \delta^{(4)}(p^\alpha + p'^\alpha - q^\alpha - q'^\alpha)
 =
 \frac{s}{4} \hat{Q}^\alpha_T\hat{Q}^\beta_T
 -
 \frac{s}{12} \Delta^{\alpha \beta}_{Q}, 
\end{equation}
which, when projected with $u_{\alpha} \Delta^{\mu \beta}$ and $\Delta^{\mu \nu}_{\alpha \beta}$ lead, respectively, to
\begin{equation}
\begin{aligned}
& \int dQ dQ' (2\pi)^5 E_{\bf p} p^{\langle \mu \rangle} \delta^{(4)}(p^\alpha + p'^\alpha - q^\alpha - q'^\alpha)
 =
\frac{1}{3} u_\alpha Q^\alpha_T Q^{\langle\mu\rangle}_T , \\
& \int dQ dQ' q^{\langle \mu}q^{\nu \rangle}  (2\pi)^5 \delta^{(4)}(p^\alpha + p'^\alpha - q^\alpha - q'^\alpha)
 =
 \frac{1}{3} Q_T^{\langle\mu} Q_T^{\nu\rangle}.
 \end{aligned}
\end{equation}

\section{Hilbert theory}
\label{appendix:Hilbert}

In this Appendix, we shall discuss the computation of transport coefficients for the Hilbert theory \cite{hilbert1912begrundung,grad:1949kinetic,cercignani:90mathematical,Rocha:2022ind}. Similarly to what was discussed for the Chapman-Enskog expansion, one converts the original Boltzmann equation into a perturbative problem by inserting  a book-keeping parameter $\epsilon$,
\begin{equation}
\label{eq:hilf}
\begin{aligned}
 \epsilon p^\mu \partial_\mu f_{\mathbf{p}}   = C[f_{\mathbf{p}}],
\end{aligned}
\end{equation}
and consider solution for $f_{\mathbf{p}}$ of the following form, 
\begin{equation}
\label{eq:chapman-expn-SPDF}
f_{\mathbf{p}} = \sum_{i=0}^{\infty} \epsilon^{i} f^{(i)}_{\mathbf{p}}.    
\end{equation}
Afterwards, the Boltzmann equation is then solved order by order in $\epsilon$. An important difference with respect to the Chapman-Enskog expansion is that the time-like derivative is not assumed to obey an independent expansion in $\epsilon$ \cite{deGroot:80relativistic,Rocha:2022ind}. This feature implies that the Chapman-Enskog theory is a resummation of the Hilbert series. One distinctive feature of Hilbert's theory is that conservation equations are obeyed order-by-order in $\epsilon$ \cite{grad1958principles,Rocha:2022ind},
\begin{equation}
\label{eq:cons-tmunu-hilb}
\begin{aligned}
&
\partial_{\mu} N^{\mu}_{(k)} = 0, \quad 
\partial_{\mu} T^{\mu \nu}_{(k)} = 0, \quad k \geq 0 .
\end{aligned}
\end{equation}
Thus, in zero-th order, the Euler equations are recovered, 
\begin{subequations}
 \label{eq:euler-eqns-hilb}
\begin{align}
 \label{eq:hydro-EoM-n0-euler}
D n_{0} + n_{0} \theta &= 0, \\
\label{eq:hydro-EoM-eps-euler}
D\varepsilon_{0} + (\varepsilon_{0}+ P_{0}) \theta &= 0, \\
\label{eq:hydro-EoM-umu-euler}
(\varepsilon_{0} + P_{0})Du^{\mu} - \nabla^{\mu}P_{0}  &= 0,
\end{align}
\end{subequations}
and considering the decomposition
\begin{subequations}
\label{eq:decompos-curr-hilb}
\begin{align}
N^{\mu}_{(k)} & \equiv \int dP p^{\mu} f^{(k)}_{\bf p}  = n_{(k)} u^{\mu} + \nu^{\mu}_{(k)}  \\
    T^{\mu \nu}_{(k)} & \equiv \int dP p^{\mu}p^{\nu} f^{(k)}_{\bf p} = \varepsilon_{(k)} u^{\mu} u^{\nu} - \Pi_{(k)} \Delta^{\mu \nu} + h^{\mu}_{(k)} u^{\nu} + h^{\nu}_{(k)} u^{\mu} + \pi^{\mu \nu}_{(k)}, \quad k \geq 0,
\end{align}    
\end{subequations}
we have, for $k=1$
\begin{subequations}
 \label{eq:hydro-EoMs-hilb}
\begin{align}
 \label{eq:hydro-EoM-n0-hilb}
D n_{(1)} + n_{(1)} \theta + \partial_{\mu} \nu^{\mu}_{(1)} &= 0, \\
\label{eq:hydro-EoM-eps-hilb}
D \varepsilon_{(1)} + (\varepsilon_{(1)} + \Pi_{(1)}) \theta - \pi^{\mu \nu}_{(1)} \sigma_{\mu \nu} + \partial_{\mu}h^{\mu}_{(1)} + u_{\mu} Dh^{\mu}_{(1)} &= 0, \\
\label{eq:hydro-EoM-umu-hilb}
(\varepsilon_{(1)} + \Pi_{(1)}) Du^{\mu} - \nabla^{\mu}\Pi_{(1)} + h^{\mu}_{(1)} \theta + h^{\alpha}_{(1)} \Delta^{\mu \nu} \partial_{\alpha}u_{\nu} +  \Delta^{\mu \nu} Dh_{(1)\nu} + \Delta^{\mu \nu} \partial_{\alpha}\pi^{\alpha}_{(1)\nu} &= 0,
\end{align}
\end{subequations}
Eqs.~\eqref{eq:euler-eqns-hilb} and \eqref{eq:hydro-EoMs-hilb} form the set of equations of motion to be obeyed at first order in Hibert expansion. The system of partial differential equations will be closed once the first order solution of the perturbative problem, $\phi_{\bf p}^{(1)}$, is known. 

The first order solution for deviation function from local equilibrium, $\phi_{\bf p}^{(1)}$, is also obtained from Eq.~\eqref{CE}. The difference being that, in Hilbert theory, the replacement of time- by space-like derivatives is exact and not perturbative, since the Euler equations are obeyed in exactly. Thus the solution is also given by
\begin{equation}
\label{eq:sol-hilb-cap8}
\phi_{\mathbf{p}} = a + b_\mu p^\mu + \frac{1}{4 \chi_{11}} L^{(3)}_{1\mathbf{p}}p_{\langle\mu\rangle}\nabla ^{\mu }\alpha 
-\frac{\beta}{\chi_{20}} 
p_{\langle \mu }p_{\nu \rangle}
\sigma^{\mu \nu }.
\end{equation}
Thus, from the decompositions \eqref{eq:decompos-curr-hilb}, we find 
\begin{subequations}
\label{eq:constrains-hilbs-ab}
\begin{align}
&
\label{eq:scalar-hilb}
n_{(1)} = a I_{1,0} + (b_{\mu} u^{\mu}) I_{2,0}, \quad \varepsilon_{(1)} = a I_{2,0} + (b_{\mu} u^{\mu}) I_{3,0}, \quad \Pi_{(1)} = a I_{2,1} + (b_{\mu} u^{\mu}) I_{3,1}, \\
&
\label{eq:vector-hilb}
\nu^{\mu}_{(1)} = - I_{2,1} b^{\langle \mu \rangle}
+
\kappa_{H} \nabla^{\mu} \alpha,
\quad h^{\mu}_{(1)} = - I_{3,1} b^{\langle \mu \rangle}
-
\lambda_{H} \nabla^{\mu} \alpha,
\\ 
&
\label{eq:shear-hilb}
\pi^{\mu \nu}_{(1)} = 2 \eta \sigma^{\mu \nu},
\end{align}    
\end{subequations}
where the Hilbert transport coefficients are given by
\begin{subequations}
\begin{align}
&
\kappa_{H} = 0, \quad \lambda_{H} 
= \frac{12}{g \beta^{3}} ,\\
&
\eta = \frac{48}{g \beta^{3}}.
\end{align}    
\end{subequations}
The fields $a$ and $b_{\mu}$ are found by employing matching conditions. However, since Eqs.~\eqref{eq:euler-eqns-hilb} and \eqref{eq:hydro-EoMs-hilb} form a system of 10 equations of motion, we need to provide only 9 further relations between the dissipative fields and derivatives of the equilibrium fields. Five of these constrains are given by the constitutive relation of the shear-stress tensor \eqref{eq:shear-hilb}. The four remaining relations stem from the constrains  
\begin{subequations}
\label{eq:constrains-H}
\begin{align}
&
\Pi_{(1)} = \frac{1}{3} \varepsilon_{(1)},
\\
&
\nu^{\mu}_{(1)} - \frac{\beta}{4} h^{\mu}_{(1)} 
= 
\left( \kappa_{H} + \frac{\beta}{4} \lambda_{H}
\right) \nabla^{\mu} \alpha
=
\frac{3}{g \beta^{2}} \nabla^{\mu} \alpha
,  
\end{align}    
\end{subequations}
which can be readily derived from Eqs.~\eqref{eq:scalar-hilb} and \eqref{eq:vector-hilb}, respectively. Thus, the system of hydrodynamic equations of motion is given by Eqs.~\eqref{eq:euler-eqns-hilb}, \eqref{eq:hydro-EoMs-hilb}, \eqref{eq:shear-hilb}, and \eqref{eq:constrains-H}.

In Bjorken flow, the equations of motion for the Hilbert theory read
\begin{subequations}
\label{eq:Hilb-EoM1-bjorken0-chap8}
\begin{align}
\label{eq:Hilb-EoM1-bjorken1-chap8}
&  \dot{n}_{0} 
 + 
 \frac{n_{0}}{\tau}
  = 0 , \\
&  \dot{\delta n} 
 + 
 \frac{\delta n}{\tau}
  = 0 , \\
&  \dot{\varepsilon}_{0} 
 + 
 \frac{4 \varepsilon_{0}}{3 \tau}
 = 0 , \\  
&  \dot{\delta \varepsilon} 
 + 
 \frac{4 \delta\varepsilon}{3 \tau}
 - \frac{64}{g \beta^{3} \tau^{2}}  = 0,
 \label{eq:Hilb-EoM-eps-2-chap8}
\end{align}
\end{subequations}
where constrain \eqref{eq:scalar-hilb} and constitutive relation \eqref{eq:shear-hilb} have been employed. It is reminded that, in Bjorken flow, 4-vectors orthogonal to $u^{\mu}$ are identically zero, due to the symmetry assumptions. Hence constrain \eqref{eq:vector-hilb} and the vector equation of motion are identically zero. The solution to Eqs.~\eqref{eq:Hilb-EoM1-bjorken0-chap8} read
\begin{equation}
\label{eq:hilbert-sol-bj-chap8}
\begin{aligned}
n_{0}(\tau) &= n_{0}(\tau_{0})\frac{\tau_{0}}{\tau}, \quad
\delta n(\tau) = \delta n(\tau_{0})\frac{\tau_{0}}{\tau}, \\
\varepsilon_{0}(\tau)  & = \varepsilon_{0}(\tau_{0}) \left(\frac{\tau_{0}}{\tau}\right)^{4/3} ,
\quad
\delta\varepsilon(\tau) = C\left(\frac{\tau_{0}}{\tau} \right)^{4/3} - \frac{3 K_{\mathrm{Hilb}}}{2} \left(\frac{\tau_{0}}{\tau}\right)^{2}  .
\end{aligned}    
\end{equation}
which is obtained making use of the equation of state $\varepsilon_{0} = 3 n_{0}/\beta$, and where
\begin{equation}
\begin{aligned}
 &
K_{\mathrm{Hilb}} 
=
\frac{64}{g \beta(\tau_{0})^{3}\tau_{0}}, 
\quad
C = 
\delta \varepsilon(\tau_{0}) 
+
\frac{3 K_{\mathrm{Hilb}}}{2}, 
\end{aligned}    
\end{equation}
Similarly to the constant relaxation time case (see e.g.~Eq.~(112) of Ref.~\cite{Rocha:2022ind}), the
late-time expansion for the Hilbert solution has the following simple form,
\begin{equation}
\begin{aligned}
& \frac{\varepsilon_{0}(\tau)+\delta \varepsilon(\tau)}{\varepsilon_{0}(\tau_{0}) + \delta \varepsilon(\tau_{0})}
=
\left(\frac{\tau_{0}}{\tau}\right)^{4/3}
\left[\left( 1 + \frac{3}{2} K_{\mathrm{Hilb}} \right)
-
\frac{3}{2} K_{\mathrm{Hilb}} \left(\frac{\tau_{0}}{\tau}\right)^{2/3}
\right],
\end{aligned}    
\end{equation}
whereas the Navier-Stokes one (see Eq.~\eqref{eq:late-time-expn-NS}) has infinitely many terms. Additionally, is noted that the Hilbert theory solution does not present the finite time divergence observed in Navier-Stokes theory (cf.~Eq.~\eqref{eq:phys-cond-NS}).  

\section{Linear stability of the BDNK theory}
\label{app:stability-BDNK}

In this appendix, we provide an initial discussion on the linear stability of BDNK theory around a global equilibrium state. In particular, we perform this analysis following the procedure outlined in Refs.~\cite{Denicol:2021,Brito_2022}, but restricting ourselves to the homogeneous limit, in which the fluid-dynamical fields do not change in space. This will provide us with necessary, yet not sufficient, conditions for the transport coefficients of BDNK theory. Furthermore, for the sake of simplicity, we only consider a fluid that is initially at rest, i.e., $u^\mu_0 = (1,0,0,0)$. 

Thus, we consider small perturbations (denoted by the symbol $\Delta$) on all fluid-dynamical variables 
\begin{equation}
\begin{aligned}
\varepsilon & \rightarrow \varepsilon_0 + \Delta\varepsilon_0 + \Delta\delta\varepsilon, \hspace{.2cm}
n \rightarrow n_0 + \Delta n_0 + \Delta\delta n,
\hspace{.2cm} P \rightarrow P_{0} + \Delta \Pi,
\hspace{.2cm} u^\mu \rightarrow u^\mu_0 + \Delta u^\mu, \hspace{.2cm}
\\
\nu^\mu & \rightarrow \Delta \nu^\mu, \hspace{.2cm}
h^\mu \rightarrow \Delta h^\mu, \hspace{.2cm}
\pi^{\mu\nu} \rightarrow \Delta \pi^{\mu\nu},
\end{aligned}
\end{equation}
The linearized fluid-dynamical equations, in the homogeneous limit, then read
\begin{subequations}
\label{eq:linear-hydro-eqs}
\begin{align}
D_0 \Delta n_0 + D_0 \Delta\delta n &= \mathcal{O}(2) \approx 0, \\
D_0 \Delta\varepsilon_0 + D_0 \Delta\delta \varepsilon &= \mathcal{O}(2) \approx 0, \\
(\varepsilon_{0} + P_{0}) D_0 \Delta u^{\mu} + D_0 \Delta h^{\mu} &= \mathcal{O}(2) \approx 0,
\end{align}
\end{subequations}
where $\mathcal{O}(2)$ refers to all the second-order terms in perturbations that were not considered and $D_0 \equiv u^\mu_0 \partial_\mu$ is the comoving derivative with respect to the background fluid 4-velocity. Furthermore, in the homogeneous limit, the constitutive relations satisfied by the dissipative currents that appear in the equations above are simplified to, cf.~Eq.~\eqref{eq:const-rel-BDNK-chap-8},
\begin{equation}
\label{eq:lin-cttv-rels}
\begin{aligned}
\Delta \delta n 
= \xi D_0 \left(\frac{\Delta n_0}{n_0} - \frac{\Delta\varepsilon_0}{\varepsilon_0} \right),
\ \
 \Delta \delta \varepsilon 
= \chi  D_0 \left(\frac{\Delta n_0}{n_0} - \frac{\Delta \varepsilon_0}{\varepsilon_0} \right), \ \
\Delta h^{\mu} 
= \lambda D_0 \Delta u^{\mu}, 
\end{aligned}
\end{equation}
where we have used the following thermodynamic relation, valid for ultra-relativistic classical gases,
\begin{equation}
\frac{\Delta \beta}{\beta_0} = \frac{\Delta n_{0}}{n_0} - \frac{\Delta \varepsilon_{0}}{\varepsilon_0}.
\end{equation}

At this point, it is useful to express the linearized fluid-dynamical fields in Fourier space. Without loss of generality, we take the following convention for the Fourier transform
\begin{eqnarray}
\tilde{X}(k^{\mu }) &=&\int d^{4}x\hspace{0.1cm}\exp \left( -ix_{\mu }k^{\mu
}\right) X(x^{\mu }), \label{transf1}\\
X(x^{\mu }) &=&\int \frac{d^{4}k}{(2\pi )^{4}}\hspace{0.1cm}\exp \left(
ix_{\mu }k^{\mu }\right) \tilde{X}(k^{\mu }),\label{transf2}
\end{eqnarray}
where $k^{\mu}=(\omega ,\mathbf{k})$, is the wave 4-vector, with $\omega$ being the frequency and $\mathbf{k}$ the wave vector. Since we are considering the homogeneous limit, terms with $\mathbf{k}$ will not appear in our analysis. Then, the linearized fluid-dynamical equations, Eqs.~\eqref{eq:linear-hydro-eqs}, in Fourier space read
\begin{subequations}
\label{eq:fourier-linear-hydro-eqs}
\begin{align}
\omega \left( \Delta \tilde{n}_0 + \Delta \tilde{\delta n} \right) &= 0, \\
\omega \left( \Delta\tilde{\varepsilon}_0 + \Delta\tilde{\delta \varepsilon} \right) &= 0, \\
\omega \left[(\varepsilon_{0} + P_{0})\Delta \tilde{u}^{\mu} + \Delta \tilde{h}^{\mu} \right] &= 0.
\end{align}
\end{subequations}
Moreover, it follows that the Fourier transform of the constitutive relations \eqref{eq:lin-cttv-rels} are given by
\begin{equation}
\Delta \tilde{\delta n} 
= i \xi \omega \left(\frac{\Delta \tilde{n}_0}{n_0} - \frac{ \Delta\tilde{\varepsilon}_0}{\varepsilon_0} \right), \ \
 \Delta \tilde{\delta \varepsilon} 
= i \chi \omega \left(\frac{\Delta \tilde{n}_0}{n_0} - \frac{\Delta\tilde {\varepsilon}_0}{\varepsilon_0} \right), \ \
\Delta \tilde{h}^{\mu} 
= i \lambda \omega \Delta \tilde{u}^{\mu}.
\label{eq:fourier-dissip-BDNK}
\end{equation}

The next step is to solve the fluid-dynamical equations in Fourier space, obtaining solutions for $\omega$. In particular, given the convention adopted for the Fourier transform, linear stability is guaranteed as long as the imaginary part of $\omega$ is positive. As a matter of fact, it can be straightforwardly seen that this leads to perturbations that decrease exponentially with time, that will thus return to their equilibrium values. From Eqs.~\eqref{eq:fourier-linear-hydro-eqs} and \eqref{eq:fourier-dissip-BDNK}, we obtain the following dispersion relation
\begin{equation}
\omega^3 \left(1 + i \omega \frac{3 \lambda}{4 \varepsilon_0} \right) \left[ 1 + i\omega \left(\frac{\xi}{n_0} - \frac{\chi}{\varepsilon_0} \right) \right] = 0.\end{equation}
In order for both nonzero modes to have a positive imaginary part, we obtain the following constraints
\begin{equation}
\begin{aligned}
\frac{\lambda}{\varepsilon_{0}} > 0, \quad \frac{\xi}{n_{0}} > \frac{\chi}{\varepsilon_{0}},   
\end{aligned}    
\end{equation}
which, when substituted in the microscopic expressions \eqref{eq:const-rel-BDNK-chap-8}, lead to the following constraints on the matching parameters  
\begin{equation}
z>1, \hspace{.45cm} (q-1)(s-1) > (q-2)(s-2).
\end{equation}
In this context, the BDNK theory \textit{can be} linearly stable for homogeneous perturbations around an equilibrium state provided that these inequalities are simultaneously satisfied.
We remark that these are the most fundamental conditions the transport coefficients must satisfy in order for the BDNK theory to be linearly stable -- in particular, they are \textit{necessary} stability conditions. Furthermore, they prohibit the so-called exotic Eckart matching conditions \cite{Bemfica:2019knx,Rocha:2021lze,Bemfica:2020zjp}, where $z=0$ and $q=1$, with the remaining parameter $s$ being free. We note that, in Ref.~\cite{Shokri:2020cxa}, which derives BDNK theory using holography techniques, transport coefficients have been reported to violate causality constrains. A thorough study of the \textit{sufficient} conditions, also including a causality analysis, will be addressed in a followup paper. 


\bibliographystyle{apsrev4-1}
\bibliography{liography}

\begin{thebibliography}{77}%
\makeatletter
\providecommand \@ifxundefined [1]{%
 \@ifx{#1\undefined}
}%
\providecommand \@ifnum [1]{%
 \ifnum #1\expandafter \@firstoftwo
 \else \expandafter \@secondoftwo
 \fi
}%
\providecommand \@ifx [1]{%
 \ifx #1\expandafter \@firstoftwo
 \else \expandafter \@secondoftwo
 \fi
}%
\providecommand \natexlab [1]{#1}%
\providecommand \enquote  [1]{``#1''}%
\providecommand \bibnamefont  [1]{#1}%
\providecommand \bibfnamefont [1]{#1}%
\providecommand \citenamefont [1]{#1}%
\providecommand \href@noop [0]{\@secondoftwo}%
\providecommand \href [0]{\begingroup \@sanitize@url \@href}%
\providecommand \@href[1]{\@@startlink{#1}\@@href}%
\providecommand \@@href[1]{\endgroup#1\@@endlink}%
\providecommand \@sanitize@url [0]{\catcode `\\12\catcode `\$12\catcode
  `\&12\catcode `\#12\catcode `\^12\catcode `\_12\catcode `\%12\relax}%
\providecommand \@@startlink[1]{}%
\providecommand \@@endlink[0]{}%
\providecommand \url  [0]{\begingroup\@sanitize@url \@url }%
\providecommand \@url [1]{\endgroup\@href {#1}{\urlprefix }}%
\providecommand \urlprefix  [0]{URL }%
\providecommand \Eprint [0]{\href }%
\providecommand \doibase [0]{http://dx.doi.org/}%
\providecommand \selectlanguage [0]{\@gobble}%
\providecommand \bibinfo  [0]{\@secondoftwo}%
\providecommand \bibfield  [0]{\@secondoftwo}%
\providecommand \translation [1]{[#1]}%
\providecommand \BibitemOpen [0]{}%
\providecommand \bibitemStop [0]{}%
\providecommand \bibitemNoStop [0]{.\EOS\space}%
\providecommand \EOS [0]{\spacefactor3000\relax}%
\providecommand \BibitemShut  [1]{\csname bibitem#1\endcsname}%
\let\auto@bib@innerbib\@empty
\bibitem [{\citenamefont {Gale}\ \emph {et~al.}(2013)\citenamefont {Gale},
  \citenamefont {Jeon},\ and\ \citenamefont {Schenke}}]{gale2013hydrodynamic}%
  \BibitemOpen
  \bibfield  {author} {\bibinfo {author} {\bibfnamefont {C.}~\bibnamefont
  {Gale}}, \bibinfo {author} {\bibfnamefont {S.}~\bibnamefont {Jeon}}, \ and\
  \bibinfo {author} {\bibfnamefont {B.}~\bibnamefont {Schenke}},\ }\href@noop
  {} {\bibfield  {journal} {\bibinfo  {journal} {International Journal of
  Modern Physics A}\ }\textbf {\bibinfo {volume} {28}},\ \bibinfo {pages}
  {1340011} (\bibinfo {year} {2013})}\BibitemShut {NoStop}%
\bibitem [{\citenamefont {Heinz}\ and\ \citenamefont
  {Snellings}(2013)}]{heinz2013collective}%
  \BibitemOpen
  \bibfield  {author} {\bibinfo {author} {\bibfnamefont {U.}~\bibnamefont
  {Heinz}}\ and\ \bibinfo {author} {\bibfnamefont {R.}~\bibnamefont
  {Snellings}},\ }\href@noop {} {\bibfield  {journal} {\bibinfo  {journal}
  {Annual Review of Nuclear and Particle Science}\ }\textbf {\bibinfo {volume}
  {63}},\ \bibinfo {pages} {123} (\bibinfo {year} {2013})}\BibitemShut
  {NoStop}%
\bibitem [{\citenamefont {Florkowski}\ \emph {et~al.}(2018)\citenamefont
  {Florkowski}, \citenamefont {Heller},\ and\ \citenamefont
  {Spali{\'n}ski}}]{florkowski2018new}%
  \BibitemOpen
  \bibfield  {author} {\bibinfo {author} {\bibfnamefont {W.}~\bibnamefont
  {Florkowski}}, \bibinfo {author} {\bibfnamefont {M.~P.}\ \bibnamefont
  {Heller}}, \ and\ \bibinfo {author} {\bibfnamefont {M.}~\bibnamefont
  {Spali{\'n}ski}},\ }\href@noop {} {\bibfield  {journal} {\bibinfo  {journal}
  {Reports on Progress in Physics}\ }\textbf {\bibinfo {volume} {81}},\
  \bibinfo {pages} {046001} (\bibinfo {year} {2018})}\BibitemShut {NoStop}%
\bibitem [{\citenamefont {Rezzolla}\ and\ \citenamefont
  {Zanotti}(2013)}]{rezzolla2013relativistic}%
  \BibitemOpen
  \bibfield  {author} {\bibinfo {author} {\bibfnamefont {L.}~\bibnamefont
  {Rezzolla}}\ and\ \bibinfo {author} {\bibfnamefont {O.}~\bibnamefont
  {Zanotti}},\ }\href@noop {} {\emph {\bibinfo {title} {Relativistic
  hydrodynamics}}}\ (\bibinfo  {publisher} {Oxford University Press},\ \bibinfo
  {year} {2013})\BibitemShut {NoStop}%
\bibitem [{\citenamefont {Chabanov}\ \emph {et~al.}(2021)\citenamefont
  {Chabanov}, \citenamefont {Rezzolla},\ and\ \citenamefont
  {Rischke}}]{chabanov:21-general}%
  \BibitemOpen
  \bibfield  {author} {\bibinfo {author} {\bibfnamefont {M.}~\bibnamefont
  {Chabanov}}, \bibinfo {author} {\bibfnamefont {L.}~\bibnamefont {Rezzolla}},
  \ and\ \bibinfo {author} {\bibfnamefont {D.~H.}\ \bibnamefont {Rischke}},\
  }\href {\doibase 10.1093/mnras/stab1384} {\bibfield  {journal} {\bibinfo
  {journal} {Monthly Notices of the Royal Astronomical Society}\ }\textbf
  {\bibinfo {volume} {505}},\ \bibinfo {pages} {5910} (\bibinfo {year}
  {2021})},\ \Eprint
  {http://arxiv.org/abs/https://academic.oup.com/mnras/article-pdf/505/4/5910/38873579/stab1384.pdf}
  {https://academic.oup.com/mnras/article-pdf/505/4/5910/38873579/stab1384.pdf}
  \BibitemShut {NoStop}%
\bibitem [{\citenamefont {Fujibayashi}\ \emph {et~al.}(2018)\citenamefont
  {Fujibayashi}, \citenamefont {Kiuchi}, \citenamefont {Nishimura},
  \citenamefont {Sekiguchi},\ and\ \citenamefont
  {Shibata}}]{Fujibayashi:2017puw}%
  \BibitemOpen
  \bibfield  {author} {\bibinfo {author} {\bibfnamefont {S.}~\bibnamefont
  {Fujibayashi}}, \bibinfo {author} {\bibfnamefont {K.}~\bibnamefont {Kiuchi}},
  \bibinfo {author} {\bibfnamefont {N.}~\bibnamefont {Nishimura}}, \bibinfo
  {author} {\bibfnamefont {Y.}~\bibnamefont {Sekiguchi}}, \ and\ \bibinfo
  {author} {\bibfnamefont {M.}~\bibnamefont {Shibata}},\ }\href {\doibase
  10.3847/1538-4357/aabafd} {\bibfield  {journal} {\bibinfo  {journal}
  {Astrophys. J.}\ }\textbf {\bibinfo {volume} {860}},\ \bibinfo {pages} {64}
  (\bibinfo {year} {2018})},\ \Eprint {http://arxiv.org/abs/1711.02093}
  {arXiv:1711.02093 [astro-ph.HE]} \BibitemShut {NoStop}%
\bibitem [{\citenamefont {Rischke}(2004)}]{rischke2004quark}%
  \BibitemOpen
  \bibfield  {author} {\bibinfo {author} {\bibfnamefont {D.~H.}\ \bibnamefont
  {Rischke}},\ }\href@noop {} {\bibfield  {journal} {\bibinfo  {journal}
  {Progress in Particle and Nuclear Physics}\ }\textbf {\bibinfo {volume}
  {52}},\ \bibinfo {pages} {197} (\bibinfo {year} {2004})}\BibitemShut
  {NoStop}%
\bibitem [{\citenamefont {Yagi}\ \emph {et~al.}(2005)\citenamefont {Yagi},
  \citenamefont {Hatsuda},\ and\ \citenamefont {Miake}}]{yagi2005quark}%
  \BibitemOpen
  \bibfield  {author} {\bibinfo {author} {\bibfnamefont {K.}~\bibnamefont
  {Yagi}}, \bibinfo {author} {\bibfnamefont {T.}~\bibnamefont {Hatsuda}}, \
  and\ \bibinfo {author} {\bibfnamefont {Y.}~\bibnamefont {Miake}},\
  }\href@noop {} {\emph {\bibinfo {title} {Quark-gluon plasma: From big bang to
  little bang}}},\ Vol.~\bibinfo {volume} {23}\ (\bibinfo  {publisher}
  {Cambridge University Press},\ \bibinfo {year} {2005})\BibitemShut {NoStop}%
\bibitem [{\citenamefont {Bemfica}\ \emph
  {et~al.}(2018{\natexlab{a}})\citenamefont {Bemfica}, \citenamefont
  {Disconzi},\ and\ \citenamefont {Noronha}}]{bemfica:18causality}%
  \BibitemOpen
  \bibfield  {author} {\bibinfo {author} {\bibfnamefont {F.~S.}\ \bibnamefont
  {Bemfica}}, \bibinfo {author} {\bibfnamefont {M.~M.}\ \bibnamefont
  {Disconzi}}, \ and\ \bibinfo {author} {\bibfnamefont {J.}~\bibnamefont
  {Noronha}},\ }\href@noop {} {\bibfield  {journal} {\bibinfo  {journal}
  {Physical Review D}\ }\textbf {\bibinfo {volume} {98}},\ \bibinfo {pages}
  {104064} (\bibinfo {year} {2018}{\natexlab{a}})}\BibitemShut {NoStop}%
\bibitem [{\citenamefont {Eckart}(1940)}]{Eckart:1940te}%
  \BibitemOpen
  \bibfield  {author} {\bibinfo {author} {\bibfnamefont {C.}~\bibnamefont
  {Eckart}},\ }\href {\doibase 10.1103/PhysRev.58.919} {\bibfield  {journal}
  {\bibinfo  {journal} {Phys. Rev.}\ }\textbf {\bibinfo {volume} {58}},\
  \bibinfo {pages} {919} (\bibinfo {year} {1940})}\BibitemShut {NoStop}%
\bibitem [{\citenamefont {Landau}\ and\ \citenamefont
  {Lifshitz}(1959)}]{landau:59fluid}%
  \BibitemOpen
  \bibfield  {author} {\bibinfo {author} {\bibfnamefont {L.}~\bibnamefont
  {Landau}}\ and\ \bibinfo {author} {\bibfnamefont {E.}~\bibnamefont
  {Lifshitz}},\ }\href@noop {} {\bibfield  {journal} {\bibinfo  {journal}
  {Course of Theoretical Physics, Pergamon Press, London}\ }\textbf {\bibinfo
  {volume} {6}} (\bibinfo {year} {1959})}\BibitemShut {NoStop}%
\bibitem [{\citenamefont {Hiscock}\ and\ \citenamefont
  {Lindblom}(1987)}]{hiscock1987linear}%
  \BibitemOpen
  \bibfield  {author} {\bibinfo {author} {\bibfnamefont {W.~A.}\ \bibnamefont
  {Hiscock}}\ and\ \bibinfo {author} {\bibfnamefont {L.}~\bibnamefont
  {Lindblom}},\ }\href@noop {} {\bibfield  {journal} {\bibinfo  {journal}
  {Physical Review D}\ }\textbf {\bibinfo {volume} {35}},\ \bibinfo {pages}
  {3723} (\bibinfo {year} {1987})}\BibitemShut {NoStop}%
\bibitem [{\citenamefont {Hiscock}\ and\ \citenamefont
  {Lindblom}(1983)}]{hiscock1983stability}%
  \BibitemOpen
  \bibfield  {author} {\bibinfo {author} {\bibfnamefont {W.~A.}\ \bibnamefont
  {Hiscock}}\ and\ \bibinfo {author} {\bibfnamefont {L.}~\bibnamefont
  {Lindblom}},\ }\href@noop {} {\bibfield  {journal} {\bibinfo  {journal}
  {Annals of Physics}\ }\textbf {\bibinfo {volume} {151}},\ \bibinfo {pages}
  {466} (\bibinfo {year} {1983})}\BibitemShut {NoStop}%
\bibitem [{\citenamefont {Olson}(1990)}]{olson1990stability}%
  \BibitemOpen
  \bibfield  {author} {\bibinfo {author} {\bibfnamefont {T.~S.}\ \bibnamefont
  {Olson}},\ }\href@noop {} {\bibfield  {journal} {\bibinfo  {journal} {Annals
  of Physics}\ }\textbf {\bibinfo {volume} {199}},\ \bibinfo {pages} {18}
  (\bibinfo {year} {1990})}\BibitemShut {NoStop}%
\bibitem [{\citenamefont {Denicol}\ \emph {et~al.}(2008)\citenamefont
  {Denicol}, \citenamefont {Kodama}, \citenamefont {Koide},\ and\ \citenamefont
  {Mota}}]{denicol2008stability}%
  \BibitemOpen
  \bibfield  {author} {\bibinfo {author} {\bibfnamefont {G.~S.}\ \bibnamefont
  {Denicol}}, \bibinfo {author} {\bibfnamefont {T.}~\bibnamefont {Kodama}},
  \bibinfo {author} {\bibfnamefont {T.}~\bibnamefont {Koide}}, \ and\ \bibinfo
  {author} {\bibfnamefont {P.}~\bibnamefont {Mota}},\ }\href@noop {} {\bibfield
   {journal} {\bibinfo  {journal} {Journal of Physics G: Nuclear and particle
  physics}\ }\textbf {\bibinfo {volume} {35}},\ \bibinfo {pages} {115102}
  (\bibinfo {year} {2008})}\BibitemShut {NoStop}%
\bibitem [{\citenamefont {Pu}\ \emph {et~al.}(2010)\citenamefont {Pu},
  \citenamefont {Koide},\ and\ \citenamefont {Rischke}}]{Pu_2010}%
  \BibitemOpen
  \bibfield  {author} {\bibinfo {author} {\bibfnamefont {S.}~\bibnamefont
  {Pu}}, \bibinfo {author} {\bibfnamefont {T.}~\bibnamefont {Koide}}, \ and\
  \bibinfo {author} {\bibfnamefont {D.~H.}\ \bibnamefont {Rischke}},\ }\href
  {\doibase 10.1103/physrevd.81.114039} {\bibfield  {journal} {\bibinfo
  {journal} {Physical Review D}\ }\textbf {\bibinfo {volume} {81}} (\bibinfo
  {year} {2010}),\ 10.1103/physrevd.81.114039}\BibitemShut {NoStop}%
\bibitem [{\citenamefont {Israel}(1976)}]{Israel:1976tn}%
  \BibitemOpen
  \bibfield  {author} {\bibinfo {author} {\bibfnamefont {W.}~\bibnamefont
  {Israel}},\ }\href {\doibase 10.1016/0003-4916(76)90064-6} {\bibfield
  {journal} {\bibinfo  {journal} {Annals Phys.}\ }\textbf {\bibinfo {volume}
  {100}},\ \bibinfo {pages} {310} (\bibinfo {year} {1976})}\BibitemShut
  {NoStop}%
\bibitem [{\citenamefont {Israel}(1979)}]{israel1979jm}%
  \BibitemOpen
  \bibfield  {author} {\bibinfo {author} {\bibfnamefont {W.}~\bibnamefont
  {Israel}},\ }in\ \href@noop {} {\emph {\bibinfo {booktitle} {Roy. Soc. Lond.
  A}}},\ Vol.\ \bibinfo {volume} {365}\ (\bibinfo {year} {1979})\ p.~\bibinfo
  {pages} {43}\BibitemShut {NoStop}%
\bibitem [{\citenamefont {Brito}\ and\ \citenamefont
  {Denicol}(2020)}]{Brito_2022}%
  \BibitemOpen
  \bibfield  {author} {\bibinfo {author} {\bibfnamefont {C.~V.}\ \bibnamefont
  {Brito}}\ and\ \bibinfo {author} {\bibfnamefont {G.~S.}\ \bibnamefont
  {Denicol}},\ }\href {\doibase 10.1103/PhysRevD.102.116009} {\bibfield
  {journal} {\bibinfo  {journal} {Phys. Rev. D}\ }\textbf {\bibinfo {volume}
  {102}},\ \bibinfo {pages} {116009} (\bibinfo {year} {2020})}\BibitemShut
  {NoStop}%
\bibitem [{\citenamefont {Sammet}\ \emph {et~al.}(2023)\citenamefont {Sammet},
  \citenamefont {Mayer},\ and\ \citenamefont {Rischke}}]{Sammet:2023bfo}%
  \BibitemOpen
  \bibfield  {author} {\bibinfo {author} {\bibfnamefont {J.}~\bibnamefont
  {Sammet}}, \bibinfo {author} {\bibfnamefont {M.}~\bibnamefont {Mayer}}, \
  and\ \bibinfo {author} {\bibfnamefont {D.~H.}\ \bibnamefont {Rischke}},\
  }\href@noop {} {\  (\bibinfo {year} {2023})},\ \Eprint
  {http://arxiv.org/abs/2302.01070} {arXiv:2302.01070 [hep-th]} \BibitemShut
  {NoStop}%
\bibitem [{\citenamefont {Bemfica}\ \emph
  {et~al.}(2019{\natexlab{a}})\citenamefont {Bemfica}, \citenamefont
  {Disconzi},\ and\ \citenamefont {Noronha}}]{Bemfica:2019cop}%
  \BibitemOpen
  \bibfield  {author} {\bibinfo {author} {\bibfnamefont {F.~S.}\ \bibnamefont
  {Bemfica}}, \bibinfo {author} {\bibfnamefont {M.~M.}\ \bibnamefont
  {Disconzi}}, \ and\ \bibinfo {author} {\bibfnamefont {J.}~\bibnamefont
  {Noronha}},\ }\href {\doibase 10.1103/PhysRevLett.122.221602} {\bibfield
  {journal} {\bibinfo  {journal} {Phys. Rev. Lett.}\ }\textbf {\bibinfo
  {volume} {122}},\ \bibinfo {pages} {221602} (\bibinfo {year}
  {2019}{\natexlab{a}})},\ \Eprint {http://arxiv.org/abs/1901.06701}
  {arXiv:1901.06701 [gr-qc]} \BibitemShut {NoStop}%
\bibitem [{\citenamefont {Bemfica}\ \emph
  {et~al.}(2021{\natexlab{a}})\citenamefont {Bemfica}, \citenamefont
  {Disconzi}, \citenamefont {Hoang}, \citenamefont {Noronha},\ and\
  \citenamefont {Radosz}}]{Bemfica:2020xym}%
  \BibitemOpen
  \bibfield  {author} {\bibinfo {author} {\bibfnamefont {F.~S.}\ \bibnamefont
  {Bemfica}}, \bibinfo {author} {\bibfnamefont {M.~M.}\ \bibnamefont
  {Disconzi}}, \bibinfo {author} {\bibfnamefont {V.}~\bibnamefont {Hoang}},
  \bibinfo {author} {\bibfnamefont {J.}~\bibnamefont {Noronha}}, \ and\
  \bibinfo {author} {\bibfnamefont {M.}~\bibnamefont {Radosz}},\ }\href
  {\doibase 10.1103/PhysRevLett.126.222301} {\bibfield  {journal} {\bibinfo
  {journal} {Phys. Rev. Lett.}\ }\textbf {\bibinfo {volume} {126}},\ \bibinfo
  {pages} {222301} (\bibinfo {year} {2021}{\natexlab{a}})},\ \Eprint
  {http://arxiv.org/abs/2005.11632} {arXiv:2005.11632 [hep-th]} \BibitemShut
  {NoStop}%
\bibitem [{\citenamefont {Kovtun}(2019{\natexlab{a}})}]{kovtun:19first}%
  \BibitemOpen
  \bibfield  {author} {\bibinfo {author} {\bibfnamefont {P.}~\bibnamefont
  {Kovtun}},\ }\href@noop {} {\bibfield  {journal} {\bibinfo  {journal}
  {Journal of High Energy Physics}\ }\textbf {\bibinfo {volume} {2019}},\
  \bibinfo {pages} {34} (\bibinfo {year} {2019}{\natexlab{a}})}\BibitemShut
  {NoStop}%
\bibitem [{\citenamefont {Bemfica}\ \emph
  {et~al.}(2019{\natexlab{b}})\citenamefont {Bemfica}, \citenamefont
  {Disconzi},\ and\ \citenamefont {Noronha}}]{bemfica2019nonlinear}%
  \BibitemOpen
  \bibfield  {author} {\bibinfo {author} {\bibfnamefont {F.~S.}\ \bibnamefont
  {Bemfica}}, \bibinfo {author} {\bibfnamefont {M.~M.}\ \bibnamefont
  {Disconzi}}, \ and\ \bibinfo {author} {\bibfnamefont {J.}~\bibnamefont
  {Noronha}},\ }\href@noop {} {\bibfield  {journal} {\bibinfo  {journal}
  {Physical Review D}\ }\textbf {\bibinfo {volume} {100}},\ \bibinfo {pages}
  {104020} (\bibinfo {year} {2019}{\natexlab{b}})}\BibitemShut {NoStop}%
\bibitem [{\citenamefont {Hoult}\ and\ \citenamefont
  {Kovtun}(2020)}]{Hoult:2020eho}%
  \BibitemOpen
  \bibfield  {author} {\bibinfo {author} {\bibfnamefont {R.~E.}\ \bibnamefont
  {Hoult}}\ and\ \bibinfo {author} {\bibfnamefont {P.}~\bibnamefont {Kovtun}},\
  }\href {\doibase 10.1007/JHEP06(2020)067} {\bibfield  {journal} {\bibinfo
  {journal} {JHEP}\ }\textbf {\bibinfo {volume} {06}},\ \bibinfo {pages} {067}
  (\bibinfo {year} {2020})},\ \Eprint {http://arxiv.org/abs/2004.04102}
  {arXiv:2004.04102 [hep-th]} \BibitemShut {NoStop}%
\bibitem [{\citenamefont {Bemfica}\ \emph {et~al.}(2022)\citenamefont
  {Bemfica}, \citenamefont {Disconzi},\ and\ \citenamefont
  {Noronha}}]{Bemfica:2020zjp}%
  \BibitemOpen
  \bibfield  {author} {\bibinfo {author} {\bibfnamefont {F.~S.}\ \bibnamefont
  {Bemfica}}, \bibinfo {author} {\bibfnamefont {M.~M.}\ \bibnamefont
  {Disconzi}}, \ and\ \bibinfo {author} {\bibfnamefont {J.}~\bibnamefont
  {Noronha}},\ }\href {\doibase 10.1103/PhysRevX.12.021044} {\bibfield
  {journal} {\bibinfo  {journal} {Phys. Rev. X}\ }\textbf {\bibinfo {volume}
  {12}},\ \bibinfo {pages} {021044} (\bibinfo {year} {2022})},\ \Eprint
  {http://arxiv.org/abs/2009.11388} {arXiv:2009.11388 [gr-qc]} \BibitemShut
  {NoStop}%
\bibitem [{\citenamefont {Bemfica}\ \emph
  {et~al.}(2021{\natexlab{b}})\citenamefont {Bemfica}, \citenamefont
  {Disconzi}, \citenamefont {Rodriguez},\ and\ \citenamefont
  {Shao}}]{DisconziBemficaRodriguezShaoSobolevConformal}%
  \BibitemOpen
  \bibfield  {author} {\bibinfo {author} {\bibfnamefont {F.~S.}\ \bibnamefont
  {Bemfica}}, \bibinfo {author} {\bibfnamefont {M.~M.}\ \bibnamefont
  {Disconzi}}, \bibinfo {author} {\bibfnamefont {C.}~\bibnamefont {Rodriguez}},
  \ and\ \bibinfo {author} {\bibfnamefont {Y.}~\bibnamefont {Shao}},\ }\href
  {\doibase 10.3934/cpaa.2021069} {\bibfield  {journal} {\bibinfo  {journal}
  {Commun. Pure Appl. Anal.}\ }\textbf {\bibinfo {volume} {20}},\ \bibinfo
  {pages} {2279} (\bibinfo {year} {2021}{\natexlab{b}})}\BibitemShut {NoStop}%
\bibitem [{\citenamefont {Bemfica}\ \emph
  {et~al.}(2021{\natexlab{c}})\citenamefont {Bemfica}, \citenamefont
  {Disconzi},\ and\ \citenamefont {Graber}}]{DisconziBemficaGraber}%
  \BibitemOpen
  \bibfield  {author} {\bibinfo {author} {\bibfnamefont {F.~S.}\ \bibnamefont
  {Bemfica}}, \bibinfo {author} {\bibfnamefont {M.~M.}\ \bibnamefont
  {Disconzi}}, \ and\ \bibinfo {author} {\bibfnamefont {P.~J.}\ \bibnamefont
  {Graber}},\ }\href {\doibase 10.3934/cpaa.2021068} {\bibfield  {journal}
  {\bibinfo  {journal} {Commun. Pure Appl. Anal.}\ }\textbf {\bibinfo {volume}
  {20}},\ \bibinfo {pages} {2885} (\bibinfo {year}
  {2021}{\natexlab{c}})}\BibitemShut {NoStop}%
\bibitem [{\citenamefont {Denicol}\ and\ \citenamefont
  {Rischke}(2021)}]{Denicol:2021}%
  \BibitemOpen
  \bibfield  {author} {\bibinfo {author} {\bibfnamefont {G.}~\bibnamefont
  {Denicol}}\ and\ \bibinfo {author} {\bibfnamefont {D.~H.}\ \bibnamefont
  {Rischke}},\ }\href@noop {} {\emph {\bibinfo {title} {Microscopic Foundations
  of Relativistic Fluid Dynamics}}}\ (\bibinfo  {publisher} {Springer},\
  \bibinfo {year} {2021})\BibitemShut {NoStop}%
\bibitem [{\citenamefont {Calzetta}\ and\ \citenamefont
  {Hu}(1988)}]{Calzetta:1986cq}%
  \BibitemOpen
  \bibfield  {author} {\bibinfo {author} {\bibfnamefont {E.}~\bibnamefont
  {Calzetta}}\ and\ \bibinfo {author} {\bibfnamefont {B.~L.}\ \bibnamefont
  {Hu}},\ }\href {\doibase 10.1103/PhysRevD.37.2878} {\bibfield  {journal}
  {\bibinfo  {journal} {Phys. Rev. D}\ }\textbf {\bibinfo {volume} {37}},\
  \bibinfo {pages} {2878} (\bibinfo {year} {1988})}\BibitemShut {NoStop}%
\bibitem [{\citenamefont {Peskin}\ and\ \citenamefont
  {Schroeder}(1995)}]{Peskin:1995ev}%
  \BibitemOpen
  \bibfield  {author} {\bibinfo {author} {\bibfnamefont {M.~E.}\ \bibnamefont
  {Peskin}}\ and\ \bibinfo {author} {\bibfnamefont {D.~V.}\ \bibnamefont
  {Schroeder}},\ }\href@noop {} {\emph {\bibinfo {title} {{An Introduction to
  quantum field theory}}}}\ (\bibinfo  {publisher} {Addison-Wesley},\ \bibinfo
  {address} {Reading, USA},\ \bibinfo {year} {1995})\BibitemShut {NoStop}%
\bibitem [{\citenamefont {Kapusta}\ and\ \citenamefont
  {Gale}(2011)}]{Kapusta:2006pm}%
  \BibitemOpen
  \bibfield  {author} {\bibinfo {author} {\bibfnamefont {J.~I.}\ \bibnamefont
  {Kapusta}}\ and\ \bibinfo {author} {\bibfnamefont {C.}~\bibnamefont {Gale}},\
  }\href {\doibase 10.1017/CBO9780511535130} {\emph {\bibinfo {title}
  {{Finite-temperature field theory: Principles and applications}}}},\
  Cambridge Monographs on Mathematical Physics\ (\bibinfo  {publisher}
  {Cambridge University Press},\ \bibinfo {year} {2011})\BibitemShut {NoStop}%
\bibitem [{\citenamefont {Denicol}\ and\ \citenamefont
  {Noronha}(2022)}]{Denicol:2022bsq}%
  \BibitemOpen
  \bibfield  {author} {\bibinfo {author} {\bibfnamefont {G.~S.}\ \bibnamefont
  {Denicol}}\ and\ \bibinfo {author} {\bibfnamefont {J.}~\bibnamefont
  {Noronha}},\ }\href@noop {} {\  (\bibinfo {year} {2022})},\ \Eprint
  {http://arxiv.org/abs/2209.10370} {arXiv:2209.10370 [nucl-th]} \BibitemShut
  {NoStop}%
\bibitem [{\citenamefont {Bjorken}(1983)}]{Bjorken:1982qr}%
  \BibitemOpen
  \bibfield  {author} {\bibinfo {author} {\bibfnamefont {J.~D.}\ \bibnamefont
  {Bjorken}},\ }\href {\doibase 10.1103/PhysRevD.27.140} {\bibfield  {journal}
  {\bibinfo  {journal} {Phys. Rev. D}\ }\textbf {\bibinfo {volume} {27}},\
  \bibinfo {pages} {140} (\bibinfo {year} {1983})}\BibitemShut {NoStop}%
\bibitem [{\citenamefont {Israel}\ and\ \citenamefont
  {Stewart}(1979)}]{israel1979annals}%
  \BibitemOpen
  \bibfield  {author} {\bibinfo {author} {\bibfnamefont {W.}~\bibnamefont
  {Israel}}\ and\ \bibinfo {author} {\bibfnamefont {J.}~\bibnamefont
  {Stewart}},\ }\href@noop {} {\bibfield  {journal} {\bibinfo  {journal}
  {Annals Phys}\ }\textbf {\bibinfo {volume} {118}},\ \bibinfo {pages} {228}
  (\bibinfo {year} {1979})}\BibitemShut {NoStop}%
\bibitem [{\citenamefont {Rocha}\ and\ \citenamefont
  {Denicol}(2021)}]{Rocha:2021lze}%
  \BibitemOpen
  \bibfield  {author} {\bibinfo {author} {\bibfnamefont {G.~S.}\ \bibnamefont
  {Rocha}}\ and\ \bibinfo {author} {\bibfnamefont {G.~S.}\ \bibnamefont
  {Denicol}},\ }\href {\doibase 10.1103/PhysRevD.104.096016} {\bibfield
  {journal} {\bibinfo  {journal} {Phys. Rev. D}\ }\textbf {\bibinfo {volume}
  {104}},\ \bibinfo {pages} {096016} (\bibinfo {year} {2021})},\ \Eprint
  {http://arxiv.org/abs/2108.02187} {arXiv:2108.02187 [nucl-th]} \BibitemShut
  {NoStop}%
\bibitem [{\citenamefont {Rocha}\ \emph {et~al.}(2022)\citenamefont {Rocha},
  \citenamefont {Denicol},\ and\ \citenamefont {Noronha}}]{Rocha:2022ind}%
  \BibitemOpen
  \bibfield  {author} {\bibinfo {author} {\bibfnamefont {G.~S.}\ \bibnamefont
  {Rocha}}, \bibinfo {author} {\bibfnamefont {G.~S.}\ \bibnamefont {Denicol}},
  \ and\ \bibinfo {author} {\bibfnamefont {J.}~\bibnamefont {Noronha}},\ }\href
  {\doibase 10.1103/PhysRevD.106.036010} {\bibfield  {journal} {\bibinfo
  {journal} {Phys. Rev. D}\ }\textbf {\bibinfo {volume} {106}},\ \bibinfo
  {pages} {036010} (\bibinfo {year} {2022})},\ \Eprint
  {http://arxiv.org/abs/2205.00078} {arXiv:2205.00078 [nucl-th]} \BibitemShut
  {NoStop}%
\bibitem [{\citenamefont {Chapman}(1916)}]{chapman1916vi}%
  \BibitemOpen
  \bibfield  {author} {\bibinfo {author} {\bibfnamefont {S.}~\bibnamefont
  {Chapman}},\ }\href@noop {} {\bibfield  {journal} {\bibinfo  {journal}
  {Philosophical Transactions of the Royal Society of London. Series A,
  Containing Papers of a Mathematical or Physical Character}\ }\textbf
  {\bibinfo {volume} {216}},\ \bibinfo {pages} {279} (\bibinfo {year}
  {1916})}\BibitemShut {NoStop}%
\bibitem [{\citenamefont {Enskog}(1917)}]{enskog1917kinetische}%
  \BibitemOpen
  \bibfield  {author} {\bibinfo {author} {\bibfnamefont {D.}~\bibnamefont
  {Enskog}},\ }\href@noop {} {\enquote {\bibinfo {title} {Kinetische theorie
  der {V}org{\"a}nge in m{\"a}ssig verd{\"u}nnten gasen. {I}. {A}llgemeiner
  teil},}\ } (\bibinfo {year} {1917})\BibitemShut {NoStop}%
\bibitem [{\citenamefont {de~Groot}\ \emph {et~al.}(1980)\citenamefont
  {de~Groot}, \citenamefont {van Leeuwen},\ and\ \citenamefont {van
  Weert}}]{deGroot:80relativistic}%
  \BibitemOpen
  \bibfield  {author} {\bibinfo {author} {\bibfnamefont {S.}~\bibnamefont
  {de~Groot}}, \bibinfo {author} {\bibfnamefont {W.}~\bibnamefont {van
  Leeuwen}}, \ and\ \bibinfo {author} {\bibfnamefont {C.}~\bibnamefont {van
  Weert}},\ }\href@noop {} {\emph {\bibinfo {title} {Relativistic Kinetic
  Theory: Principles and Applications}}}\ (\bibinfo  {publisher} {North-Holland
  Publishing Co.},\ \bibinfo {year} {1980})\BibitemShut {NoStop}%
\bibitem [{\citenamefont {Cercignani}\ and\ \citenamefont
  {Kremer}(2002)}]{cercignani:02relativistic}%
  \BibitemOpen
  \bibfield  {author} {\bibinfo {author} {\bibfnamefont {C.}~\bibnamefont
  {Cercignani}}\ and\ \bibinfo {author} {\bibfnamefont {G.~M.}\ \bibnamefont
  {Kremer}},\ }\href@noop {} {\emph {\bibinfo {title} {The Relativistic
  {B}oltzmann Equation: Theory and Applications}}}\ (\bibinfo  {publisher}
  {Springer},\ \bibinfo {year} {2002})\BibitemShut {NoStop}%
\bibitem [{\citenamefont {Gradshteyn}\ and\ \citenamefont
  {Ryzhik}(2014)}]{gradshteyn2014table}%
  \BibitemOpen
  \bibfield  {author} {\bibinfo {author} {\bibfnamefont {I.~S.}\ \bibnamefont
  {Gradshteyn}}\ and\ \bibinfo {author} {\bibfnamefont {I.~M.}\ \bibnamefont
  {Ryzhik}},\ }\href@noop {} {\emph {\bibinfo {title} {Table of integrals,
  series, and products}}}\ (\bibinfo  {publisher} {Academic press},\ \bibinfo
  {year} {2014})\BibitemShut {NoStop}%
\bibitem [{\citenamefont {Gavassino}(2022)}]{Gavassino:2021owo}%
  \BibitemOpen
  \bibfield  {author} {\bibinfo {author} {\bibfnamefont {L.}~\bibnamefont
  {Gavassino}},\ }\href {\doibase 10.1103/PhysRevX.12.041001} {\bibfield
  {journal} {\bibinfo  {journal} {Phys. Rev. X}\ }\textbf {\bibinfo {volume}
  {12}},\ \bibinfo {pages} {041001} (\bibinfo {year} {2022})},\ \Eprint
  {http://arxiv.org/abs/2111.05254} {arXiv:2111.05254 [gr-qc]} \BibitemShut
  {NoStop}%
\bibitem [{\citenamefont {Bemfica}\ \emph
  {et~al.}(2018{\natexlab{b}})\citenamefont {Bemfica}, \citenamefont
  {Disconzi},\ and\ \citenamefont {Noronha}}]{Bemfica:2017wps}%
  \BibitemOpen
  \bibfield  {author} {\bibinfo {author} {\bibfnamefont {F.~S.}\ \bibnamefont
  {Bemfica}}, \bibinfo {author} {\bibfnamefont {M.~M.}\ \bibnamefont
  {Disconzi}}, \ and\ \bibinfo {author} {\bibfnamefont {J.}~\bibnamefont
  {Noronha}},\ }\href {\doibase 10.1103/PhysRevD.98.104064} {\bibfield
  {journal} {\bibinfo  {journal} {Phys. Rev. D}\ }\textbf {\bibinfo {volume}
  {98}},\ \bibinfo {pages} {104064} (\bibinfo {year} {2018}{\natexlab{b}})},\
  \Eprint {http://arxiv.org/abs/1708.06255} {arXiv:1708.06255 [gr-qc]}
  \BibitemShut {NoStop}%
\bibitem [{\citenamefont {Bemfica}\ \emph
  {et~al.}(2019{\natexlab{c}})\citenamefont {Bemfica}, \citenamefont {Bemfica},
  \citenamefont {Disconzi}, \citenamefont {Disconzi}, \citenamefont {Noronha},\
  and\ \citenamefont {Noronha}}]{Bemfica:2019knx}%
  \BibitemOpen
  \bibfield  {author} {\bibinfo {author} {\bibfnamefont {F.~S.}\ \bibnamefont
  {Bemfica}}, \bibinfo {author} {\bibfnamefont {F.~S.}\ \bibnamefont
  {Bemfica}}, \bibinfo {author} {\bibfnamefont {M.~M.}\ \bibnamefont
  {Disconzi}}, \bibinfo {author} {\bibfnamefont {M.~M.}\ \bibnamefont
  {Disconzi}}, \bibinfo {author} {\bibfnamefont {J.}~\bibnamefont {Noronha}}, \
  and\ \bibinfo {author} {\bibfnamefont {J.}~\bibnamefont {Noronha}},\ }\href
  {\doibase 10.1103/PhysRevD.100.104020} {\bibfield  {journal} {\bibinfo
  {journal} {Phys. Rev. D}\ }\textbf {\bibinfo {volume} {100}},\ \bibinfo
  {pages} {104020} (\bibinfo {year} {2019}{\natexlab{c}})},\ \bibinfo {note}
  {[Erratum: Phys.Rev.D 105, 069902 (2022)]},\ \Eprint
  {http://arxiv.org/abs/1907.12695} {arXiv:1907.12695 [gr-qc]} \BibitemShut
  {NoStop}%
\bibitem [{\citenamefont {Kovtun}(2019{\natexlab{b}})}]{Kovtun:2019hdm}%
  \BibitemOpen
  \bibfield  {author} {\bibinfo {author} {\bibfnamefont {P.}~\bibnamefont
  {Kovtun}},\ }\href {\doibase 10.1007/JHEP10(2019)034} {\bibfield  {journal}
  {\bibinfo  {journal} {JHEP}\ }\textbf {\bibinfo {volume} {10}},\ \bibinfo
  {pages} {034} (\bibinfo {year} {2019}{\natexlab{b}})},\ \Eprint
  {http://arxiv.org/abs/1907.08191} {arXiv:1907.08191 [hep-th]} \BibitemShut
  {NoStop}%
\bibitem [{\citenamefont {Denicol}\ \emph {et~al.}(2012)\citenamefont
  {Denicol}, \citenamefont {Niemi}, \citenamefont {Molnar},\ and\ \citenamefont
  {Rischke}}]{Denicol:2012cn}%
  \BibitemOpen
  \bibfield  {author} {\bibinfo {author} {\bibfnamefont {G.~S.}\ \bibnamefont
  {Denicol}}, \bibinfo {author} {\bibfnamefont {H.}~\bibnamefont {Niemi}},
  \bibinfo {author} {\bibfnamefont {E.}~\bibnamefont {Molnar}}, \ and\ \bibinfo
  {author} {\bibfnamefont {D.~H.}\ \bibnamefont {Rischke}},\ }\href {\doibase
  10.1103/PhysRevD.85.114047} {\bibfield  {journal} {\bibinfo  {journal} {Phys.
  Rev. D}\ }\textbf {\bibinfo {volume} {85}},\ \bibinfo {pages} {114047}
  (\bibinfo {year} {2012})},\ \bibinfo {note} {[Erratum: Phys.Rev.D 91, 039902
  (2015)]},\ \Eprint {http://arxiv.org/abs/1202.4551} {arXiv:1202.4551
  [nucl-th]} \BibitemShut {NoStop}%
\bibitem [{\citenamefont {Brito}\ and\ \citenamefont
  {Denicol}(2022)}]{Brito:2021iqr}%
  \BibitemOpen
  \bibfield  {author} {\bibinfo {author} {\bibfnamefont {C.~V.}\ \bibnamefont
  {Brito}}\ and\ \bibinfo {author} {\bibfnamefont {G.~S.}\ \bibnamefont
  {Denicol}},\ }\href {\doibase 10.1103/PhysRevD.105.096026} {\bibfield
  {journal} {\bibinfo  {journal} {Phys. Rev. D}\ }\textbf {\bibinfo {volume}
  {105}},\ \bibinfo {pages} {096026} (\bibinfo {year} {2022})},\ \Eprint
  {http://arxiv.org/abs/2107.10319} {arXiv:2107.10319 [nucl-th]} \BibitemShut
  {NoStop}%
\bibitem [{\citenamefont {de~Brito}\ and\ \citenamefont
  {Denicol}(2023)}]{deBrito:2023tgb}%
  \BibitemOpen
  \bibfield  {author} {\bibinfo {author} {\bibfnamefont {C.~V.~P.}\
  \bibnamefont {de~Brito}}\ and\ \bibinfo {author} {\bibfnamefont {G.~S.}\
  \bibnamefont {Denicol}},\ }\href@noop {} {\  (\bibinfo {year} {2023})},\
  \Eprint {http://arxiv.org/abs/2302.09097} {arXiv:2302.09097 [nucl-th]}
  \BibitemShut {NoStop}%
\bibitem [{\citenamefont {Bhatnagar}\ \emph {et~al.}(1954)\citenamefont
  {Bhatnagar}, \citenamefont {Gross},\ and\ \citenamefont
  {Krook}}]{bhatnagar:54model}%
  \BibitemOpen
  \bibfield  {author} {\bibinfo {author} {\bibfnamefont {P.~L.}\ \bibnamefont
  {Bhatnagar}}, \bibinfo {author} {\bibfnamefont {E.~P.}\ \bibnamefont
  {Gross}}, \ and\ \bibinfo {author} {\bibfnamefont {M.}~\bibnamefont
  {Krook}},\ }\href@noop {} {\bibfield  {journal} {\bibinfo  {journal}
  {Physical review}\ }\textbf {\bibinfo {volume} {94}},\ \bibinfo {pages} {511}
  (\bibinfo {year} {1954})}\BibitemShut {NoStop}%
\bibitem [{\citenamefont {Marle}(1969{\natexlab{a}})}]{marle:69etab}%
  \BibitemOpen
  \bibfield  {author} {\bibinfo {author} {\bibfnamefont {C.}~\bibnamefont
  {Marle}},\ }in\ \href@noop {} {\emph {\bibinfo {booktitle} {Annales de l'IHP
  Physique th{\'e}orique}}},\ Vol.~\bibinfo {volume} {10}\ (\bibinfo {year}
  {1969})\ pp.\ \bibinfo {pages} {67--126}\BibitemShut {NoStop}%
\bibitem [{\citenamefont
  {Marle}(1969{\natexlab{b}})}]{marle:69-2-etablissement}%
  \BibitemOpen
  \bibfield  {author} {\bibinfo {author} {\bibfnamefont {C.}~\bibnamefont
  {Marle}},\ }in\ \href@noop {} {\emph {\bibinfo {booktitle} {Annales de l'IHP
  Physique th{\'e}orique}}},\ Vol.~\bibinfo {volume} {10}\ (\bibinfo {year}
  {1969})\ pp.\ \bibinfo {pages} {127--194}\BibitemShut {NoStop}%
\bibitem [{\citenamefont {Anderson}\ and\ \citenamefont
  {Witting}(1974)}]{andersonRTA:74}%
  \BibitemOpen
  \bibfield  {author} {\bibinfo {author} {\bibfnamefont {J.~L.}\ \bibnamefont
  {Anderson}}\ and\ \bibinfo {author} {\bibfnamefont {H.}~\bibnamefont
  {Witting}},\ }\href@noop {} {\bibfield  {journal} {\bibinfo  {journal}
  {Physica}\ }\textbf {\bibinfo {volume} {74}},\ \bibinfo {pages} {466}
  (\bibinfo {year} {1974})}\BibitemShut {NoStop}%
\bibitem [{\citenamefont {Rocha}\ \emph {et~al.}(2021)\citenamefont {Rocha},
  \citenamefont {Denicol},\ and\ \citenamefont {Noronha}}]{Rocha:2021zcw}%
  \BibitemOpen
  \bibfield  {author} {\bibinfo {author} {\bibfnamefont {G.~S.}\ \bibnamefont
  {Rocha}}, \bibinfo {author} {\bibfnamefont {G.~S.}\ \bibnamefont {Denicol}},
  \ and\ \bibinfo {author} {\bibfnamefont {J.}~\bibnamefont {Noronha}},\ }\href
  {\doibase 10.1103/PhysRevLett.127.042301} {\bibfield  {journal} {\bibinfo
  {journal} {Phys. Rev. Lett.}\ }\textbf {\bibinfo {volume} {127}},\ \bibinfo
  {pages} {042301} (\bibinfo {year} {2021})},\ \Eprint
  {http://arxiv.org/abs/2103.07489} {arXiv:2103.07489 [nucl-th]} \BibitemShut
  {NoStop}%
\bibitem [{\citenamefont {Hu}\ and\ \citenamefont {Xu}(2023)}]{Hu:2022mvl}%
  \BibitemOpen
  \bibfield  {author} {\bibinfo {author} {\bibfnamefont {J.}~\bibnamefont
  {Hu}}\ and\ \bibinfo {author} {\bibfnamefont {Z.}~\bibnamefont {Xu}},\ }\href
  {\doibase 10.1103/PhysRevD.107.016010} {\bibfield  {journal} {\bibinfo
  {journal} {Phys. Rev. D}\ }\textbf {\bibinfo {volume} {107}},\ \bibinfo
  {pages} {016010} (\bibinfo {year} {2023})},\ \Eprint
  {http://arxiv.org/abs/2205.15755} {arXiv:2205.15755 [hep-ph]} \BibitemShut
  {NoStop}%
\bibitem [{\citenamefont {Bazow}\ \emph
  {et~al.}(2016{\natexlab{a}})\citenamefont {Bazow}, \citenamefont {Denicol},
  \citenamefont {Heinz}, \citenamefont {Martinez},\ and\ \citenamefont
  {Noronha}}]{Bazow:2015dha}%
  \BibitemOpen
  \bibfield  {author} {\bibinfo {author} {\bibfnamefont {D.}~\bibnamefont
  {Bazow}}, \bibinfo {author} {\bibfnamefont {G.~S.}\ \bibnamefont {Denicol}},
  \bibinfo {author} {\bibfnamefont {U.}~\bibnamefont {Heinz}}, \bibinfo
  {author} {\bibfnamefont {M.}~\bibnamefont {Martinez}}, \ and\ \bibinfo
  {author} {\bibfnamefont {J.}~\bibnamefont {Noronha}},\ }\href {\doibase
  10.1103/PhysRevLett.116.022301} {\bibfield  {journal} {\bibinfo  {journal}
  {Phys. Rev. Lett.}\ }\textbf {\bibinfo {volume} {116}},\ \bibinfo {pages}
  {022301} (\bibinfo {year} {2016}{\natexlab{a}})},\ \Eprint
  {http://arxiv.org/abs/1507.07834} {arXiv:1507.07834 [hep-ph]} \BibitemShut
  {NoStop}%
\bibitem [{\citenamefont {Bazow}\ \emph
  {et~al.}(2016{\natexlab{b}})\citenamefont {Bazow}, \citenamefont {Denicol},
  \citenamefont {Heinz}, \citenamefont {Martinez},\ and\ \citenamefont
  {Noronha}}]{Bazow:2016oky}%
  \BibitemOpen
  \bibfield  {author} {\bibinfo {author} {\bibfnamefont {D.}~\bibnamefont
  {Bazow}}, \bibinfo {author} {\bibfnamefont {G.~S.}\ \bibnamefont {Denicol}},
  \bibinfo {author} {\bibfnamefont {U.}~\bibnamefont {Heinz}}, \bibinfo
  {author} {\bibfnamefont {M.}~\bibnamefont {Martinez}}, \ and\ \bibinfo
  {author} {\bibfnamefont {J.}~\bibnamefont {Noronha}},\ }\href {\doibase
  10.1103/PhysRevD.94.125006} {\bibfield  {journal} {\bibinfo  {journal} {Phys.
  Rev. D}\ }\textbf {\bibinfo {volume} {94}},\ \bibinfo {pages} {125006}
  (\bibinfo {year} {2016}{\natexlab{b}})},\ \Eprint
  {http://arxiv.org/abs/1607.05245} {arXiv:1607.05245 [hep-ph]} \BibitemShut
  {NoStop}%
\bibitem [{\citenamefont {Mullins}\ \emph {et~al.}(2022)\citenamefont
  {Mullins}, \citenamefont {Denicol},\ and\ \citenamefont
  {Noronha}}]{Mullins:2022fbx}%
  \BibitemOpen
  \bibfield  {author} {\bibinfo {author} {\bibfnamefont {N.}~\bibnamefont
  {Mullins}}, \bibinfo {author} {\bibfnamefont {G.~S.}\ \bibnamefont
  {Denicol}}, \ and\ \bibinfo {author} {\bibfnamefont {J.}~\bibnamefont
  {Noronha}},\ }\href {\doibase 10.1103/PhysRevD.106.056024} {\bibfield
  {journal} {\bibinfo  {journal} {Phys. Rev. D}\ }\textbf {\bibinfo {volume}
  {106}},\ \bibinfo {pages} {056024} (\bibinfo {year} {2022})},\ \Eprint
  {http://arxiv.org/abs/2207.07786} {arXiv:2207.07786 [hep-ph]} \BibitemShut
  {NoStop}%
\bibitem [{\citenamefont {Struchtrup}(2004)}]{struchtrup2004stable}%
  \BibitemOpen
  \bibfield  {author} {\bibinfo {author} {\bibfnamefont {H.}~\bibnamefont
  {Struchtrup}},\ }\href@noop {} {\bibfield  {journal} {\bibinfo  {journal}
  {Physics of Fluids}\ }\textbf {\bibinfo {volume} {16}},\ \bibinfo {pages}
  {3921} (\bibinfo {year} {2004})}\BibitemShut {NoStop}%
\bibitem [{\citenamefont {Fotakis}\ \emph {et~al.}(2022)\citenamefont
  {Fotakis}, \citenamefont {Moln\'ar}, \citenamefont {Niemi}, \citenamefont
  {Greiner},\ and\ \citenamefont {Rischke}}]{Fotakis:2022usk}%
  \BibitemOpen
  \bibfield  {author} {\bibinfo {author} {\bibfnamefont {J.~A.}\ \bibnamefont
  {Fotakis}}, \bibinfo {author} {\bibfnamefont {E.}~\bibnamefont {Moln\'ar}},
  \bibinfo {author} {\bibfnamefont {H.}~\bibnamefont {Niemi}}, \bibinfo
  {author} {\bibfnamefont {C.}~\bibnamefont {Greiner}}, \ and\ \bibinfo
  {author} {\bibfnamefont {D.~H.}\ \bibnamefont {Rischke}},\ }\href {\doibase
  10.1103/PhysRevD.106.036009} {\bibfield  {journal} {\bibinfo  {journal}
  {Phys. Rev. D}\ }\textbf {\bibinfo {volume} {106}},\ \bibinfo {pages}
  {036009} (\bibinfo {year} {2022})},\ \Eprint
  {http://arxiv.org/abs/2203.11549} {arXiv:2203.11549 [nucl-th]} \BibitemShut
  {NoStop}%
\bibitem [{\citenamefont {Denicol}\ \emph {et~al.}(2010)\citenamefont
  {Denicol}, \citenamefont {Koide},\ and\ \citenamefont
  {Rischke}}]{Denicol:2010xn}%
  \BibitemOpen
  \bibfield  {author} {\bibinfo {author} {\bibfnamefont {G.~S.}\ \bibnamefont
  {Denicol}}, \bibinfo {author} {\bibfnamefont {T.}~\bibnamefont {Koide}}, \
  and\ \bibinfo {author} {\bibfnamefont {D.~H.}\ \bibnamefont {Rischke}},\
  }\href {\doibase 10.1103/PhysRevLett.105.162501} {\bibfield  {journal}
  {\bibinfo  {journal} {Phys. Rev. Lett.}\ }\textbf {\bibinfo {volume} {105}},\
  \bibinfo {pages} {162501} (\bibinfo {year} {2010})},\ \Eprint
  {http://arxiv.org/abs/1004.5013} {arXiv:1004.5013 [nucl-th]} \BibitemShut
  {NoStop}%
\bibitem [{\citenamefont {Denicol}\ \emph {et~al.}(2014)\citenamefont
  {Denicol}, \citenamefont {Jeon},\ and\ \citenamefont
  {Gale}}]{Denicol:2014vaa}%
  \BibitemOpen
  \bibfield  {author} {\bibinfo {author} {\bibfnamefont {G.~S.}\ \bibnamefont
  {Denicol}}, \bibinfo {author} {\bibfnamefont {S.}~\bibnamefont {Jeon}}, \
  and\ \bibinfo {author} {\bibfnamefont {C.}~\bibnamefont {Gale}},\ }\href
  {\doibase 10.1103/PhysRevC.90.024912} {\bibfield  {journal} {\bibinfo
  {journal} {Phys. Rev. C}\ }\textbf {\bibinfo {volume} {90}},\ \bibinfo
  {pages} {024912} (\bibinfo {year} {2014})},\ \Eprint
  {http://arxiv.org/abs/1403.0962} {arXiv:1403.0962 [nucl-th]} \BibitemShut
  {NoStop}%
\bibitem [{\citenamefont {Muronga}(2004)}]{Muronga:2003ta}%
  \BibitemOpen
  \bibfield  {author} {\bibinfo {author} {\bibfnamefont {A.}~\bibnamefont
  {Muronga}},\ }\href {\doibase 10.1103/PhysRevC.69.034903} {\bibfield
  {journal} {\bibinfo  {journal} {Phys. Rev. C}\ }\textbf {\bibinfo {volume}
  {69}},\ \bibinfo {pages} {034903} (\bibinfo {year} {2004})},\ \Eprint
  {http://arxiv.org/abs/nucl-th/0309055} {arXiv:nucl-th/0309055} \BibitemShut
  {NoStop}%
\bibitem [{\citenamefont {Denicol}\ and\ \citenamefont
  {Noronha}(2019)}]{Denicol:2018pak}%
  \BibitemOpen
  \bibfield  {author} {\bibinfo {author} {\bibfnamefont {G.~S.}\ \bibnamefont
  {Denicol}}\ and\ \bibinfo {author} {\bibfnamefont {J.}~\bibnamefont
  {Noronha}},\ }\href {\doibase 10.1103/PhysRevD.99.116004} {\bibfield
  {journal} {\bibinfo  {journal} {Phys. Rev. D}\ }\textbf {\bibinfo {volume}
  {99}},\ \bibinfo {pages} {116004} (\bibinfo {year} {2019})},\ \Eprint
  {http://arxiv.org/abs/1804.04771} {arXiv:1804.04771 [nucl-th]} \BibitemShut
  {NoStop}%
\bibitem [{\citenamefont {Liddle}\ \emph {et~al.}(1994)\citenamefont {Liddle},
  \citenamefont {Parsons},\ and\ \citenamefont {Barrow}}]{Liddle_1994}%
  \BibitemOpen
  \bibfield  {author} {\bibinfo {author} {\bibfnamefont {A.~R.}\ \bibnamefont
  {Liddle}}, \bibinfo {author} {\bibfnamefont {P.}~\bibnamefont {Parsons}}, \
  and\ \bibinfo {author} {\bibfnamefont {J.~D.}\ \bibnamefont {Barrow}},\
  }\href {\doibase 10.1103/PhysRevD.50.7222} {\bibfield  {journal} {\bibinfo
  {journal} {Phys. Rev. D}\ }\textbf {\bibinfo {volume} {50}},\ \bibinfo
  {pages} {7222} (\bibinfo {year} {1994})}\BibitemShut {NoStop}%
\bibitem [{\citenamefont {Heller}\ and\ \citenamefont
  {Spalinski}(2015)}]{Heller:2015dha}%
  \BibitemOpen
  \bibfield  {author} {\bibinfo {author} {\bibfnamefont {M.~P.}\ \bibnamefont
  {Heller}}\ and\ \bibinfo {author} {\bibfnamefont {M.}~\bibnamefont
  {Spalinski}},\ }\href {\doibase 10.1103/PhysRevLett.115.072501} {\bibfield
  {journal} {\bibinfo  {journal} {Phys. Rev. Lett.}\ }\textbf {\bibinfo
  {volume} {115}},\ \bibinfo {pages} {072501} (\bibinfo {year} {2015})},\
  \Eprint {http://arxiv.org/abs/1503.07514} {arXiv:1503.07514 [hep-th]}
  \BibitemShut {NoStop}%
\bibitem [{\citenamefont {Denicol}\ and\ \citenamefont
  {Noronha}(2018)}]{Denicol:2017lxn}%
  \BibitemOpen
  \bibfield  {author} {\bibinfo {author} {\bibfnamefont {G.~S.}\ \bibnamefont
  {Denicol}}\ and\ \bibinfo {author} {\bibfnamefont {J.}~\bibnamefont
  {Noronha}},\ }\href {\doibase 10.1103/PhysRevD.97.056021} {\bibfield
  {journal} {\bibinfo  {journal} {Phys. Rev. D}\ }\textbf {\bibinfo {volume}
  {97}},\ \bibinfo {pages} {056021} (\bibinfo {year} {2018})},\ \Eprint
  {http://arxiv.org/abs/1711.01657} {arXiv:1711.01657 [nucl-th]} \BibitemShut
  {NoStop}%
\bibitem [{\citenamefont {Marrochio}\ \emph {et~al.}(2015)\citenamefont
  {Marrochio}, \citenamefont {Noronha}, \citenamefont {Denicol}, \citenamefont
  {Luzum}, \citenamefont {Jeon},\ and\ \citenamefont
  {Gale}}]{Marrochio:2013wla}%
  \BibitemOpen
  \bibfield  {author} {\bibinfo {author} {\bibfnamefont {H.}~\bibnamefont
  {Marrochio}}, \bibinfo {author} {\bibfnamefont {J.}~\bibnamefont {Noronha}},
  \bibinfo {author} {\bibfnamefont {G.~S.}\ \bibnamefont {Denicol}}, \bibinfo
  {author} {\bibfnamefont {M.}~\bibnamefont {Luzum}}, \bibinfo {author}
  {\bibfnamefont {S.}~\bibnamefont {Jeon}}, \ and\ \bibinfo {author}
  {\bibfnamefont {C.}~\bibnamefont {Gale}},\ }\href {\doibase
  10.1103/PhysRevC.91.014903} {\bibfield  {journal} {\bibinfo  {journal} {Phys.
  Rev. C}\ }\textbf {\bibinfo {volume} {91}},\ \bibinfo {pages} {014903}
  (\bibinfo {year} {2015})},\ \Eprint {http://arxiv.org/abs/1307.6130}
  {arXiv:1307.6130 [nucl-th]} \BibitemShut {NoStop}%
\bibitem [{\citenamefont {Behtash}\ \emph {et~al.}(2019)\citenamefont
  {Behtash}, \citenamefont {Kamata}, \citenamefont {Martinez},\ and\
  \citenamefont {Shi}}]{Behtash:2019txb}%
  \BibitemOpen
  \bibfield  {author} {\bibinfo {author} {\bibfnamefont {A.}~\bibnamefont
  {Behtash}}, \bibinfo {author} {\bibfnamefont {S.}~\bibnamefont {Kamata}},
  \bibinfo {author} {\bibfnamefont {M.}~\bibnamefont {Martinez}}, \ and\
  \bibinfo {author} {\bibfnamefont {H.}~\bibnamefont {Shi}},\ }\href {\doibase
  10.1103/PhysRevD.99.116012} {\bibfield  {journal} {\bibinfo  {journal} {Phys.
  Rev. D}\ }\textbf {\bibinfo {volume} {99}},\ \bibinfo {pages} {116012}
  (\bibinfo {year} {2019})},\ \Eprint {http://arxiv.org/abs/1901.08632}
  {arXiv:1901.08632 [hep-th]} \BibitemShut {NoStop}%
\bibitem [{\citenamefont {Zaitsev}\ and\ \citenamefont
  {Polyanin}(2002)}]{zaitsev2002handbook}%
  \BibitemOpen
  \bibfield  {author} {\bibinfo {author} {\bibfnamefont {V.~F.}\ \bibnamefont
  {Zaitsev}}\ and\ \bibinfo {author} {\bibfnamefont {A.~D.}\ \bibnamefont
  {Polyanin}},\ }\href@noop {} {\emph {\bibinfo {title} {Handbook of exact
  solutions for ordinary differential equations}}}\ (\bibinfo  {publisher} {CRC
  press},\ \bibinfo {year} {2002})\BibitemShut {NoStop}%
\bibitem [{\citenamefont {Panayotounakos}(2005)}]{panayotounakos2005exact}%
  \BibitemOpen
  \bibfield  {author} {\bibinfo {author} {\bibfnamefont {D.~E.}\ \bibnamefont
  {Panayotounakos}},\ }\href@noop {} {\bibfield  {journal} {\bibinfo  {journal}
  {Applied mathematics letters}\ }\textbf {\bibinfo {volume} {18}},\ \bibinfo
  {pages} {155} (\bibinfo {year} {2005})}\BibitemShut {NoStop}%
\bibitem [{\citenamefont {Wagner}\ \emph {et~al.}(2022)\citenamefont {Wagner},
  \citenamefont {Palermo},\ and\ \citenamefont {Ambru\c{s}}}]{Wagner:2022ayd}%
  \BibitemOpen
  \bibfield  {author} {\bibinfo {author} {\bibfnamefont {D.}~\bibnamefont
  {Wagner}}, \bibinfo {author} {\bibfnamefont {A.}~\bibnamefont {Palermo}}, \
  and\ \bibinfo {author} {\bibfnamefont {V.~E.}\ \bibnamefont {Ambru\c{s}}},\
  }\href {\doibase 10.1103/PhysRevD.106.016013} {\bibfield  {journal} {\bibinfo
   {journal} {Phys. Rev. D}\ }\textbf {\bibinfo {volume} {106}},\ \bibinfo
  {pages} {016013} (\bibinfo {year} {2022})},\ \Eprint
  {http://arxiv.org/abs/2203.12608} {arXiv:2203.12608 [nucl-th]} \BibitemShut
  {NoStop}%
\bibitem [{\citenamefont {Hilbert}(1912)}]{hilbert1912begrundung}%
  \BibitemOpen
  \bibfield  {author} {\bibinfo {author} {\bibfnamefont {D.}~\bibnamefont
  {Hilbert}},\ }\href@noop {} {\bibfield  {journal} {\bibinfo  {journal}
  {Mathematische Annalen}\ }\textbf {\bibinfo {volume} {72}},\ \bibinfo {pages}
  {562} (\bibinfo {year} {1912})}\BibitemShut {NoStop}%
\bibitem [{\citenamefont {Grad}(1949)}]{grad:1949kinetic}%
  \BibitemOpen
  \bibfield  {author} {\bibinfo {author} {\bibfnamefont {H.}~\bibnamefont
  {Grad}},\ }\href@noop {} {\bibfield  {journal} {\bibinfo  {journal}
  {Communications on pure and applied mathematics}\ }\textbf {\bibinfo {volume}
  {2}},\ \bibinfo {pages} {331} (\bibinfo {year} {1949})}\BibitemShut {NoStop}%
\bibitem [{\citenamefont {Cercignani}(1990)}]{cercignani:90mathematical}%
  \BibitemOpen
  \bibfield  {author} {\bibinfo {author} {\bibfnamefont {C.}~\bibnamefont
  {Cercignani}},\ }\href@noop {} {\emph {\bibinfo {title} {Mathematical methods
  in kinetic theory}}}\ (\bibinfo  {publisher} {Springer},\ \bibinfo {year}
  {1990})\BibitemShut {NoStop}%
\bibitem [{\citenamefont {Grad}(1958)}]{grad1958principles}%
  \BibitemOpen
  \bibfield  {author} {\bibinfo {author} {\bibfnamefont {H.}~\bibnamefont
  {Grad}},\ }in\ \href@noop {} {\emph {\bibinfo {booktitle} {Thermodynamik der
  Gase/Thermodynamics of Gases}}}\ (\bibinfo  {publisher} {Springer},\ \bibinfo
  {year} {1958})\ pp.\ \bibinfo {pages} {205--294}\BibitemShut {NoStop}%
\bibitem [{\citenamefont {Shokri}\ and\ \citenamefont
  {Taghinavaz}(2020)}]{Shokri:2020cxa}%
  \BibitemOpen
  \bibfield  {author} {\bibinfo {author} {\bibfnamefont {M.}~\bibnamefont
  {Shokri}}\ and\ \bibinfo {author} {\bibfnamefont {F.}~\bibnamefont
  {Taghinavaz}},\ }\href {\doibase 10.1103/PhysRevD.102.036022} {\bibfield
  {journal} {\bibinfo  {journal} {Phys. Rev. D}\ }\textbf {\bibinfo {volume}
  {102}},\ \bibinfo {pages} {036022} (\bibinfo {year} {2020})},\ \Eprint
  {http://arxiv.org/abs/2002.04719} {arXiv:2002.04719 [hep-th]} \BibitemShut
  {NoStop}%
\end{thebibliography}%

\end{document}